\documentclass[preprint,floats,aps,12pt,superscriptaddress,nofootinbib,floatfix]{revtex4}
\usepackage{natbib} 
\usepackage{times} 
\usepackage{amssymb,amsmath}
\usepackage{csquotes}
\usepackage{prelim2e}\usepackage[none,bottom]{draftcopy}
\draftcopyName{preprint / }{1.3} %ADAPT TEXT
\ifx\pdfoutput\undefined
\usepackage[usenames,dvips]{color}
\usepackage{graphicx}     % Include figure files
\else
\usepackage[usenames,dvipsnames]{color}
\usepackage{epstopdf}
\usepackage[pdftex]{graphicx}
\usepackage[pdftex]{hyperref}
\hypersetup{a4paper,
pdfproducer={lateX},
pdfview=FitV,       % FitH
pdfstartview=FitB,
linkcolor=blue,     % links to same page
citecolor=blue,     % citations
urlcolor=red,      % links to URLs
breaklinks=true,    % links may be split onto 2 lines
colorlinks=true,
citebordercolor=0 0 0,  % color for \cite
filebordercolor=0 0 0,
linkbordercolor=0 0 0,
menubordercolor=0 0 0,
urlbordercolor=0 0 0,
pdfhighlight=/I,
pdfborder=0 0 0,   % no box around links
bookmarksopen=true,
bookmarksnumbered=true
}
\fi
\ifx\pdfoutput\undefined
\DeclareGraphicsExtensions{.eps}
\else
\DeclareGraphicsExtensions{.jpg, .pdf, .tif, .png}
\fi
\graphicspath{{.},{./Figures/}} %$ %ADAPT DIRECTORY
\usepackage[usenames,dvipsnames]{color}
\usepackage[normalem]{ulem}

\begin{document}
%
%----------------------------------------------------------------%
\title{Bifurcation study for a surface-acoustic-wave driven meniscus}
\author{Kevin David Joachim Mitas}
\email{kevin.mitas@uni-muenster.de}
\affiliation{Institut f\"ur Theoretische Physik, Westf\"alische Wilhelms-Universit\"at M\"unster, Wilhelm-Klemm-Strasse\ 9, 48149 M\"unster, Germany}
\author{Ofer Manor}
\affiliation{Wolfson Department of Chemical Engineering, Technion - Israel Institute of Technology, Haifa, Israel 32000}
\email{manoro@technion.ac.il}
\thanks{ORCID ID: 0000-0003-1526-5266}
\author{Uwe Thiele}
\email{u.thiele@uni-muenster.de}
\homepage{http://www.uwethiele.de}
\thanks{ORCID ID: 0000-0001-7989-9271}
\affiliation{Institut f\"ur Theoretische Physik, Westf\"alische Wilhelms-Universit\"at M\"unster, Wilhelm-Klemm-Strasse\ 9, 48149 M\"unster, Germany}
\affiliation{Center of Nonlinear Science (CeNoS), Westf{\"a}lische Wilhelms-Universit\"at M\"unster, Corrensstr.\ 2, 48149 M\"unster, Germany}
\begin{abstract}
A thin-film model for a meniscus driven by Rayleigh surface acoustic waves (SAW) is analysed, a problem closely related to the classical Landau-Levich or dragged-film problem where a plate is withdrawn at constant speed from a bath. We consider a mesoscopic hydrodynamic model for a partially wetting liquid, were wettability is incorporated via a Derjaguin (or disjoining) pressure and combine SAW driving with the elements known from the dragged-film problem.
For a one-dimensional substrate, i.e., neglecting transverse perturbations, we employ numerical path continuation to investigate in detail how the various occurring steady and time-periodic states depend on relevant control parameters like the Weber number and SAW strength. The bifurcation structure related to qualitative transitions caused by the SAW is analysed with particular attention on the {appearance and interplay of Hopf bifurcations where branches of time-periodic states emerge. The latter correspond to the regular shedding of liquid ridges from the meniscus. The obtained information is relevant to the entire class of dragged-film problems.}
\end{abstract}
\maketitle
\section{Introduction}
\label{sec:intro}
Over recent years, the interest of industry has grown in applications that control the deposition of a surface coating via the transfer of simple or complex liquids from a bath. Examples are dip-coating processes and Langmuir-Blodgett transfer processes of liquids, solutions and suspensions \cite{WeRu2004arfm,JiAM2005pf,SADF2007jfm,MRRQ2011jcis,WTGK2015mmnp,GLFD2016jfm}.

The transfer of a simple viscous liquid from a bath onto a continuously withdrawn plate is an important reference case for all investigations of coating phenomena. Where the bath contacts the plate, a dynamic liquid meniscus is formed resulting from the acting forces of capillarity due to surface tension, the viscous drag force of liquid advection and the wettability of the plate. The latter results from the interaction of the liquid, the solid and the ambient gas at the three-phase contact line.
Overall, {the transfer of liquid from a bath to a moving plate} is known as the Landau-Levich problem as it was first theoretically analysed by Landau and Levich \cite{LaLe1942apu} and Derjaguin \cite{Derj1945apu} in the case of a simple ideally wetting nonvolatile liquid. They show, that a homogeneous macroscopic liquid layer is deposited on the plate whose thickness $h_c$ depends on the capillary number $\mathrm{Ca}= \mu v_p / \gamma$ in the form of a power law, $h_c\propto \mathrm{Ca}^{2/3}$. Here, $v_p$ is the velocity of the plate withdrawal, and $\mu$ and $\gamma$ are the viscosity and surface tension of the liquid, respectively. For partially wetting liquids, a transition from a meniscus state where a macroscopically dry plate emerges from the bath to the Landau-Levich regime, {where a macroscopic liquid film is deposited on the plate (Landau-Levich state),} occurs at a finite plate velocity \cite{SADF2007jfm,ziegler2009film,GTLT2014prl,GLFD2016jfm}. Several different modi of transition are described and are classified as dynamic unbinding transitions \cite{GTLT2014prl}. In the course of these transitions beside meniscus states and Landau-Levich states, so-called foot states may occur, {where a steady finite-length foot-like protrusion covers part of the moving plate \cite{SADF2007jfm,ziegler2009film,GTLT2014prl,GLFD2016jfm,BeZu2008jfm}.} Time-periodic states corresponding to the periodic shedding of liquid ridges, oriented orthogonally to the withdrawal direction, from the liquid meniscus are also described \cite{TWGT2019prf}, similar to line deposition occurring in Langmuir-Blodgett transfer \cite{koepf2010pattern,KGFT2012njp,KoTh2014n}. Here, we also refer to this phenomenon as line deposition.

The classic Landau-Levich or dragged-plate system is a prominent example of a wide class of related coating systems where a meniscus deforms and film deposition occurs under lateral driving forces. {For instance, plate withdrawal may be replaced by a temperature gradient along the plate \cite{CHTC1990n,CaCa1993jcis,Muen2003prl,MuEv2005pd,BSBB2005pd,BMFC1998prl,ScCa2000lb,Wils1995jem},} by a pressure gradient when propelling an air bubble through a tube (Landau-Levich-Bretherton problem) \cite{bretherton1961motion,KSPK2018prf,HoZM2020prf} and by Rayleigh surface acoustic waves (SAWs), which propagate in a resting plate in contact with a bath of liquid \cite{RMYF2014prsapes,AlMa2016pf}. Here, we are particularly interested in the latter. {Note that
the Landau-Levich-Bretherton problem is also reconsidered under the influence of SAW \cite{HoMM2017pre,HoZM2020prf}.}

Briefly, simple vibrations, flexural bulk waves, and Rayleigh surface acoustic waves (SAWs) in solid substrates are capable of displacing, manipulating, and deforming liquid films and drops {situated on the substrate} \cite{biwersi2000displacement,alzuaga2005motion,matveyRev}. In particular, Rayleigh SAWs, which are of MHz frequency and propagate in a solid substrate, may cause the dynamic wetting of the solid surface by liquid films. This was first shown by Rezk et al.~\cite{RMYF2014prsapes,RMFY2012nc} in the case of silicon oil. They named the phenomenon ``acoustowetting''. It was thoroughly explained in a later study \cite{AlMa2015pf}. Similar observations are made in the case of water, albeit there capillary stresses compete with and even diminish the SAW-induced dynamic wetting effect \cite{MRFY2015pre,AlMa2016pf}. Generally, the direction of the SAW-induced dynamic wetting is determined by the thickness of the liquid film and the wavelength of the SAW. This is consistent with an interplay between radiation pressure and acoustic flow in oil films. Liquid film thicknesses, which are small compared to the wavelength of sound leakage {from the SAW into the liquid,} dynamically wet the solid substrate along the propagation direction of the SAW. The wavelength of the sound leakage is approximately the wavelength of the SAW times the ratio between the phase velocities of the SAW and the velocity of sound in the liquid. It is in the range of tens to hundreds of microns in the case of MHz-frequency SAWs. If the thickness of the liquid film is comparable to the wavelength of the sound leakage, acoustic resonance effects are found to enhance the pressure in the liquid film, so that the film may even spread opposite to the direction of the SAW. In addition, when the thickness of the liquid film is greater than a few wavelengths of the sound leakage, a different mechanism, Eckart streaming, governs transport in the liquid. This again yields dynamic wetting along the direction of the SAW. The capillary stress at the free surface of the liquid is capable of arresting the dynamic wetting process, especially, in the case of water. The opposite contributions of capillarity and SAW to film dynamics is quantitatively weighed in an acoustic Weber number, which governs the onset of dynamic wetting and is further discussed in this paper.

Results obtained for these Landau-Levich-type systems are also of relevance for other, possibly more complicated, coating geometries \cite{CaKh2000aj,WeRu2004arfm,Wilczek_2016}. Furthermore, they provide a basis for the understanding of the behaviour of more complex liquids in similar settings \cite{CrMa2009rmp,ASKS2008jem,RKKF2010jcis,SDDR2010el,Thie2014acis,WTGK2015mmnp,abo2018connecting}. Often, Landau-Levich-type systems are modelled based on mesoscopic hydrodynamic thin-film models (also called long-wave or lubrication models) \cite{CrMa2009rmp,OrDB1997rmp}. It is assumed that length scales orthogonal to the substrate are small compared to the ones parallel to the substrate, e.g., cases of small contact angles and small interface slopes are considered. This allows for the derivation of an evolution equation for the film thickness profile. Thereby the motion of three-phase contact lines is either modeled through a slip model \cite{SADF2007jfm,ziegler2009film} or a precursor film model \cite{GTLT2014prl,TWGT2019prf}. Here, we employ the latter, i.e., wettability is modelled by an additional pressure term, the so-called Derjaguin (or disjoining) pressure that allows for a precursor film or adsorption layer on the macroscopically dry substrate \cite{de1985wetting,derjaguin1975untersuchungen,Thie2007,StVe2009jpm}.

Employing such a precursor film model, Ref.~\cite{GTLT2014prl} shows that depending on plate inclination angle (and equilibrium contact angle), different transition scenarios occur with increasing capillary number (dimensionless plate velocity). These are, in particular, four different continuous and discontinuous unbinding (or dynamic wetting and emptying) transitions that are out-of-equilibrium equivalents of equilibrium transitions, which were discussed earlier (see \cite{GTLT2014prl} and references therein). 

Here, we consider how the picture is amended when employing SAW driving to force the meniscus, instead of driving by dragging a plate. {SAW driving for this purpose was recently investigated experimentally \cite{RMFY2012nc,RMYF2014prsapes,AlMa2015pf,AlMa2016pf,MZAM2016l} and theoretically \cite{MRFY2015pre,AlMa2015pf,AlMa2016pf,MoMa2017jfm}. In all considered cases, a drop of liquid or a liquid meniscus rests on a smooth solid plate. When a propagating SAW in the MHz-frequency range is excited within the plate, the propagating high-frequency substrate vibration induces static menisci to deform and give rise to a spreading film \cite{RMFY2012nc,AlMa2015pf} or to become unstable with respect to fingering where liquid protrusions spread from the meniscus \cite{RMFY2012nc}. Individual small drops on the substrate slide and deform \cite{AlMa2016pf}. Experiments are performed with simple liquids \cite{RMFY2012nc}, simple liquids with surfactant (to control capillarity and wettability) \cite{AlMa2015pf,AlMa2016pf} and solutions \cite{MZAM2016l}. In the latter case, the SAW influence the deposition of line patterns of solute from the moving contact line. Various thin-film models are developed in \cite{MRFY2015pre,AlMa2015pf,AlMa2016pf,MoMa2017jfm}. They mainly differ in the form of the resulting effective SAW driving term and the particular further physical effects that are taken into account.} 

{Among other mechanisms, all experimentally observed effects} are caused by a high-frequency periodic flow in a boundary layer close to the substrate within the liquid that induces a mean convective flow on hydrodynamic length and time scales. This flow may overcome the resistence of capillarity and wettability in the contact line region where the meniscus meets the plate. Then, at a critical strength, the acoustically driven flow results in an advance of the contact line. This is very similar to the transitions, which occurs at critical capillary numbers in the Landau-Levich problem for partially wetting liquids. It remains an open question whether SAW driving can be employed to control deposition of structured coatings, which may be a simpler coating technique with no moving parts, compared to driving the coating film by moving the plate \cite{MZAM2016l,ly2019effects}.

In particular, here we combine the precursor-based thin-film model for the Landau-Levich problem for partially wetting liquids in Refs.~\cite{GTLT2014prl,TsGT2014epje,TWGT2019prf} with the thin-film model for the SAW-driven meniscus {of an ideally wetting liquid} in Ref.~\cite{MoMa2017jfm}. With other words, the SAW model of \cite{MoMa2017jfm} is expanded by introducing lateral forces due to the withdrawal of an inclined plate and substrate wettability. This allows us to investigate the behavior of the SAW-driven system in direct comparison to results obtained for the dragged-plate system. {The systems studied in Refs.~\cite{GTLT2014prl,TsGT2014epje,TWGT2019prf} and Ref.~\cite{MoMa2017jfm}, respectively,
provide reference cases for our study that focuses on a SAW-driven meniscus for partially wetting liquid. We chose a nondimensionalization and parametrization that allows us to discuss both reference cases as limiting cases.} The model is numerically investigated employing a combination of path continuation techniques \cite{KrauskopfOsingaGalan-Vioque2007,DWCD2014ccp,EGUW2019springer} bundled in the software package \textsc{pde2path} \cite{uecker2014pde2path,uecker2017hopf} and direct time simulations using the software package \textsc{oomph-lib} \cite{HeHa2006}. 

We consider a one-dimensional substrate, i.e., we neglect transverse perturbations, and investigate in detail how the various occurring steady and time-periodic states depend on relevant control parameters like the Weber number and SAW strength. {We analyze the full bifurcation structure of steady and time-periodic states and discuss the occurring qualitative transitions in the system behaviour as well as in the entire bifurcation structure. Particular attention is paid to Hopf bifurcations where time-periodic states emergence that correspond to the {regular shedding of liquid ridges} from the meniscus. Up to now they have received little attention in the literature on Landau-Levich-type systems \cite{TWGT2019prf}. The emergence and interplay of the Hopf bifurcations and emerging branches of  time-periodic states is investigated by tracking the bifurcations in selected parameter planes. A discussion of the special transition points (so called codimension-2 bifurcations) where the individual bifurcations emerge and disappear allows us to extract detailed information which is relevant to the entire class of dragged-film problems.}

The structure of our work is as follows: In section~\ref{sec:model}, we briefly introduce the governing mathematical model, provide a short overview of the employed continuation method, and discuss the used solution measures. {Next, in section~\ref{sec:draggedfilmwithouSAW} we reproduce selected bifurcation diagrams for the classical Landau-Levich system \cite{GTLT2014prl}. Such bifurcation diagrams are presented in section~\ref{sec:dependenceonthewebernumber} for SAW-driven menisci of completely wetting liquids~\cite{MoMa2017jfm}. This provides us with the two central reference cases for our study. Subsequently, in section~\ref{sec:dependenceonSAWstrength} we consider SAW-driven menisci of partially wetting liquids,} first focusing on steady states and their bifurcations. The behavior of the occurring Hopf bifurcations is analyzed in detail in section~\ref{sec:hopfbranches} together with properties of the emerging branches of stable and unstable time-periodic states. Finally, we conclude in section~\ref{sec:conclusion} with a summary and outlook.
%
%%%%%%%%%%%%%%%%%%%%%%%%%%%%%%%%%%%%%%%%%%%%%%%%%%%%%%%%%%%%%%%%%%%%%%%%%%%%%%%%
\section{Model}
\label{sec:model}
%%%%%%%%%%%%%%%%%%%%%%%%%%%%%%%%%%%%%%%%%%%%%%%%%%%%%%%%%%%%%%%%%%%%%%%%%%%%%%%%
%
%%%%%%%%%%%%%%%%%%%%%%%%%%%%%%%%%%%%%%%%%%%%%%%%%%%%%%%%%%%%%%%%%%%%%%%%%%%%%%%%
%
\begin{figure}[hbt]
\center
\includegraphics[width=0.475\textwidth]{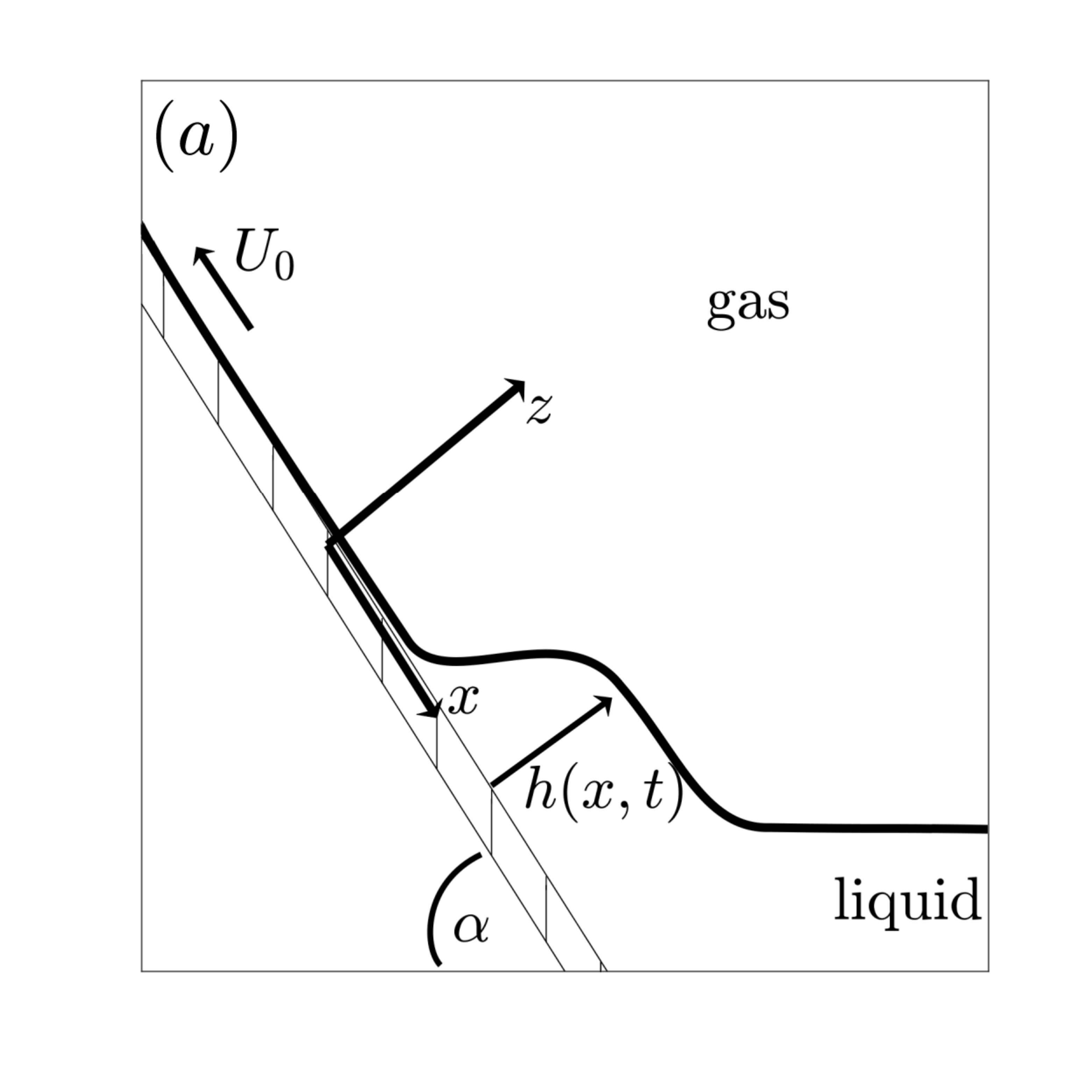}
\includegraphics[angle=0,width=0.475\textwidth]{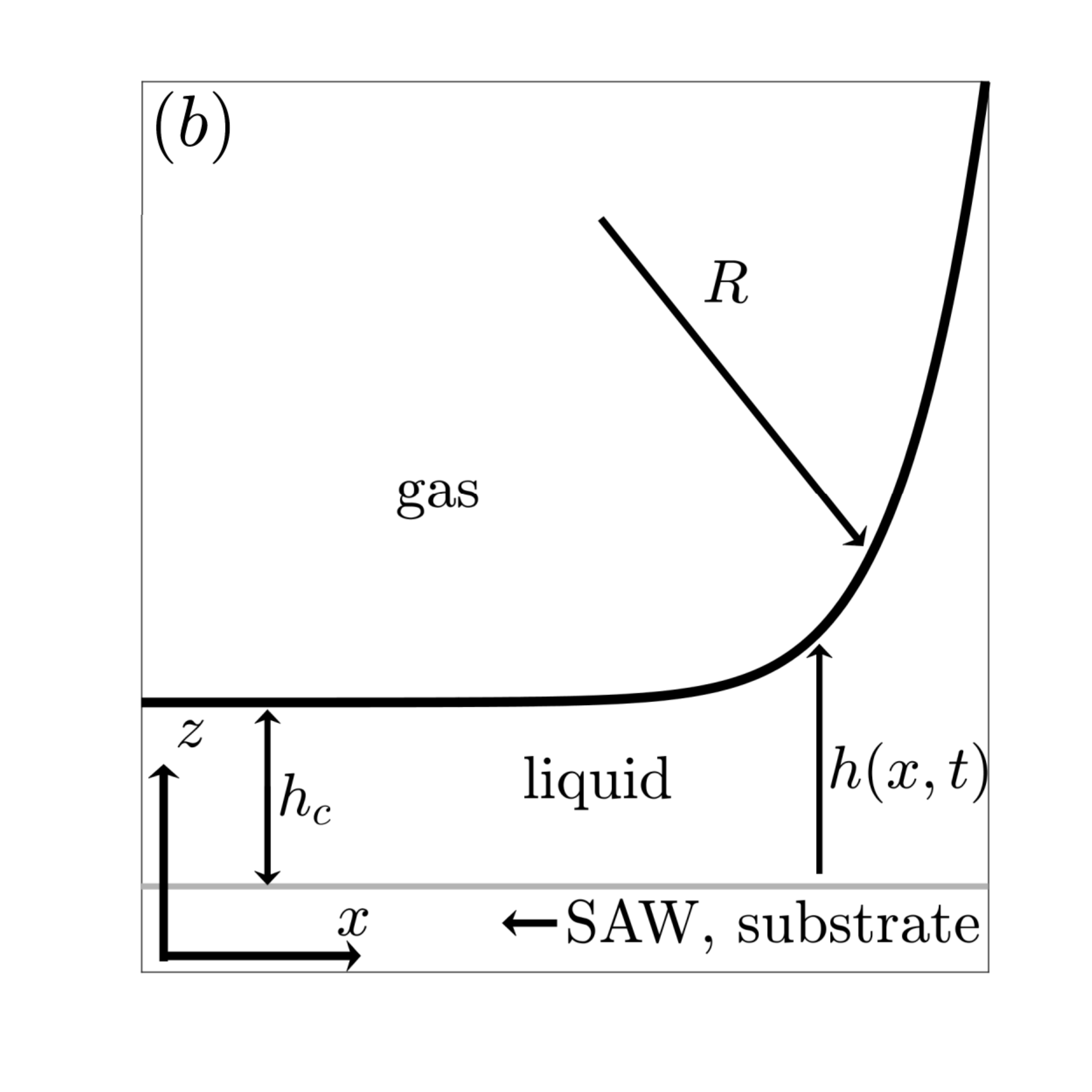}
\caption{Sketches of the meniscus regions of the two considered Landau-Levich-type systems: (a) in a dip-coating system liquid is deposited on a plate, which is withdrawn at velocity $U_0$ and angle $\alpha$ from a liquid bath; (b) in a surface acoustic wave (SAW)-driven system liquid is deposited from a liquid meniscus with radius of curvature $R$ onto a horizontal plate.}
\label{model:fig:Skizzen}
\end{figure}

In our study we combine the two Landau-Levich-type systems sketched in Fig.~\ref{model:fig:Skizzen}, namely, a simple nonvolatile partially wetting liquid in the classical Landau-Levich geometry of dip-coating \cite{GTLT2014prl,TWGT2019prf} and a liquid meniscus driven by surface acoustic waves (SAW) \cite{MoMa2017jfm}. Overall the set-up for the two systems is very similar with the main difference beside the driving force being the condition on the bath/meniscus side.
The kinetic equation that describes the development of the film thickness profile $h(x,t)$ when the two systems are combined reads in nondimensional form in the case of a one-dimensional substrate
\begin{align}
 \partial_t h(x,t)&=-\partial_x\left\{Q(h)\partial_x\left[\frac{1}{\mathrm{We}_\mathrm{s}}\partial_{xx} h-f'(h)\right]-\chi(h) 
 + h\left[\epsilon_\mathrm{s} v_{\mathrm{s}}(h) + U_0\right]\right\}, \label{eq:LL}\\
 \mathrm{where}\qquad\chi(h)&=G\,Q(h)\left(\partial_x h - \alpha\right),\qquad Q(h)=\frac{h^3}{3}.\nonumber
\end{align}
Above, $\alpha$ is the substrate inclination angle in long-wave scaling, $Q(h)$ is the mobility function resulting from a no-slip boundary condition at the substrate, $\mathrm{We}_\mathrm{s}$ is the Weber number, and $G$ is a dimensionless gravity number. The wetting potential \cite{Pism2001pre,Thie2010jpcm}
\begin{equation}
 f(h) = \mathrm{Ha} \left( \frac{h_p^3}{5 h^5} - \frac{1}{2 h^2}\right)
\label{eq:disjoin}
\end{equation}
describes wettability for a partially wetting liquid and results in the Derjaguin (or disjoining) pressure $\Pi=-f'(h)=\mathrm{Ha}(h_p^3/h^6-1/h^3)$ \cite{StVe2009jpm,Thie2010jpcm}. $\mathrm{Ha}$ is a nondimensional Hamaker number that controls the long-wave equilibrium contact angle. 

\begin{figure}[tbh]
\centering
\includegraphics[angle=0,width=0.6\textwidth]{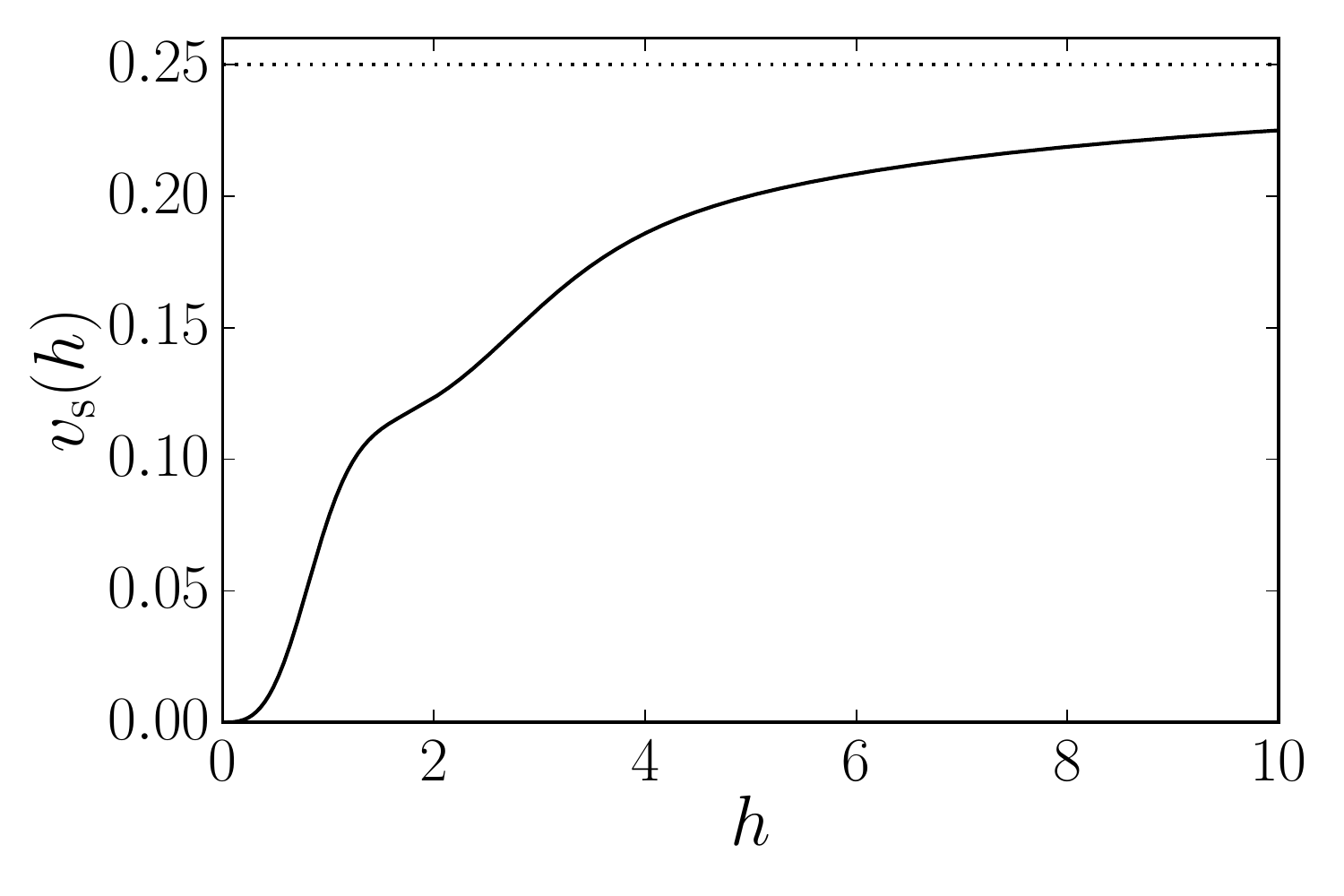}
\caption{Dependence of SAW driving $v_{\mathrm{s}}(h)$ on the thickness of the liquid film (solid line). The thin horizontal dotted line corresponds to the asymptotic value of $v_{\mathrm{s}}=1/4$ at large $h$.}
\label{ills:fig:different_SAW_term_alone_norm}
\end{figure}

In addition, capillarity and wettability give rise to the Laplace and Derjaguin pressure contributions in the first bracket of Eq.~(\ref{eq:LL}). The next term [$\chi(h)$] accounts for gravitational contributions that balance for the bath where $\partial_x h = \alpha$ and the final bracket gives the driving force for the coating process. It combines the effect of the plate moving at a velocity $U_0$ and of the SAW at an intensity of $\epsilon_\mathrm{s}$. The corresponding dependence of SAW driving on the thickness of the liquid films is \cite{MoMa2017jfm} 
\begin{eqnarray}
  v_{\mathrm{s}}(h) &=& \frac{1}{4 h} \left( \frac{h \sinh(2h)- h \sin(2h) + 2 \cos(h) \cosh(h)}{ \cos(2h) + \cosh(2h)}-1\right)  \label{ills:eq:SAW}\\
                    &\mathrel{\underset{h\to 0}{\approx}}& \frac{h^3}{8} +\mathcal{O}(h^{7})
                                \nonumber
\end{eqnarray}
which is illustrated in Fig.~\ref{ills:fig:different_SAW_term_alone_norm}. {Note that the thickness scale results from a property of the SAW driving (see Appendix~\ref{sec:appendix}).} The influence of SAW driving on the coating process increases with increasing film thickness and saturates in the limit of large film heights with $v_{\mathrm{s}}(h\to\infty)=1/4$. {The second line of Eq.~\eqref{ills:eq:SAW} shows that at small $h$ one has $v_{\mathrm{s}}(h)\sim h^3$, i.e., SAW driving and dragged-plate driving represent different functional dependencies at small $h$ but become similar at large $h$.}

Equation~(\ref{eq:LL}) is derived using the well-established lubrication approximation, which is also known as long-wave or thin-film approximation \cite{OrDB1997rmp,Thie2007,CrMa2009rmp}, i.e., it is assumed that all relevant length scales orthogonal to the substrate are small compared to the ones parallel to the substrate, e.g., one considers the case of small physical contact angle, small plate inclination angle and, in general, small interface slopes. {Note, that in consequence of long-wave scaling the corresponding quantities are of order one in the nondimensional long-wave model.}

The derivation starts from the Navier-Stokes and continuity equations with the usual stress-free conditions at the free surface of the liquid film and with no-slip and no-penetration conditions at the solid-liquid interface. In addition, the solid-liquid interface undergoes SAW-induced high-frequency travelling wave oscillations. This implies that fast and slow time-scales can be separated in a multi-scale approach. This results in the particular effective SAW driving term $v_{\mathrm{s}}$ that acts on the viscous time scale \cite{AlMa2015pf,AlMa2016pf,MoMa2017jfm}. A similar approach is employed for liquid film on vertically vibrated substrates where it is used to stabilize various interface instabilities \cite{LaMV2001pre,ThVK2006jfm}. 

{The employed scaling is discussed in Appendix~\ref{sec:appendix}. It is chosen because it combines aspects of the scalings employed in Refs.~\cite{MoMa2017jfm} and \cite{GTLT2014prl,TWGT2019prf} and allows us to use both, the Weber number $\mathrm{We}_\mathrm{s}$ and the SAW strength $\epsilon_\mathrm{s}$, as control parameters. This facilitates comparison to both limiting cases studied in the literature: (i) The model in Ref.~\cite{MoMa2017jfm} is recovered from Eq.~(\ref{eq:LL}) when neglecting wettability (Ha$=0$ in Eq.~(\ref{eq:disjoin})) and gravity ($G=0$). (ii) The model in Refs.~\cite{GTLT2014prl,TWGT2019prf} is recovered when neglecting the SAW driving ($\epsilon_\mathrm{s}=0$). The control parameters remaining in the limiting cases then coincide with the ones in the respective references.}

We set specific boundary conditions (BC) for $h(x,t)$ to investigate the dragged-plate and SAW-driven coating systems. We consider 
the domain $\Omega = [-L/2 , L/2]$, where
\begin{eqnarray}
 h &=& h_m, \quad\mathrm{and}\quad \partial_{xx} h = 1 \quad\mathrm{at}\quad x = L/2\\
 \partial_x h &=& 0 \quad\mathrm{and}\quad \partial_{xx} h = 0 \quad\mathrm{at}\quad x=  -L/2 \nonumber
 \label{num:eq:bc_dragged}
\end{eqnarray}
i.e., on the right hand side ($x = L/2$) the film profile approaches a meniscus of constant curvature exactly as in Ref.~\cite{MoMa2017jfm}. Note that the nondimensional radius of curvature is equal to one. For the dragged-plate system this BC is amended to $\partial_{xx} h|_{x=  L/2} = 0$ as the profile approaches a straight line on the bath side.
The other BC on the bath/meniscus side fixes the film thickness on the boundary of the numerical domain to a value $h_m$. {The value is not particularly important as it essentially pins the translational degree of freedom and controls the position of the contact line region within the numerical domain. The BC on the opposite side ($x = -L/2$) allow for a flat film of arbitrary $h_c$ -- the coating height. As the lubrication Eq.~\eqref{eq:LL} is parabolic and corresponds to a overdamped dynamics, no reflection of waves occurs even when inhomogeneous profiles, e.g., liquid ridges, cross the boundary. The effect of the particular chosen outflow boundary conditions on the profile is minor: when a ridge reaches the boundary its profile adjusts and directly on the boundary its slope becomes zero. The effect is very small and its relative importance can always be decreased by increasing the domain size. This has been further discussed in open flow systems involving sliding droplets, see e.g.~\cite{EnTh2019el}.}

We numerically investigate the model Eq.~(\ref{eq:LL}) with the introduced BC by employing a combination of pseudo-arclength path continuation \cite{KrauskopfOsingaGalan-Vioque2007,DWCD2014ccp,EGUW2019springer} as provided by the package \textsc{pde2path} \cite{uecker2014pde2path,uecker2017hopf} and of direct time simulations employing the package \textsc{oomph-lib} \cite{HeHa2006}.

Briefly, the continuation technique is employed to determine steady solutions ($\partial_t \phi= 0$) of Eq.~(\ref{eq:LL}). This is written as the operator $G[\phi, \lambda]=0$, which is discretized using a finite element scheme. Here, $\lambda$ stands for a control parameter or a set of them. One follows a discretized solution $\phi(x)$ in parameter space varying $\lambda$ using a prediction-correction scheme:  in the prediction step, one uses the tangent of the solution curve to advance from the known solution at parameter value $\lambda$ to a first guess of a solution at a new parameter value $\lambda + \Delta \lambda$, i.e., the control parameter is used as continuation parameter. Then, in the correction step a Newton procedure is employed to converge from the guess to a solution of the PDE at $\lambda + \Delta \lambda$. This procedure is iterated to advance step by step along a solution branch. However this simple 
continuation scheme fails when the solution branch undergoes a saddle-node bifurcation (or fold). Therefore one employs pseudo-arclength continuation, where the arclength along the branch is employed as continuation parameter while the original control parameter $\lambda$ is in the correction step adjusted at fixed arclength in parallel to the state $\phi$. The arclength itself is determined through an approximation, hence the name ``pseudo-arclength continuation''. This allows one to follow solution branches through saddle-node bifurcations. The package also allows one to detect various bifurcations, track them in parameter planes through two-parameter continuations, switch to bifurcating branches of other steady states or to branches of time-periodic states \cite{uecker2014pde2path,uecker2017hopf}. For other recent examples where path continuation is applied to thin-film and {closely related Cahn-Hilliard-type equations} see 
\cite{KoTh2014n,EWGT2016prf,LRTT2016pf,LTBK2018aml,TWGT2019prf,EGUW2019springer,TFEK2019njp,TSJT2020pre,FrWT2021pre}.

Here, we use continuation to investigate in detail how the various occurring steady and time-periodic states depend on relevant control parameters like the Weber number and SAW strength. The results are presented in terms of bifurcation diagrams employing as main solution measures (i) the thickness of the coating layer, i.e., the value of $h$ measured on the boundary at $x = -L/2$, and (ii) the excess volume $V_\mathrm{ex} = V - V_0$ of liquid dynamically extracted from the bath/meniscus. The volume is calculated as the integral $V = \int_\Omega h(x) \mathrm{d}x$ and $V_0$ is the reference volume without driving. In the case of time-periodic states the mean value of the measure over one period is shown and the period becomes an additional solution measure. 
%
%%%%%%%%%%%%%%%%%%%%%%%%%%%%%%%%%%%%%%%%%%%%%%%%%%%%%%%%%%%%%%%%%%%%%%%%%%%%%%%%
\section{Dragged film without SAW}
\label{sec:draggedfilmwithouSAW}
\begin{figure}[hbt]
\centering
    \includegraphics[width=0.45\textwidth]{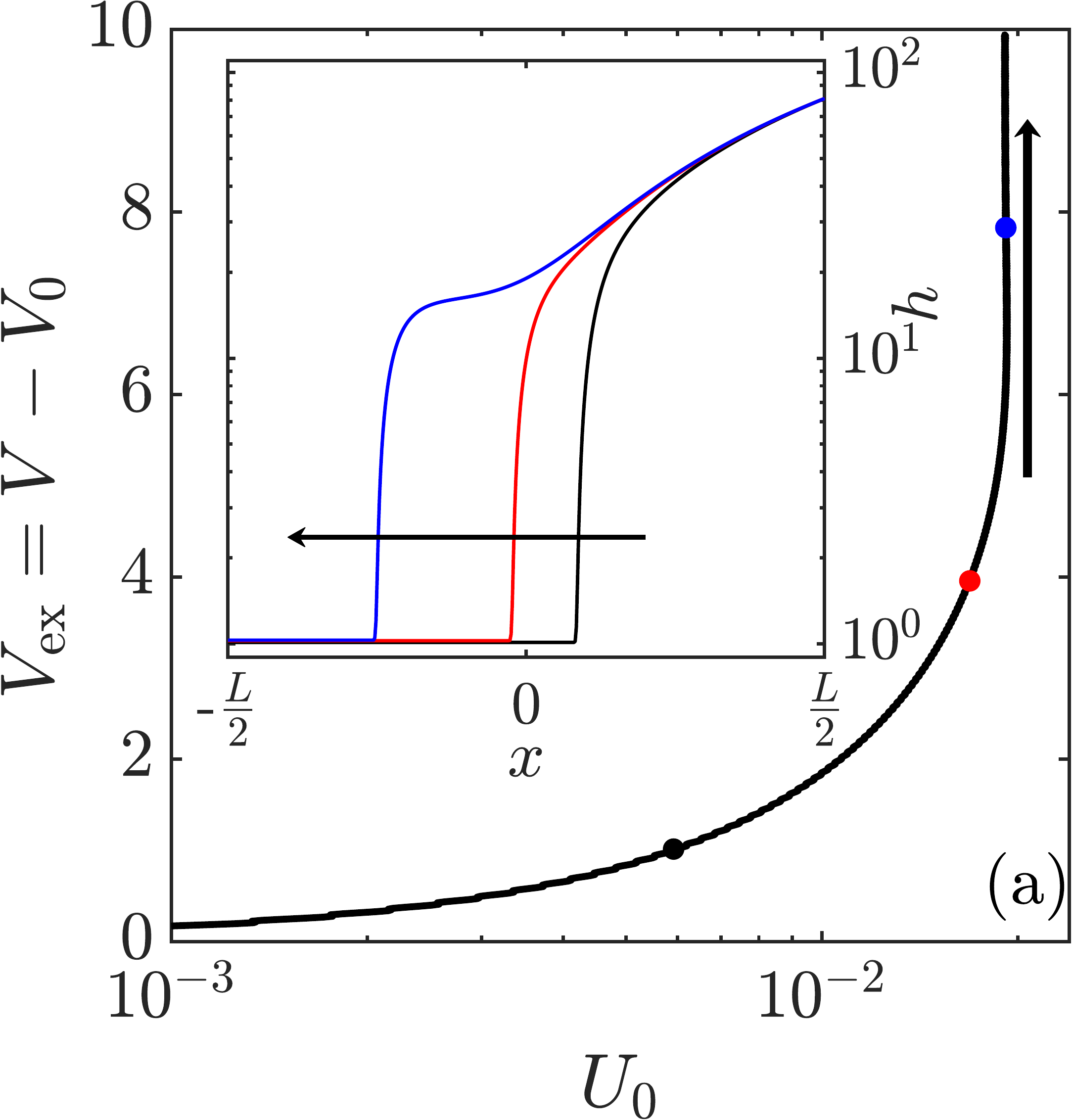}
    \includegraphics[width=0.45\textwidth]{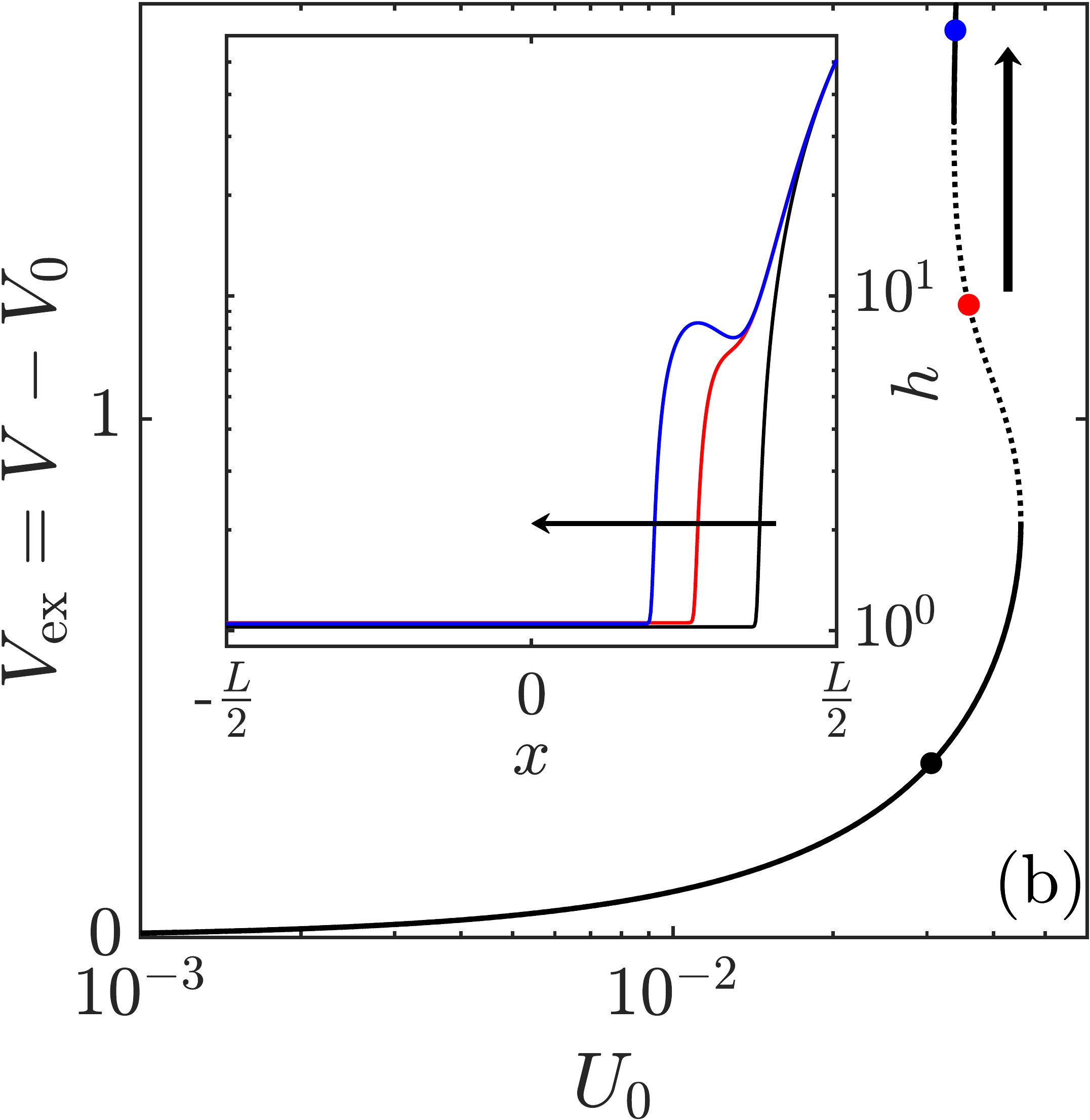}
    \includegraphics[width=0.45\textwidth]{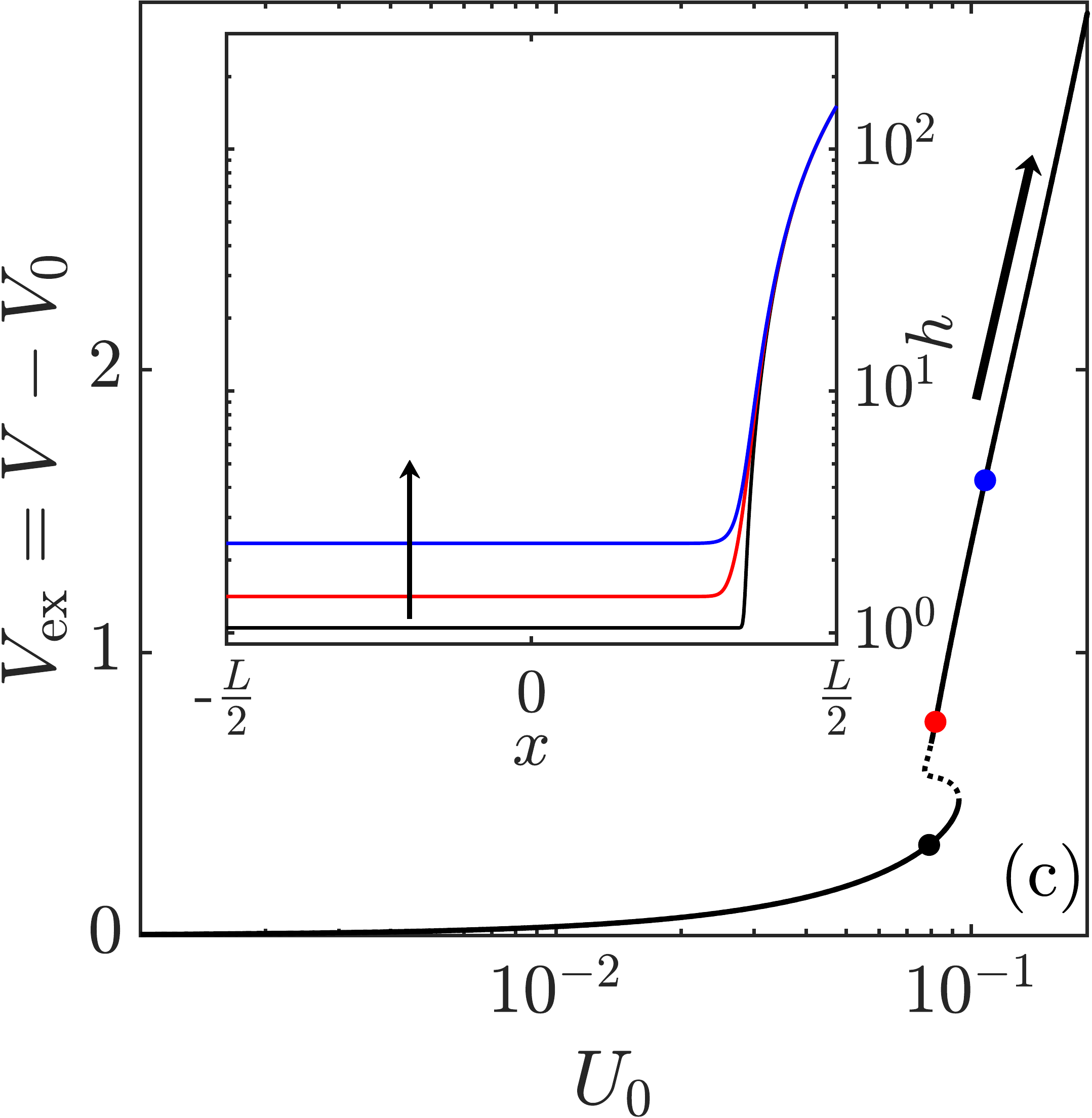}
    \includegraphics[width=0.45\textwidth]{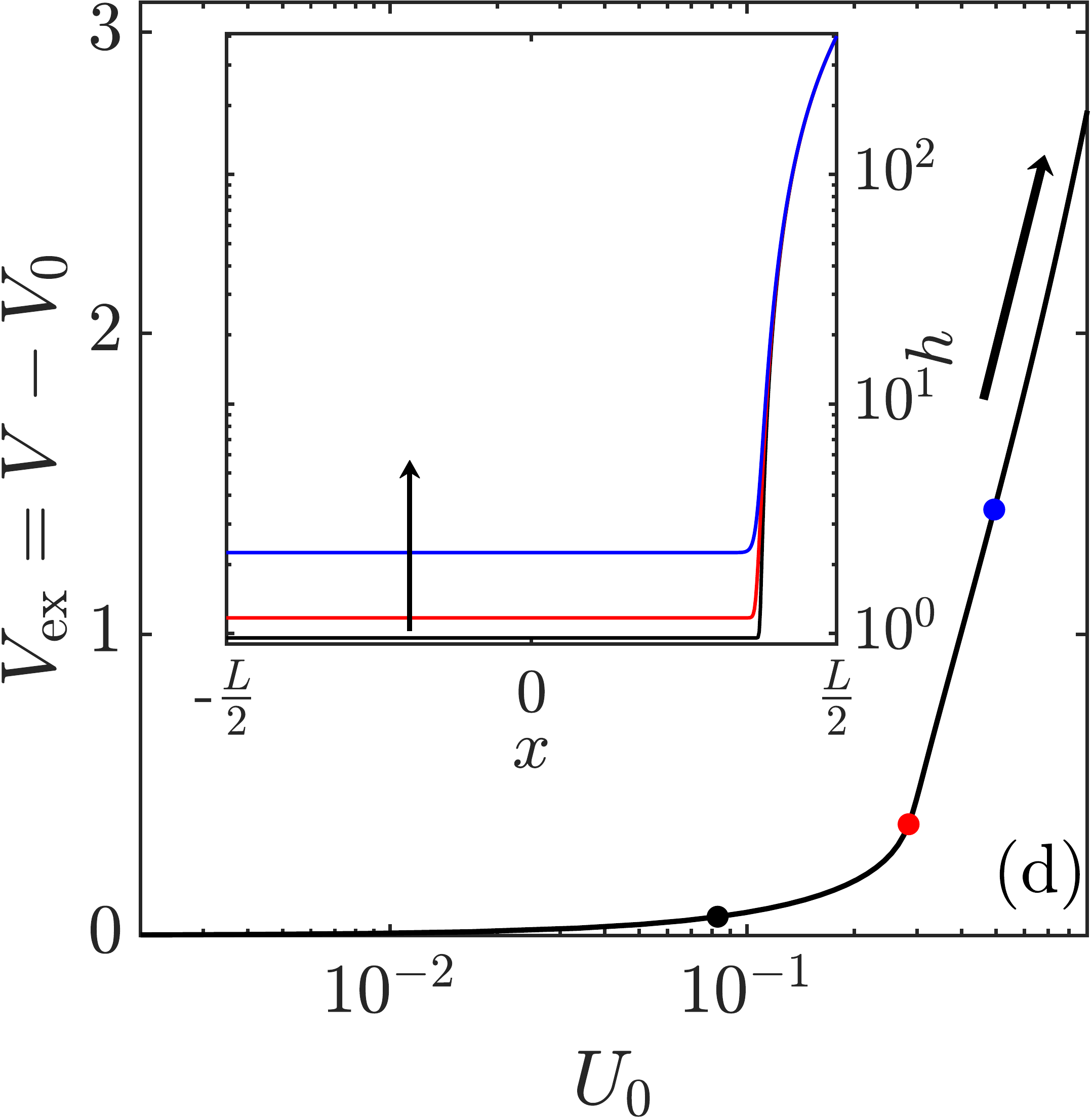}
\caption {Bifurcation diagrams of steady meniscus states in dependence of plate velocity $U_0$ for the {classical Landau-Levich system ($\epsilon_\mathrm{s} = 0$)} at different inclination angles (a) $\alpha=0.2$, (b) $\alpha=1.0$, (c) $\alpha=3.0$ and (d) $\alpha=10.0$ [Eq.~\eqref{eq:LL}]. The solution measure is the excess volume $V_\mathrm{ex}=V-V_0$. Dotted and solid lines represent unstable and stable states, respectively. The insets show examples of steady state profiles at loci indicated on the bifurcation curves by correspondingly coloured, filled circles. Note that the thickness scale is logarithmic. The arrows along the bifurcation curves and in the insets indicate corresponding directions of change. The remaining parameters are $\mathrm{Ha}=1.0$, $\mathrm{We}_\mathrm{s}= 1.0$, $G=0.001$, $h_p=1$, and $L=800$.
}

\label{dfws:fig:dragged_angles_}
\end{figure}

%%%%%%%%%%%%%%%%%%%%%%%%%%%%%%%%%%%%%%%%%%%%%%%%%%%%%%%%%%%%%%%%%%%%%%%%%%%%%%%%
%
Before we present our main results for the SAW-driven case, we first briefly review the transitions occurring for the classical Landau-Levich system (dip coating) \cite{SADF2007jfm,ziegler2009film,GTLT2014prl}. Employing Eq.~(\ref{eq:LL}) with $\mathrm{We}_\mathrm{s}=1$, $\mathrm{Ha}=1$ and $\epsilon_\mathrm{s} = 0$, we obtain the thin-film equation analysed in \cite{GTLT2014prl} with only a small change in the boundary conditions as discussed above. As explained in the Appendix~\ref{sec:appendix}, this is equivalent to the setting in Refs.~\cite{GTLT2014prl,TWGT2019prf}. The remaining main control parameters are the plate velocity $U_0$ and inclination angle $\alpha$. The geometry is as given in Fig.~\ref{model:fig:Skizzen}~(a). The bifurcation diagrams in the four main panels of Fig.~\ref{dfws:fig:dragged_angles_} indicate how the dependence of the meniscus state on $U_0$ changes with increasing $\alpha$. The insets give a selection of particular stable and unstable steady profiles.

In all cases, for $U_0\to0$ the steady meniscus approaches the reference state of zero excess volume. The profile shows a smooth continuous transition from the bath to the coating layer of thickness $h_c$. It nearly corresponds to the adsorption layer of equilibrium thickness  $h_c=h_p=1$ given by $\Pi(h)=0$. The behaviour for increasing $U_0$ qualitatively depends on the value of $\alpha$: Fig.~\ref{dfws:fig:dragged_angles_}~(a) shows that at small $\alpha$, the excess volume $V_\mathrm{ex}$ monotonically increases with $U_0$, first slowly than faster until it diverges at a critical value $U_{c_1} \approx 0.021$ of the velocity at $U_0$. All states on the branch are linearly stable. The steady meniscus profile deforms as the velocity increases and a foot-like structure of increasing length is dragged out of the bath with its length diverging for $U_0 \to U_{c_1}$ [see inset of Fig.~\ref{dfws:fig:dragged_angles_}~(a)]. For $U_0 > U_{c_1}$, time simulations show that a foot is continuously drawn out of the bath, and the system eventually settles on a Landau-Levich film state (not shown here). The transition at $U_{c_1}$ is termed ``dynamic continuous emptying transition'' \cite{GTLT2014prl} in analogy to the equilibrium emptying transition described in \cite{PRJA2012prl}.

The first qualitative change in the bifurcation curve occurs at $\alpha=\alpha_{c_1} \approx 0.5$, see Fig.~\ref{dfws:fig:dragged_angles_}~(b) for an example at $\alpha=1.0>\alpha_{c_1}$. As before, at 
small $U_0$ the excess volume monotonically increases with $U_0$. However, then at $U_0\approx0.5$ a saddle-node bifurcation occurs and the bifurcation curve continues towards smaller $U_0$ now consisting of  unstable states. At a second saddle-node bifurcation at $U_0 \approx 0.3$ the states becomes stable again, and the curve folds back towards larger $U_0$. Overall the curve undergoes an exponential (or collapsed) snaking \cite{ma2010defect} about a vertical asymptote at a critical velocity $U_0=U_{c_2} \approx 0.3$. We emphasize that at each fold there occurs a change in stability. This is analysed in detail in \cite{TsGT2014epje}. Looking at the inset of Fig.~\ref{dfws:fig:dragged_angles_}~(b), the steady profiles again develop a foot whose span now shows undulations. This is related to the snaking bifurcation curve. In consequence, only certain ranges of foot length correspond to stable states. The transition at $U_0=U_{c_2}$ is termed ``dynamic discontinuous emptying transition'' \cite{GTLT2014prl}. Note, that our results qualitatively agree with the ones of \cite{GTLT2014prl}, although the different BC and domain sizes result in slightly different critical values.

The second qualitative change in the bifurcation curve occurs at $\alpha=\alpha_{c_2} \approx 2.52$, see Fig.~\ref{dfws:fig:dragged_angles_}~(c) for an example at $\alpha=3.0>\alpha_{c_2}$. At 
small $U_0$ the curve behaves as before, then undergoes a single pair of saddle-node bifurcations; both destabilize the states, i.e., two eigenvalues have positive real parts after the second one. Further increasing $U_0$, the steady states become stable again at a Hopf bifurcation, where two complex conjugated eigenvalues cross the imaginary axis. This bifurcation was only recently described in Ref.~\cite{TWGT2019prf} and is discussed in detail for the SAW-driven system in section~\ref{sec:hopfbranches}. The branch of stable states continues to arbitrarily large $U_0$ following the classical Landau-Levich power law $V_\mathrm{ex} \propto U_0^{2/3}$. Note that in contrast to the cases shown in Fig.~\ref{dfws:fig:dragged_angles_}~(a) and (b), no asymptotic value of $U_0$ exists. An inspection of the steady profiles in the inset of Fig.~\ref{dfws:fig:dragged_angles_}~(c) shows that the transition does not involve an advancing foot. Instead, the coating thickness $h_c$ increases homogeneously. As the film surface after ``unbinding''  from the substrate  homogeneously increases similar to an equilibrium wetting transition, the transition is called a ``discontinuous dynamic wetting transition'' \cite{GTLT2014prl}. 

The final qualitative change in the bifurcation curve can be appreciated when comparing Fig.~\ref{dfws:fig:dragged_angles_}~(c) and Fig.~\ref{dfws:fig:dragged_angles_}~(d). In Fig.~\ref{dfws:fig:dragged_angles_}~(d) the pair of bifurcations has annihilated and the curve increases monotonically. Otherwise the behaviour is as in Fig.~\ref{dfws:fig:dragged_angles_}~(c) - also see the profiles in the inset. In consequence, this behaviour is termed ``continuous dynamic wetting transition'' \cite{GTLT2014prl}.

This brief overview of the classical Landau-Levich system in the case of partially wetting liquid provides our first reference case.

% %%%%%%%%%%%%%%%%%%%%%%%%%%%%%%%%%%%%%%%%%%%%%%%%%%%%%%%%%%%%%%%%%%%%%%%%%%%%%%%%%%%%%%
\section{SAW-driven meniscus}\label{sec:saw}
% %%%%%%%%%%%%%%%%%%%%%%%%%%%%%%%%%%%%%%%%%%%%%%%%%%%%%%%%%%%%%%%%%%%%%%%%%%%%%%%%%%%%%%
%
After the brief revision of the transition behaviour in the case of the classical Landau-Levich system, we next consider the SAW-driven system. {First, we investigate in section~\ref{sec:dependenceonthewebernumber} the case without wettability as in Ref.~\cite{MoMa2017jfm}, before considering partially wetting liquids in section~\ref{sec:dependenceonSAWstrength}.}

% %%%%%%%%%%%%%%%%%%%%%%%%%%%%%%%%%%%%%%%%%%%%%%%%%%%%%%%%%%%%%%%%%%%%%%%%%%%%%%%%%%%%%%
\subsection{Without wettability}
\label{sec:dependenceonthewebernumber}
% %%%%%%%%%%%%%%%%%%%%%%%%%%%%%%%%%%%%%%%%%%%%%%%%%%%%%%%%%%%%%%%%%%%%%%%%%%%%%%%%%%%%%%
%
We begin our analysis by reproducing the case studied in Ref.~\cite{MoMa2017jfm} where wettability and gravity are not considered and the meniscus connects to a resting horizontal plate. In other words, we consider Eq.~(\ref{eq:LL}) with $\mathrm{Ha} = U_0 = \alpha = G = 0$, fix $\epsilon_\mathrm{s}= 1$, and employ the  Weber number $\mathrm{We}_\mathrm{s}$ as control parameter. As explained in Appendix~\ref{sec:appendix}, this is equivalent to the setting in \cite{MoMa2017jfm}. There, steady profiles are obtained using a shooting method, while here we employ path continuation. 

\begin{figure}[hbt]
\centering
\includegraphics[angle=0,width=1.\textwidth]{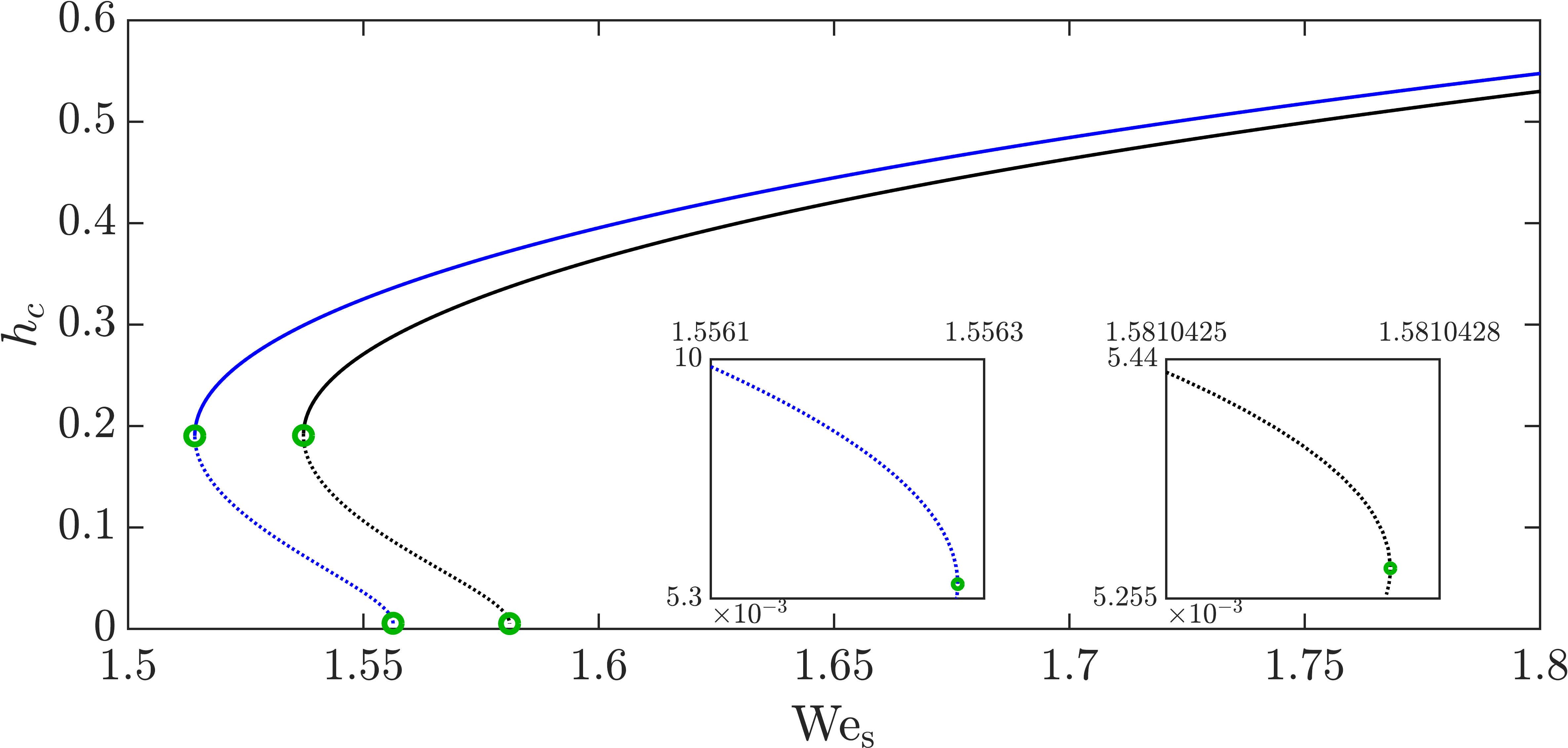}
\caption {Bifurcation diagram for a SAW-driven meniscus {without wettability} ($\mathrm{Ha}=0$) on a resting horizontal plate ($U_0 = \alpha = 0$), which shows the coating film thickness $h_c$ in dependence of the Weber number $\mathrm{We}_\mathrm{s}$ at fixed SAW strength ($\epsilon_\mathrm{s} =1$). The black and blue curve show cases without the contribution of gravity ($G=0$, as in Ref.~\cite{MoMa2017jfm}) and with the contribution of gravity ($G=10^{-3}$), respectively. Dotted and solid lines represent unstable and stable states, respectively. Saddle-node bifurcations (folds) are marked by green circles. The inset magnifies the region at very small $h_c$ where the saddle-node bifurcation occurs. The remaining parameters are $h_\text{m}= 8.1$ and $L = 40$.
}
\label{down:fig:momaplot}
\end{figure}

Resulting bifurcation diagrams with and without hydrostatic pressure are presented in Fig.~\ref{down:fig:momaplot}. The steady profiles on these branches correspond to a meniscus, which is smoothly and monotonically connected to a homogeneous coating layer. At large $\mathrm{We}_\mathrm{s}$, the film thickness increases monotonically following the power law $ h_c \propto \mathrm{We}_\mathrm{s}^{2/3}$ identical to the power law dependence on capillary number for a classical Landau-Levich film.

However, the behaviour of the coating system qualitatively changes at smaller $\mathrm{We}_\mathrm{s}$. The dependence of $h_c$ on $\mathrm{We}_\mathrm{s}$ becomes multivalued as the branch features two saddle-node bifurcations. The one occurring at very small $h_c$ was not detected in Ref.~\cite{MoMa2017jfm}. No film deposition occurs below $\mathrm{We}_\mathrm{s} \approx 1.54$ -- the threshold value where the crucial saddle-node bifurcation occurs. The meniscus profile then ends on the truly dry substrate, a situation not captured by the model in \cite{MoMa2017jfm} as it does not allow for slip at the contact line. 
When following the bifurcation curve from large to small values of $\mathrm{We}_\mathrm{s}$, one first observes a gradual decrease in the coating thickness. Then, passing the saddle-node bifurcation, the film deposition state collapses abruptly from the finite thickness $h_c\approx0.2$ (cf.~Fig.~\ref{down:fig:momaplot}). Incorporating gravity, the saddle-node bifurcation is shifted towards smaller values of $\mathrm{We}_\mathrm{s}$, but the corresponding $h_c$ is nearly constant.

Following the branch through the saddle-node bifurcation, a sub-branch of unstable solutions continues towards larger $\mathrm{We}_\mathrm{s}$ values, until turning back again (and becoming more unstable) at another saddle-node bifurcation at rather small $h_c$ values (see insets of Fig.~\ref{down:fig:momaplot}). The subsequent tiny sub-branch of unstable states ends when $h_c$ approaches zero at a critical Weber number. Note that this is not a bifurcation point. Mathematically, the alternative state of a meniscus ending at a true microscopic contact point is a finite support solution. It has a different topology than the Landau-Levich film state and can not be obtained with the numerical methods employed here. However, this situation can be amended by explicitly incorporating a description of wettability, as in the following section.
%
%%%%%%%%%%%%%%%%%%%%%%%%%%%%%%%%%%%%%%%%%%%%%%%%%%%%%%%%%%%%%%%%%%%%%%%%%%%%%%%%%%%%%%
\subsection{Partially wetting liquid}
\label{sec:dependenceonSAWstrength}
% %%%%%%%%%%%%%%%%%%%%%%%%%%%%%%%%%%%%%%%%%%%%%%%%%%%%%%%%%%%%%%%%%%%%%%%%%%%%%%%%%%%%%%
%
\begin{figure}[hbt]
\centering
\includegraphics[angle=0,width=1.\textwidth]{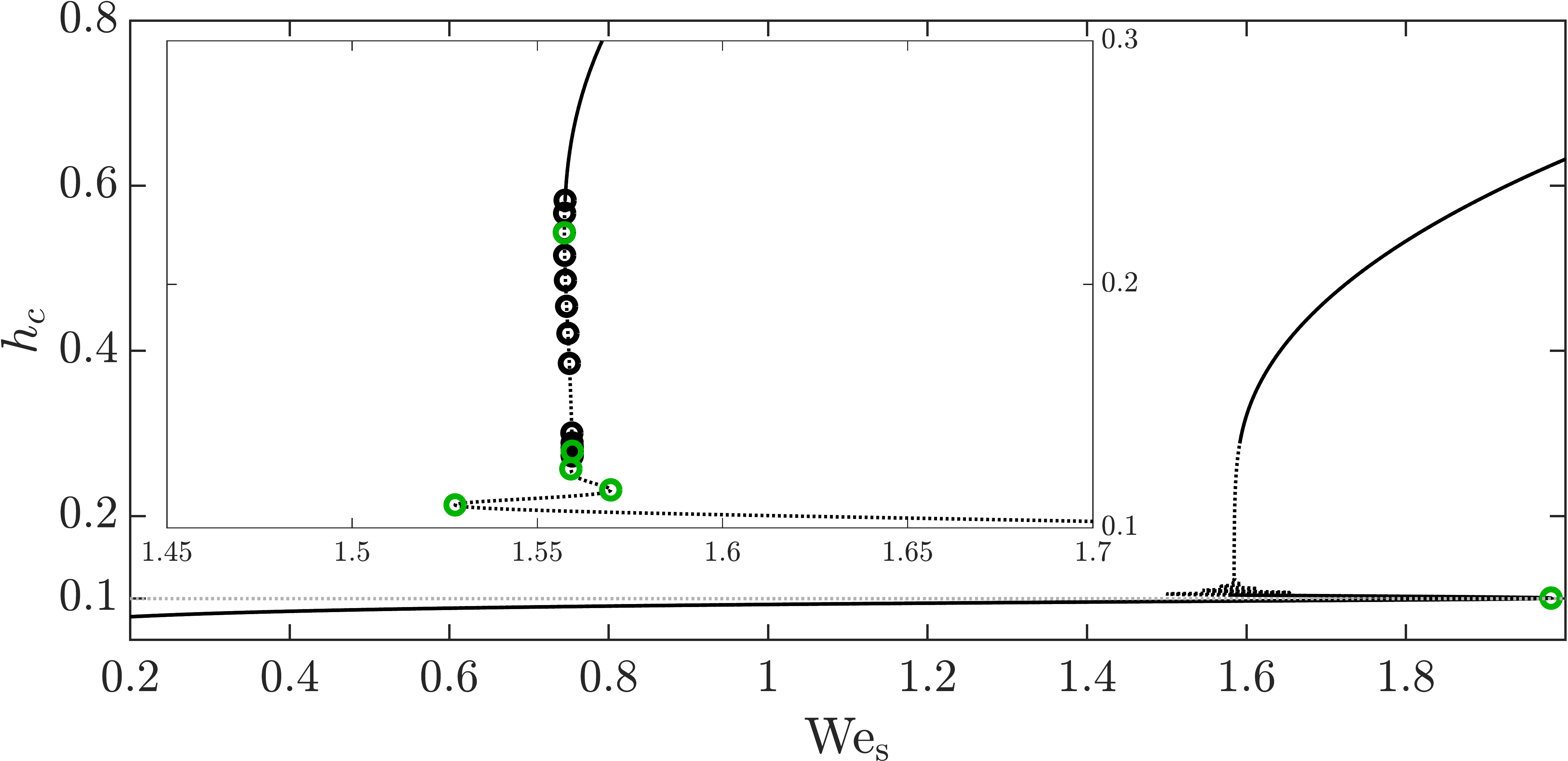}
\caption{
Bifurcation diagram for a SAW-driven meniscus of partially wetting liquid on a resting plate ($U_0 = 0$). Shown is the coating thickness $h_c$ in dependence of the Weber number $\mathrm{We}_\mathrm{s}$ at fixed SAW strength $\epsilon_\mathrm{s} =1$. Dotted and solid lines represent unstable and stable states, respectively. Saddle-node and Hopf bifurcations are marked by green and black circles, respectively. The inset zooms onto the region where the curve undergoes ``snaking''. The remaining parameters are $h_\text{m}= 8.1$, $G = 10^{-3}$, $\alpha=0.2$, $\mathrm{Ha}=0.002$, $h_p = 0.1$ and $L = 40$.
}
\label{down:fig:ha2_we2_hopffold_eps1_hp}
\end{figure}

{Next, we consider the full evolution equation~(\ref{eq:LL}), while accounting for partially wetting liquids 
but keep the substrate at rest and nearly horizontal.
\footnote{A truly horizontal substrate gives very similar results but renders details of the continuation procedure more cumbersome.} In this way, we consider the influence of partial wettability on the results in \cite{MoMa2017jfm} using $\mathrm{We}_\mathrm{s}$ as control parameter.} Inspecting the bifurcation diagram in Fig.~\ref{down:fig:ha2_we2_hopffold_eps1_hp}, we notice that the incorporation of partial wettability results in strong changes. For partially wetting liquids described with a wetting energy, the macroscopically dry substrate is always covered by a thin adsorption layer of liquid. This implies that the Landau-Levich film state and the finite-support meniscus state, discussed in section~\ref{sec:dependenceonthewebernumber}, are not anymore topologically different. Instead the former finite-support state becomes a state where the meniscus smoothly connects to the adsorption layer. Therefore, decreasing $\mathrm{We}_\mathrm{s}$ from large values, the branch of steady states in Fig.~\ref{down:fig:ha2_we2_hopffold_eps1_hp} does not end at finite $\mathrm{We}_\mathrm{s}$ as in Fig.~\ref{down:fig:momaplot}. Instead, it continues towards $\mathrm{We}_\mathrm{s}=0$ (not shown).
Starting at $\mathrm{We}_\mathrm{s}\approx0.2$ as in Fig.~\ref{down:fig:ha2_we2_hopffold_eps1_hp}, the coating thickness $h_c$ slowly increases with increasing $\mathrm{We}_\mathrm{s}$ until a first saddle-node bifurcation occurs at $\mathrm{We}_\mathrm{s} \approx 1.93$. The thickness $h_c$ continues to increase steadily when following the curve through all bifurcations. Namely, at the first saddle-node bifurcation the curve folds back and the steady states lose stability as a first real eigenvalue crosses the imaginary axis. Following the branch, $\mathrm{We}_\mathrm{s}$ decreases until a second saddle-node bifurcation occurs where an additional real eigenvalue becomes positive similar to the case in Fig.~\ref{dfws:fig:dragged_angles_}~(c). Subsequently, the branch wiggles through another eight saddle-node bifurcations resulting in states that are more and more unstable. This could still be called exponential snaking as the distance (in $\mathrm{We}_\mathrm{s}$) between subsequent bifurcations exponentially decreases. However, in contrast to \cite{GTLT2014prl} the snaking stops after ten saddle-node bifurcations, while in standard exponential snaking, the behaviour continues ad infinitum (domain size permitting). {We emphasize that the increasing instability strongly differs from the classical dip-coating system \cite{GTLT2014prl}, where at each saddle-node bifurcation a change between linearly stable and unstable states occurs. However, the finding resembles the behaviour, which is encountered in a Cahn-Hilliard-type model of Langmuir-Blodgett (LB) transfer \cite{KGFT2012njp}.} Beyond the final saddle-node bifurcation, on the scale of Fig.~\ref{down:fig:ha2_we2_hopffold_eps1_hp}, $h_c$ first increases at nearly constant $\mathrm{We}_\mathrm{s}\approx 1.56$. Then, the curve bends towards larger $\mathrm{We}_\mathrm{s}$ values (see inset of Fig.~\ref{down:fig:ha2_we2_hopffold_eps1_hp}). On this part of the branch, eight Hopf bifurcations occur in close succession. Each results in a further coating film destabilization. Following the branch further, we observe another sequence of 13 Hopf bifurcations that successively stabilize the branch, i.e., beyond the final Hopf bifurcation all states are stable.
To summarize, the first ten saddle-node bifurcations and eight Hopf bifurcations destabilize the steady states so that 26 eigenvalues cross the imaginary axis; another 13 Hopf bifurcations stabilize the states again. Further increasing $\mathrm{We}_\mathrm{s}$, the branch follows the power law $ h_c\sim\mathrm{We}_\mathrm{s}^{2/3}$ as described above.

{We have analysed how the bifurcation behaviour changes when switching from an ideally wetting liquid (section~\ref{sec:dependenceonthewebernumber} and Ref.~\cite{MoMa2017jfm}) to the case of partial wetting liquid using in both cases $\mathrm{We}_\mathrm{s}$ as control parameter. Next, we consider the SAW strength $\epsilon_\mathrm{s}$ as control parameter at fixed wettability. This allows us to directly compare to the Landau-Levich system where the plate velocity $U_0$ is used as control parameter (cf.~section~\ref{sec:draggedfilmwithouSAW}).}

\begin{figure}[tbh]
\centering
\includegraphics[angle=0,width=1.\textwidth]{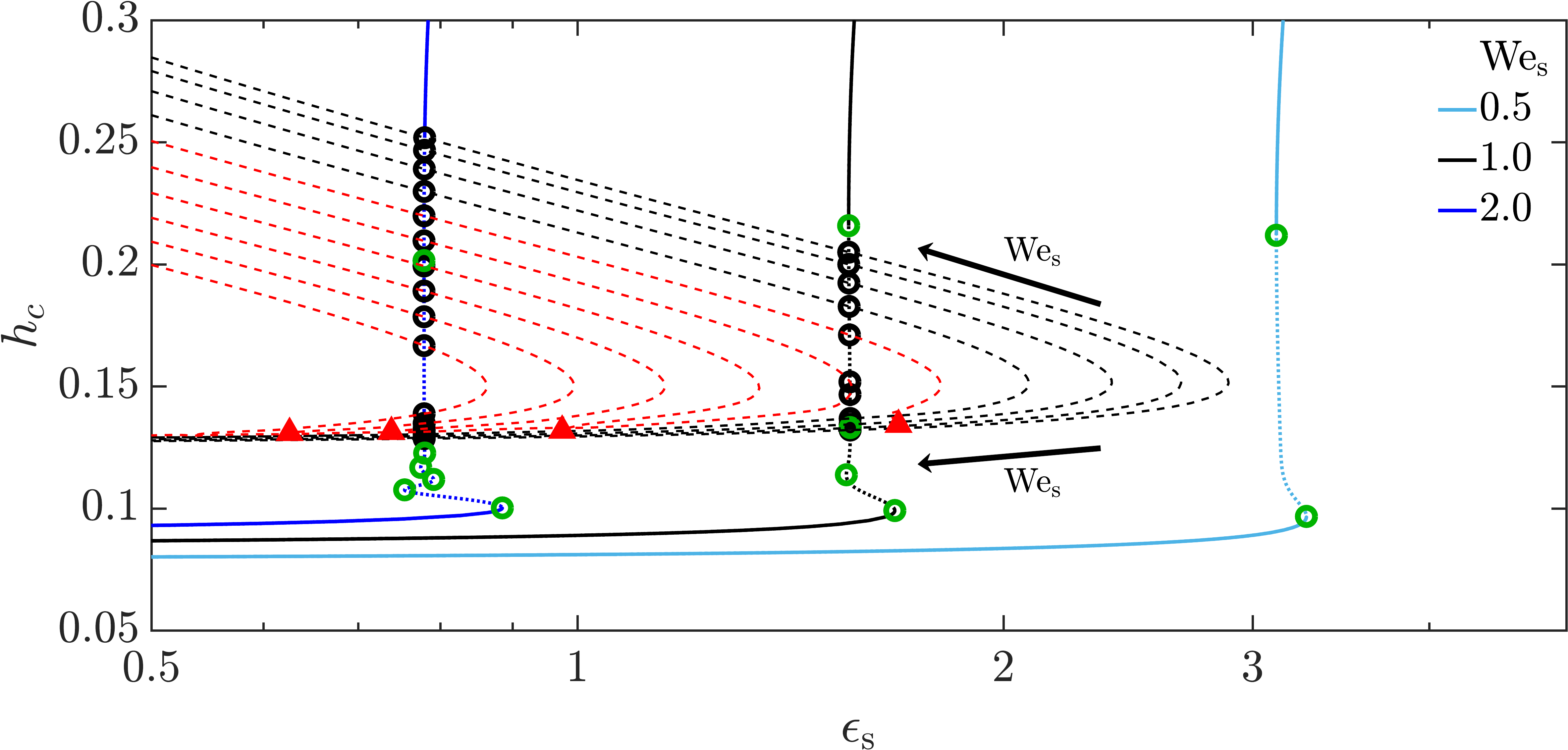}
\caption {{A bifurcation diagram showing showing as solid (stable states) and dotted lines (unstable states) the coating thickness $h_c$ in dependence of the SAW strength $\epsilon_\mathrm{s}$, for three different Weber numbers $\mathrm{We}_\mathrm{s}$ at fixed $\mathrm{Ha}=0.001$. All remaining parameters and the circle symbols are as in Fig.~\ref{down:fig:ha2_we2_hopffold_eps1_hp}. Additionally, we show as black and red dashed lines how the loci of the Hopf bifurcations change when $\mathrm{We}_\mathrm{s}$ is changed. The black arrows indicate the direction of increasing $\mathrm{We}_\mathrm{s}$. The red triangles mark Bogdanov-Takens bifurcations. For further details see main text.}
}
\label{doss:fig:we_ha1_hopfbif_eps_hopfcont}
\end{figure}

{Corresponding bifurcation curves for three different fixed values of $\mathrm{We}_\mathrm{s}$ and fixed $\mathrm{Ha}= 0.001$ are given in Fig.~\ref{doss:fig:we_ha1_hopfbif_eps_hopfcont}. Overall, these curves are similar to Fig.~\ref{down:fig:ha2_we2_hopffold_eps1_hp}, indicating that $\mathrm{We}_\mathrm{s}$ and $\epsilon_\mathrm{s}$ have a similar influence on the behavior. Decreasing $\mathrm{We}_\mathrm{s}$ in Fig.~\ref{doss:fig:we_ha1_hopfbif_eps_hopfcont}, the entire curve shifts towards larger $\epsilon_\mathrm{s}$. In other words, the system needs more SAW power to spread over the substrate for a smaller ratio of convective and capillary stress at the surface. Roughly speaking, at larger $\mathrm{We}_\mathrm{s}$ a lower SAW strength is needed to obtain the same excess volume.
The number of saddle-node and Hopf bifurcations decreases with decreasing $\mathrm{We}_\mathrm{s}$. All bifurcations with the exception of two saddle-node bifurcations have vanished at $\mathrm{We}_\mathrm{s} = 0.5$. This process can be well appreciated when tracking the loci of bifurcations. Fig.~\ref{doss:fig:we_ha1_hopfbif_eps_hopfcont} shows the traces of all Hopf bifurcations as black and red dashed lines. This and the behavior in the parameter plane spanned by $\mathrm{We}_\mathrm{s}$ and $\epsilon_\mathrm{s}$ is discussed further below.}

\begin{figure}[hbt]
\centering
    \includegraphics[width=0.48\textwidth]{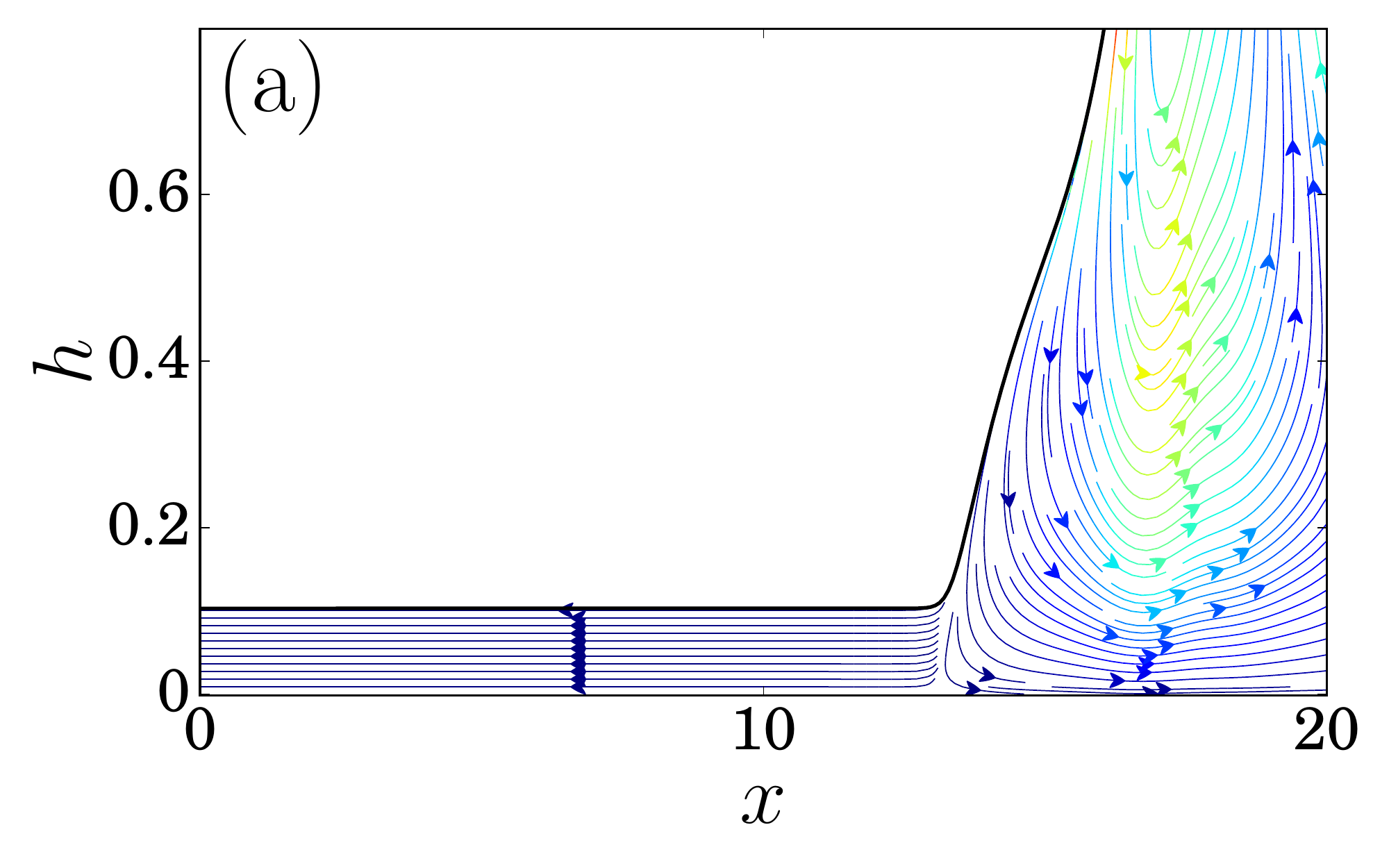}
    \includegraphics[width=0.48\textwidth]{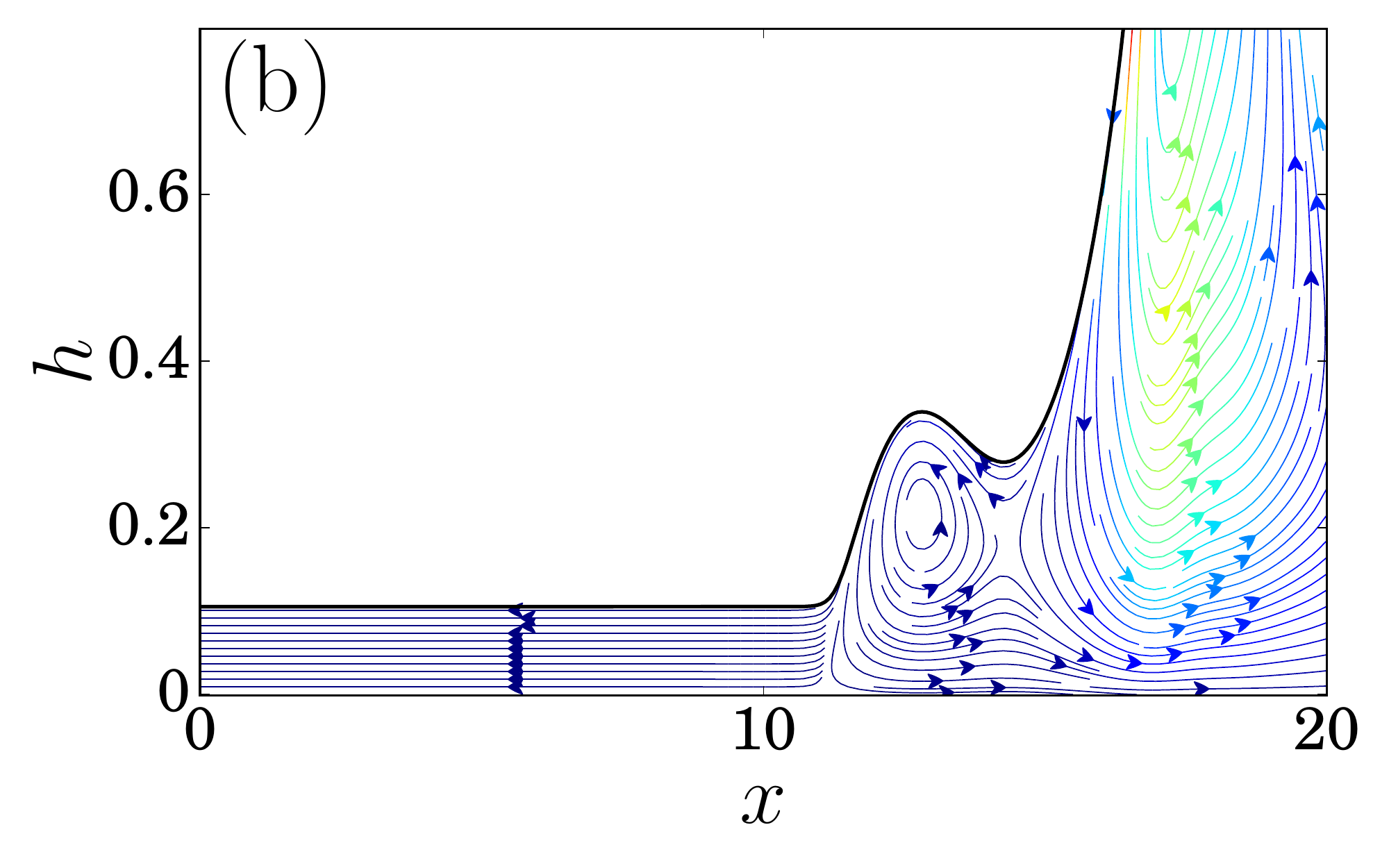}\\
    \includegraphics[width=0.48\textwidth]{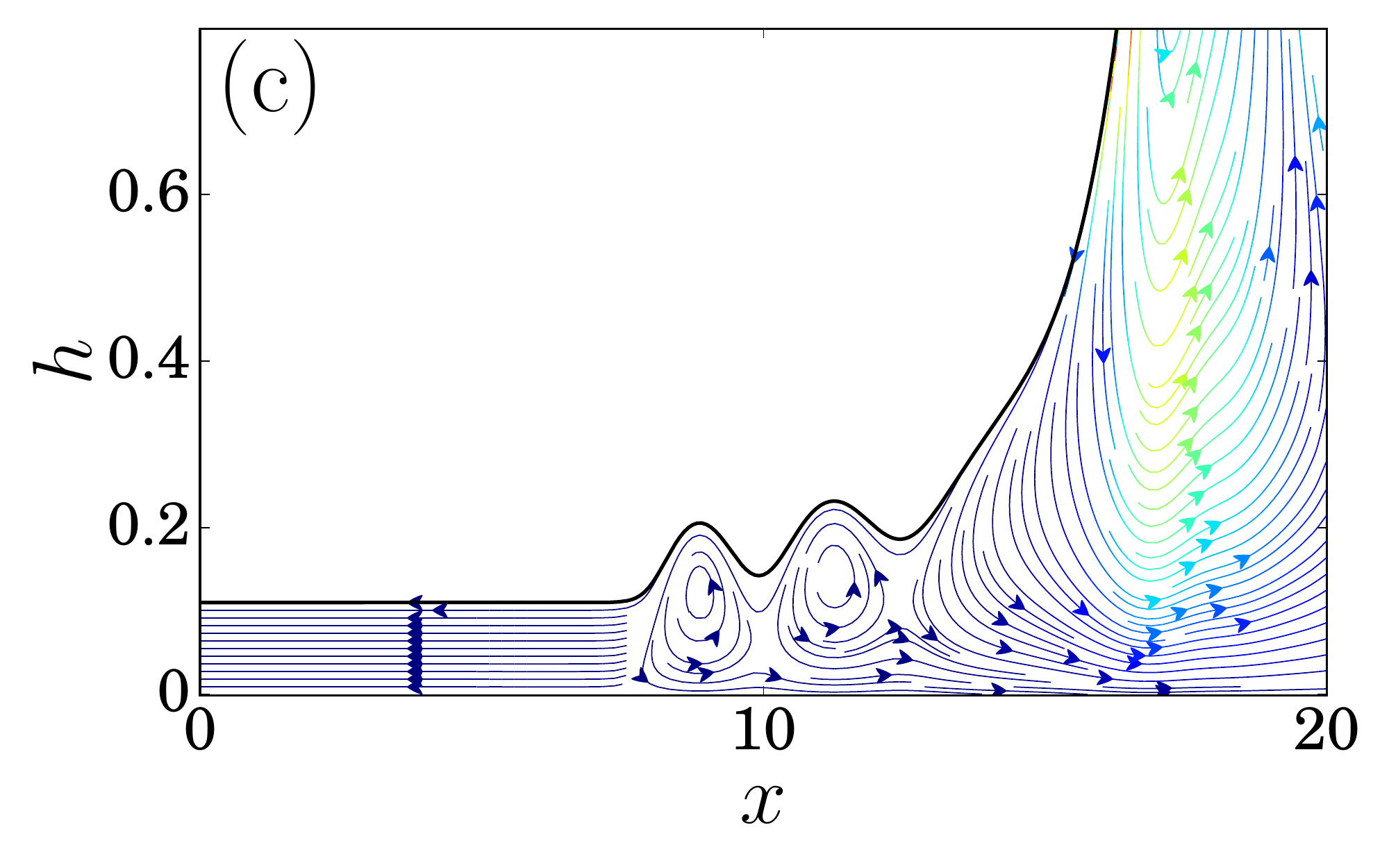}
    \includegraphics[width=0.48\textwidth]{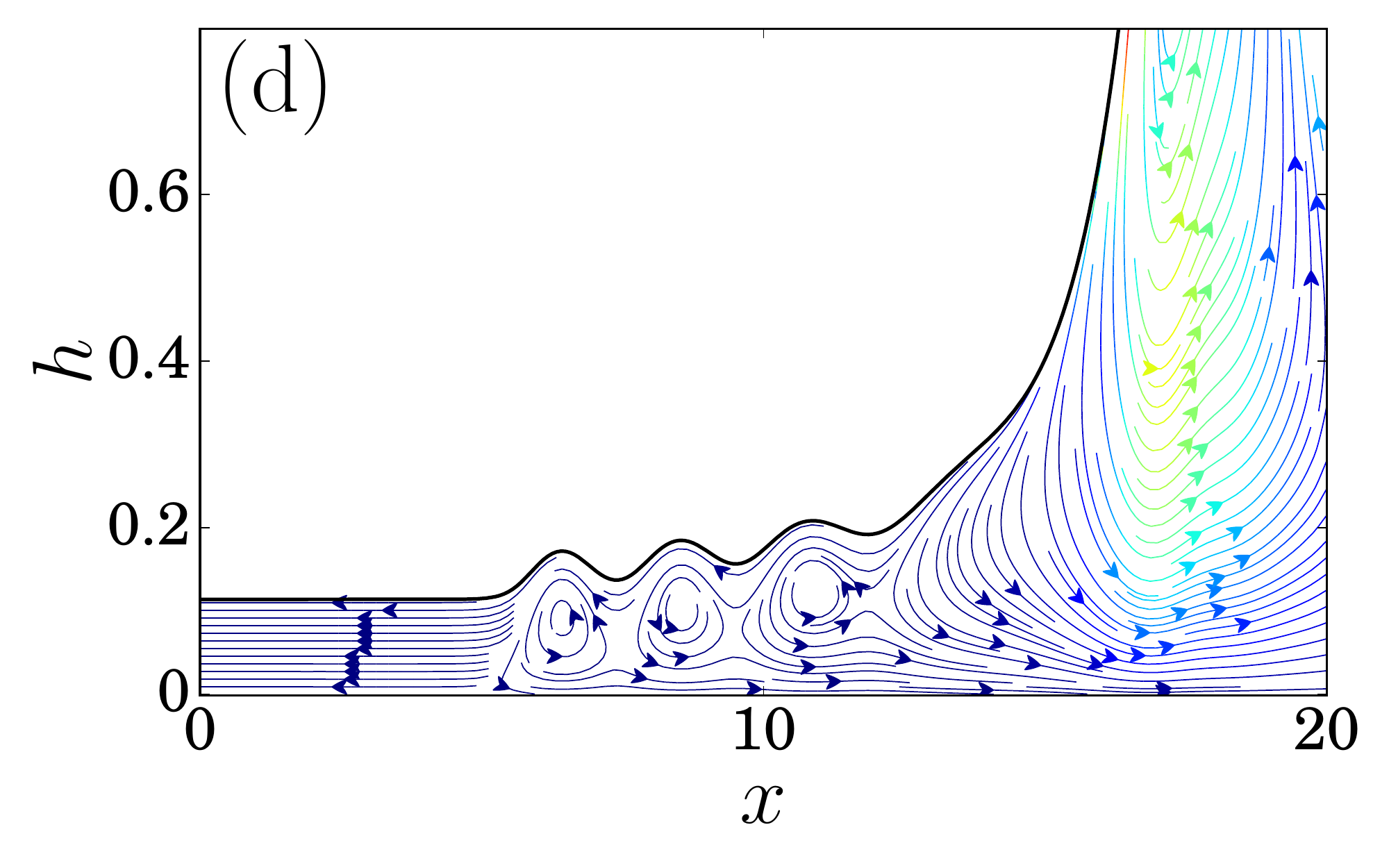}\\
    \includegraphics[width=0.48\textwidth]{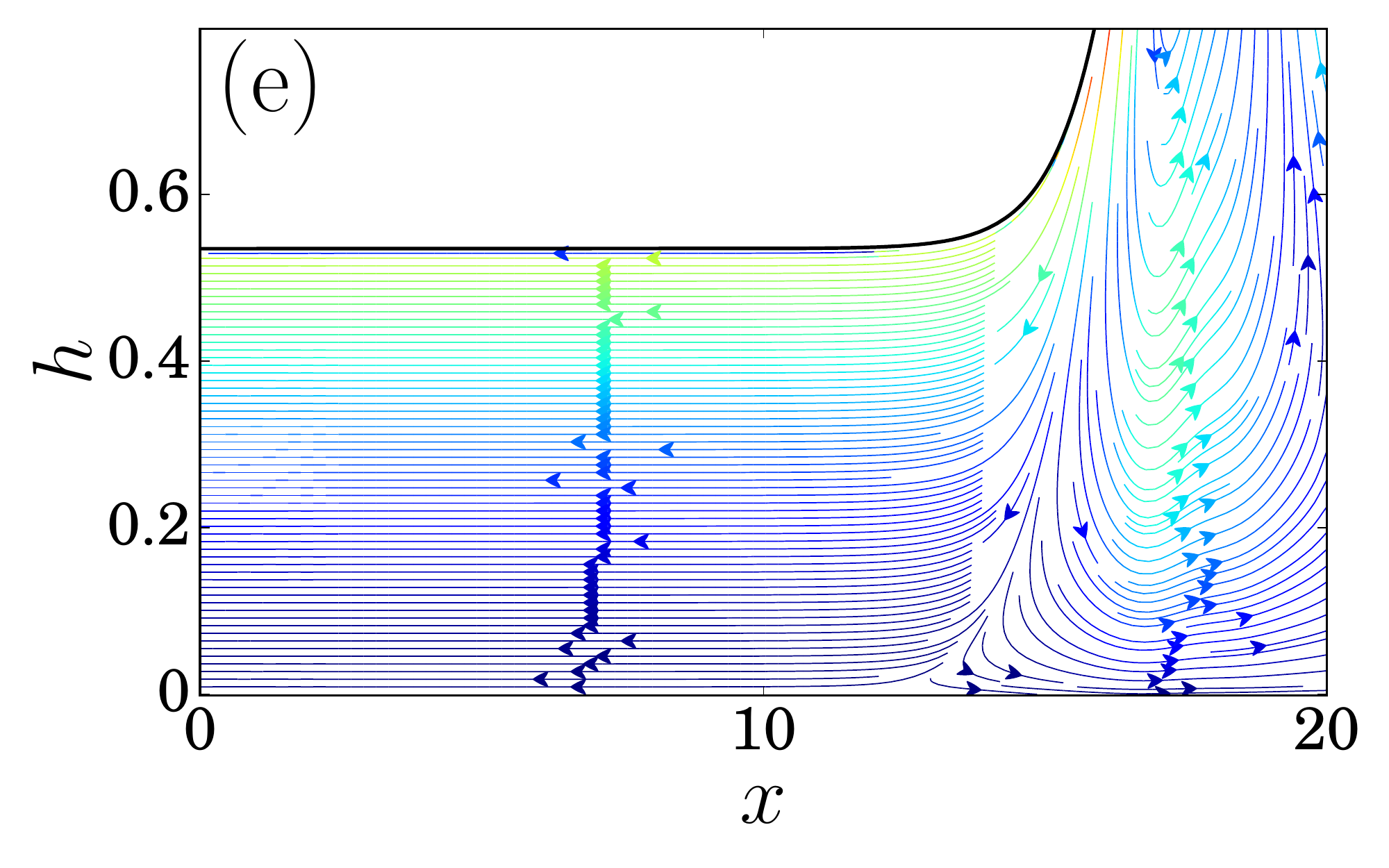}
    \includegraphics[width=0.48\textwidth]{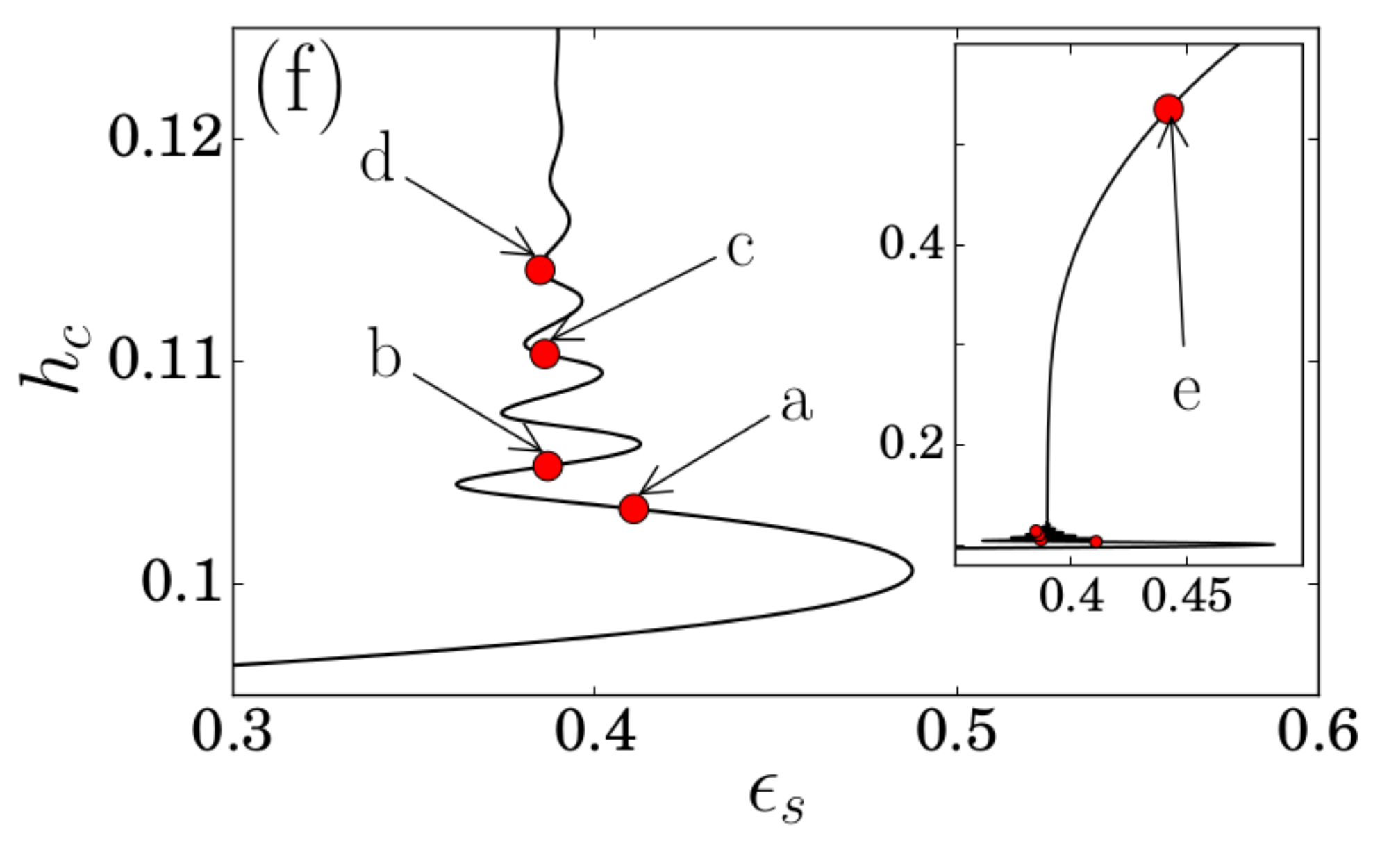}
\caption{
{Panels (a) to (e) show steady thickness profiles obtained by continuation and the corresponding streamlines within the fluid at parameters indicated by symbols marked ``a'' to ``e'' in panel (f), respectively. Panel (f) presents the snaking region of the corresponding bifurcation diagram at $\mathrm{We}_\mathrm{s}=4$. In particular, (a) presents a meniscus solution, (b) to (d)  modulated foot solutions and (e) a Landau-Levich film state. The remaining parameters are as in Fig.~\ref{doss:fig:we_ha1_hopfbif_eps_hopfcont}. Shown is only part of the computational domain of $L = 40$.}
}
\label{s:fig:saw_velo_field_steady_states}
\end{figure}

First, however, we present steady thickness profiles and the underlying flow field as characterized by the streamlines. {In particular, Fig.~\ref{s:fig:saw_velo_field_steady_states}~(a) to (e) presents profiles from the snaking region of the bifurcation diagram in Fig.~\ref{s:fig:saw_velo_field_steady_states}~(f).
Panel~(a) corresponds to a meniscus solution, i.e., the profile smoothly and monotonically connects the meniscus on the right with the adsorption layer on the left, similar to the black profiles in Figs.~\ref{dfws:fig:dragged_angles_}~(a) to~(d). 
Then, following the snaking curve in panel~(f) from point ``a'' to point ``d'',} a foot-like structure develops and expands. One can see that the SAW induces strong modulations in the foot thickness, related to convection rolls within the foot. At the same time, the single large-scale convection roll in the meniscus is nearly unchanged. Following the bifurcation curve further, one observes the addition of more modulations. However, in contrast to the Landau-Levich case in Figs.~\ref{dfws:fig:dragged_angles_}~(a) and~(b), the foot does not possess a clearly defined thickness, but becomes thinner with increasing distance from the meniscus. As a result, the transition from the foot to the adsorption layer is continuous. In parallel, the convection rolls in the foot seem to weaken and partially fuse. {The profile in panel~(e) is beyond the snaking region, see inset of (f), and corresponds to a standard Landau-Levich film similar to profiles in Figs.~\ref{dfws:fig:dragged_angles_}~(c) and (d).}

\begin{figure}[hbt]
\centering
    \includegraphics[width=0.48\textwidth]{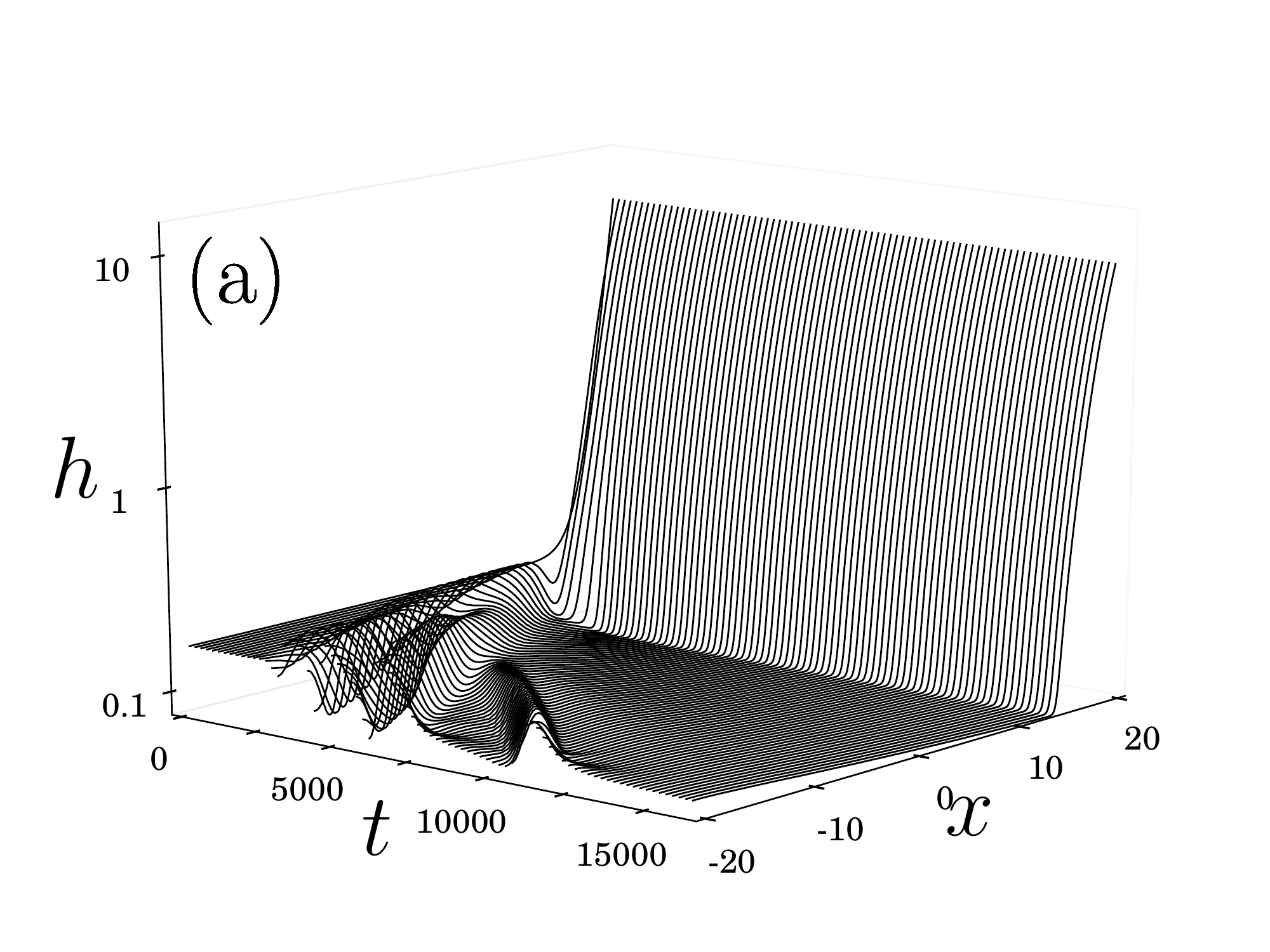}
    \includegraphics[width=0.48\textwidth]{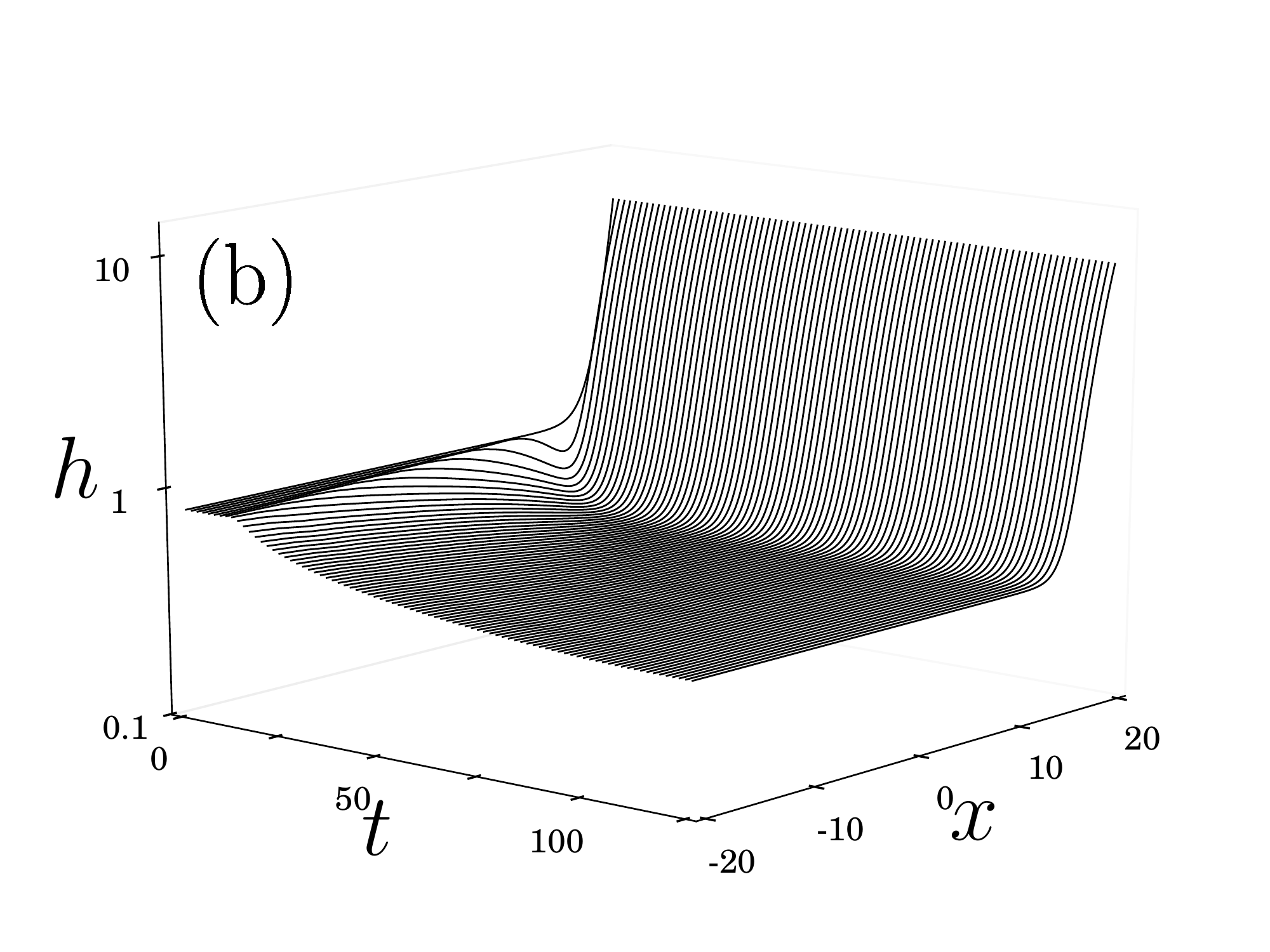}
\caption{
{Space-time plots of time evolutions are shown that indicate bistability of different steady states. The case of $\mathrm{We}_\mathrm{s} = 1$ in Fig.~\ref{doss:fig:we_ha1_hopfbif_eps_hopfcont} is used as example, i.e., all remaining parameters are given there. Panels (a) and (b) use, at $\epsilon_\mathrm{s}=1.6$, initial conditions with relatively small ($h_c = 0.2$) and large ($h_c = 0.8$) coating film thicknesses, respectively. After a transient two different steady states emerge, in (a) a macroscopically dry state and in (b) a Landau-levich film.}
}
\label{s:fig:time_simulations_same_values}
\end{figure}

{A snaking shape of bifurcation curves as in in Figs.~\ref{down:fig:ha2_we2_hopffold_eps1_hp} and~\ref{doss:fig:we_ha1_hopfbif_eps_hopfcont} often indicates bi- or multistability of different states. This implies that time simulations may depending on initial conditions converge toward different states. Here, this is illustrated for the case of $\mathrm{We}_\mathrm{s} = 1$ in Fig.~\ref{doss:fig:we_ha1_hopfbif_eps_hopfcont} by the space-time plots in Fig.~\ref{s:fig:time_simulations_same_values}. At identical $\epsilon_\mathrm{s}=1.6$, initially relatively small and large coating film thickness indeed result in different linearly stable states, namely macroscopically dry and Landau-Levich film state, respectively. This indicates that  the transition between these two states shows hysteresis. Interestingly, the relaxation process in Fig.~\ref{doss:fig:we_ha1_hopfbif_eps_hopfcont}~(a) shows transient oscillations related to dewetting and advection of the resulting drops, while in (b) the transient is monotonic. Note that the oscillatory transient in (a) takes much longer than the process in (b).}

\begin{figure}[hbt]
\centering
\includegraphics[angle=0,width=1.\textwidth]{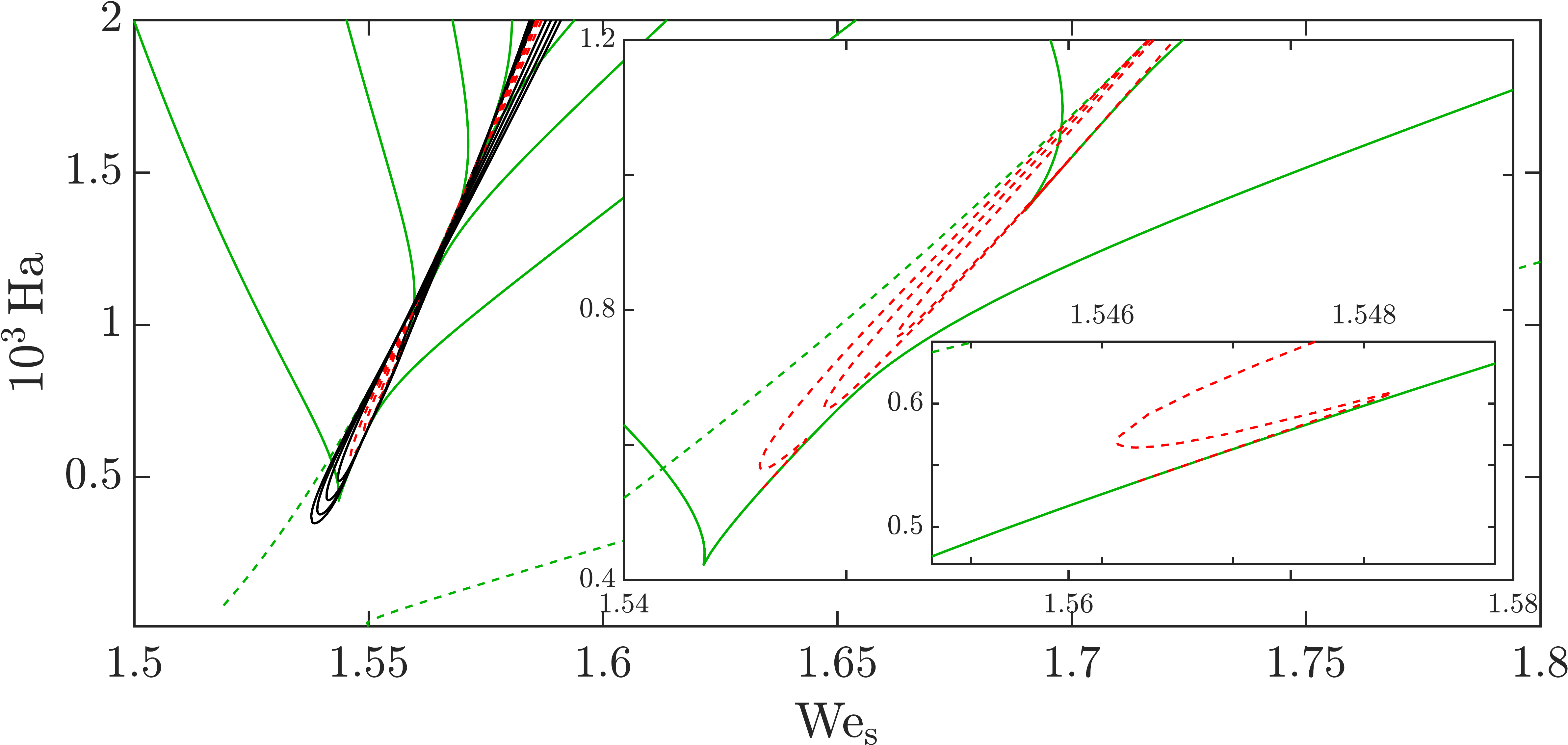}
\caption {
Shown are the loci of saddle-node (green lines) and Hopf bifurcations {(black solid and red dashed lines)} in the parameter plane spanned by Weber number $\mathrm{We}_\mathrm{s}$ and Hamaker number $\mathrm{Ha}$. The insets focus on the Hopf bifurcations tracked as red dashed lines. The inner inset magnifies the region where the leftmost of these Hopf bifurcations emerges from the saddle-node bifurcation, i.e., where a Bogdanov-Takens bifurcation occurs. Further explanations on line styles are given in the main text. The remaining parameters are as in Fig.~\ref{down:fig:ha2_we2_hopffold_eps1_hp}. }
\label{down:fig:map_ha2_vs_we2_foldhopf_eps1_2}
\end{figure}

\begin{figure}[hbt]
\centering
\includegraphics[angle=0,width=.9\textwidth]{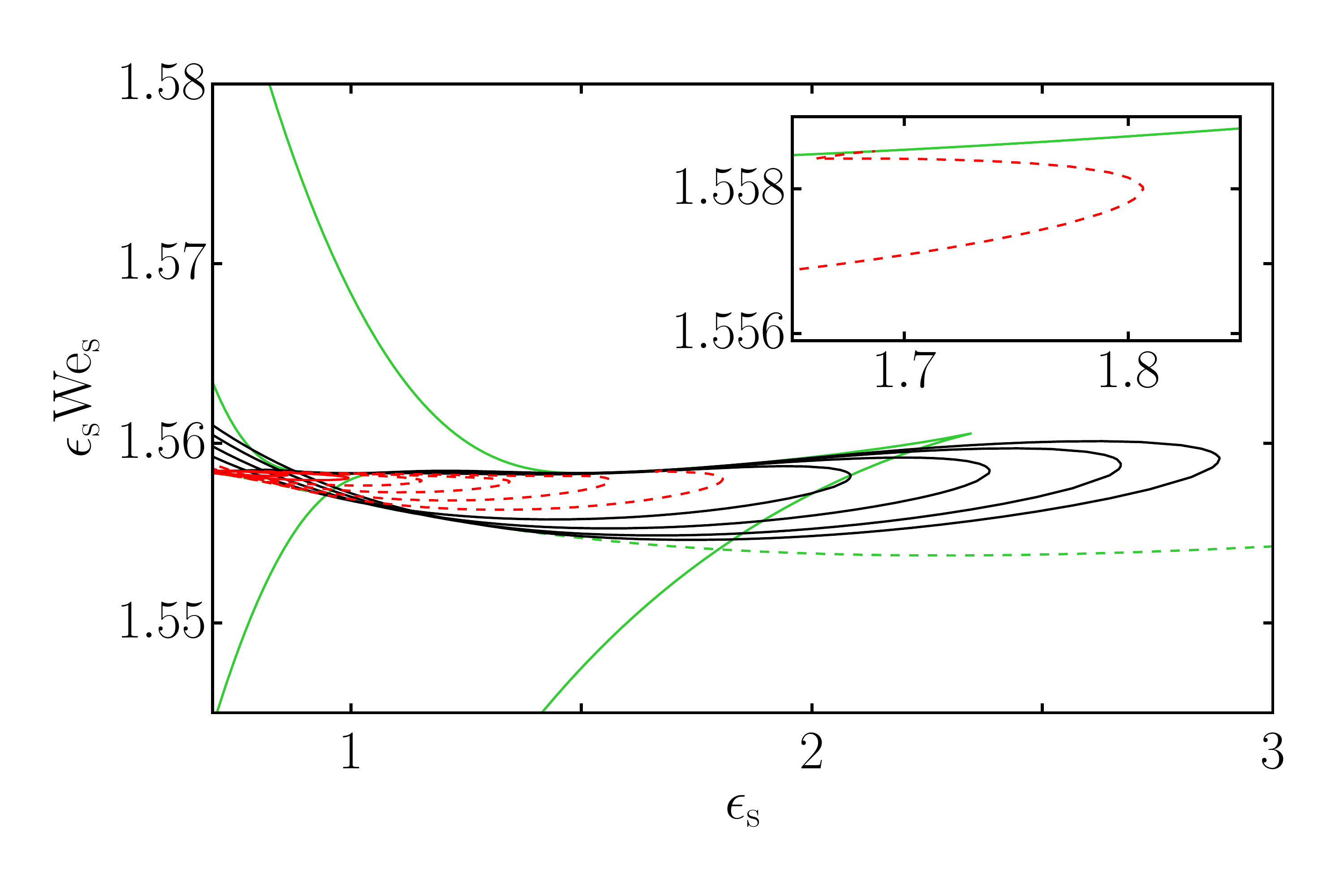}
\caption {Loci of saddle-node and Hopf bifurcations are shown in the ($\epsilon_\mathrm{s}$, $\epsilon_\mathrm{s}\mathrm{We}_\mathrm{s}$)-parameter plane.
The inset magnifies the parameter region where a Bogdanov-Takens bifurcation occurs. Line styles and remaining parameters are as in Fig.~\ref{doss:fig:we_ha1_hopfbif_eps_hopfcont}.
}
\label{fig:map_weeps_vs_eps_foldhopf}
\end{figure}

Up to here, we have investigated individual bifurcation diagrams by employing either the Weber number $\mathrm{We}_\mathrm{s}$ or the SAW strength $\epsilon_\mathrm{s}$ as control parameter.  {Thereby, we have detected many saddle-node and Hopf bifurcations. As their exact number strongly depends on $\mathrm{We}_\mathrm{s}$ we investigate next, how this comes about by tracking} all bifurcations visible in Figs.~\ref{down:fig:ha2_we2_hopffold_eps1_hp} and \ref{doss:fig:we_ha1_hopfbif_eps_hopfcont} in the ($\epsilon_\mathrm{s}$,$\mathrm{Ha}$)- and ($\epsilon_\mathrm{s}$,$\mathrm{We}_\mathrm{s}$)-parameter planes.

First, Fig.~\ref{down:fig:map_ha2_vs_we2_foldhopf_eps1_2} gives the loci of saddle-node (green lines) and Hopf bifurcations (black solid and red dashed lines) in the ($\mathrm{We}_\mathrm{s}$,$\mathrm{Ha}$)-plane, 
i.e., we investigate the interplay of capillarity and wettability. The solid green lines denote saddle-node bifurcations that emerge in pairs in codimension-2 hysteresis bifurcations that are cusp-like in the 
($\mathrm{We}_\mathrm{s}, \mathrm{Ha}$)-plane (see larger-scale inset). In contrast, the two dashed green lines correspond to saddle-node bifurcations that persist in the entire studied parameter range nearly down to $\mathrm{Ha}=0$. 
These two lines correspond to the saddle-node bifurcations visible in the SAW reference case in Fig.~\ref{down:fig:momaplot}.

Inspecting the loci of the Hopf bifurcations in Fig.~\ref{down:fig:map_ha2_vs_we2_foldhopf_eps1_2}, we note that increasing $\mathrm{We}_\mathrm{s}$, the first pair of Hopf bifurcations appears together in a codimension-2 double Hopf bifurcation.
The next three pairs appear in the same way. All of them are shown as black lines. However, the next Hopf bifurcation emerges alone from a line of saddle-node bifurcations (see smaller-scale inset of Fig.~\ref{down:fig:map_ha2_vs_we2_foldhopf_eps1_2}). 
There, a codimension-2 Bogdanov-Takens bifurcation occurs, where the Hopf bifurcation and a global (homoclinic) bifurcation emerge together at a saddle-node bifurcation. 
Following the corresponding line of Hopf bifurcations, it separates from the line of saddle-node bifurcations and eventually annihilates with another Hopf bifurcation that itself emerged from a fifth double Hopf bifurcation. 
All further Hopf bifurcations appear in a similar way. Red dashed lines in the figure mark Hopf bifurcations with this type of emergence.

In general, one can say that the number of occurring bifurcations increases with the Weber number (i.e., with decreasing surface tension) and also increases with the Hamaker number $\mathrm{Ha}$ (i.e., with {decreasing wettability}). The emergence of Hopf bifurcations is further scrutinized in section~\ref{sec:hopfbranches}.

Next, we track the bifurcations in the ($\epsilon_\mathrm{s}, \mathrm{We}_\mathrm{s}$)-plane. An initial inspection of the curves shows that for all tracked bifurcations, $\mathrm{We}_\mathrm{s}\sim 1/\epsilon_\mathrm{s}$, 
and the curves practically coincide when presented in the ($\epsilon_\mathrm{s}, \mathrm{We}_\mathrm{s}$)-plane. Therefore, in Fig.~\ref{fig:map_weeps_vs_eps_foldhopf} we present them in the ($\epsilon_\mathrm{s}, \epsilon_\mathrm{s} \mathrm{We}_\mathrm{s}$)-plane where they are readily discernible.
Furthermore, we illustrate the appearance of the Hopf bifurcations in a different representations by including their loci as black and red dashed lines in Fig.~\ref{doss:fig:we_ha1_hopfbif_eps_hopfcont} above. In this way we visualize how the three bifurcation curves shown there are related. The line styles correspond to the ones in Fig.~\ref{fig:map_weeps_vs_eps_foldhopf}.
%
%%%%%%%%%%%%%%%%%%%%%%%%%%%%%%%%%%%%%%%%%%%%%%%%%%%%%%%%%%%%%%%%
\subsection{Time-periodic states}
\label{sec:hopfbranches}
%%%%%%%%%%%%%%%%%%%%%%%%%%%%%%%%%%%%%%%%%%%%%%%%%%%%%%%%%%%%%%%%%
%
In our investigation of the bifurcation behaviour of the 1D SAW system we have encountered a large number of Hopf bifurcations when using the Weber number and SAW strength as main control parameters. Tracking the loci of the bifurcations in parameter planes has indicated that the Hopf bifurcations are created either in double Hopf bifurcations or Bogdanov-Takens bifurcations. Next, we employ \textsc{pde2path} to investigate the branches of time-periodic states (TPS), which emerge at the Hopf bifurcations. Note that recently such branches were described for the dragged-plate system \cite{TWGT2019prf} (also compare section~\ref{sec:draggedfilmwithouSAW}). Here, we provide a deeper analysis by discussing how the single-parameter bifurcation diagram (e.g., with  $\epsilon_\mathrm{s}$ as control parameter) transforms when a second parameter is changed. We are particularly interested in potential reconnections between the different branches of TPS. This is relevant for the present system, but also provides valuable information for the behaviour of the entire class of coating systems.

\begin{figure}[hbt]
\begin{minipage}{0.7\textwidth}
\center
\includegraphics[width=.9\textwidth]{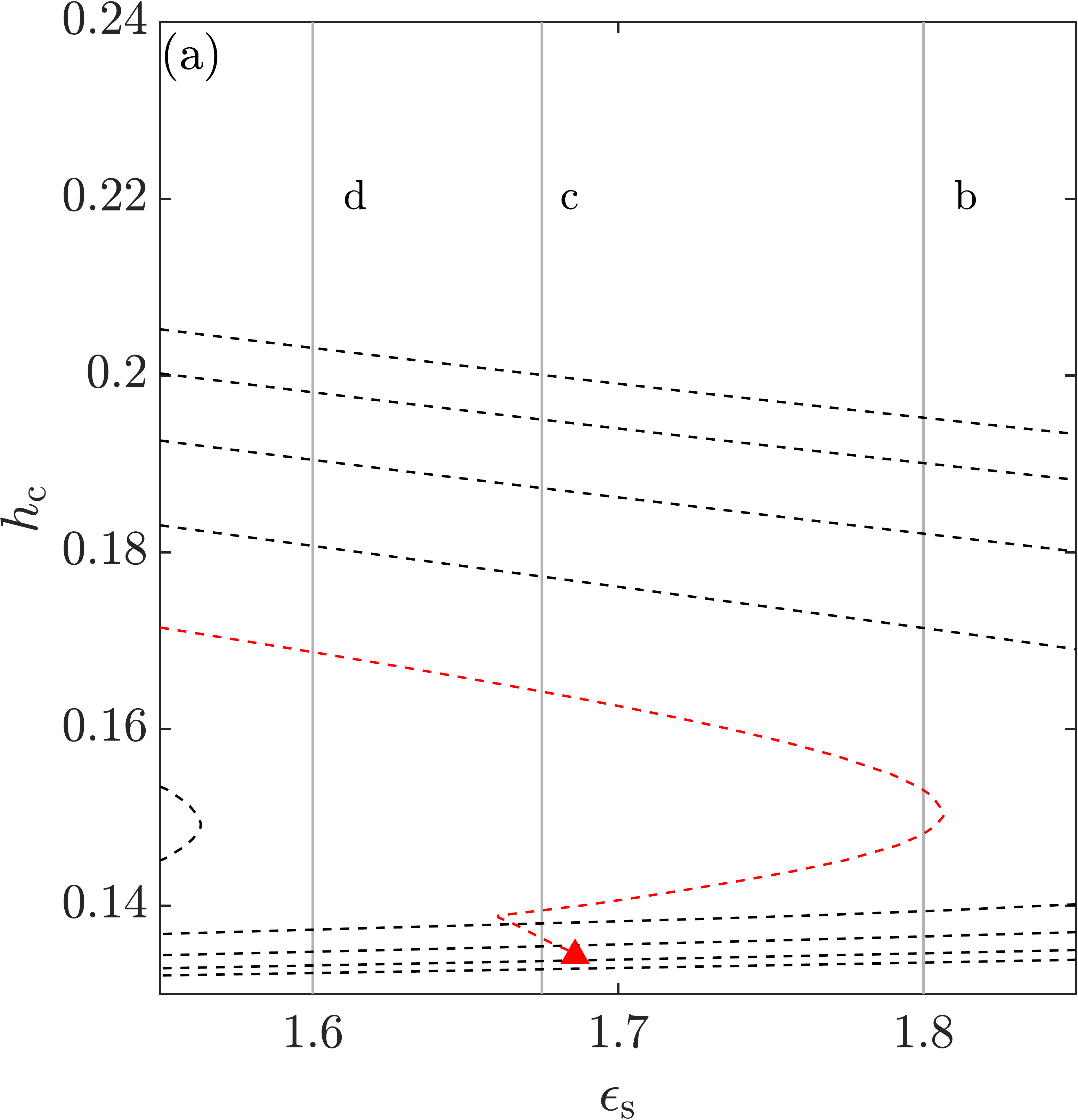}
\end{minipage}
\hspace{-0.025\hsize}
\begin{minipage}{0.2\textwidth}
\vspace{-0.3cm}
\includegraphics[width=1.0\textwidth]{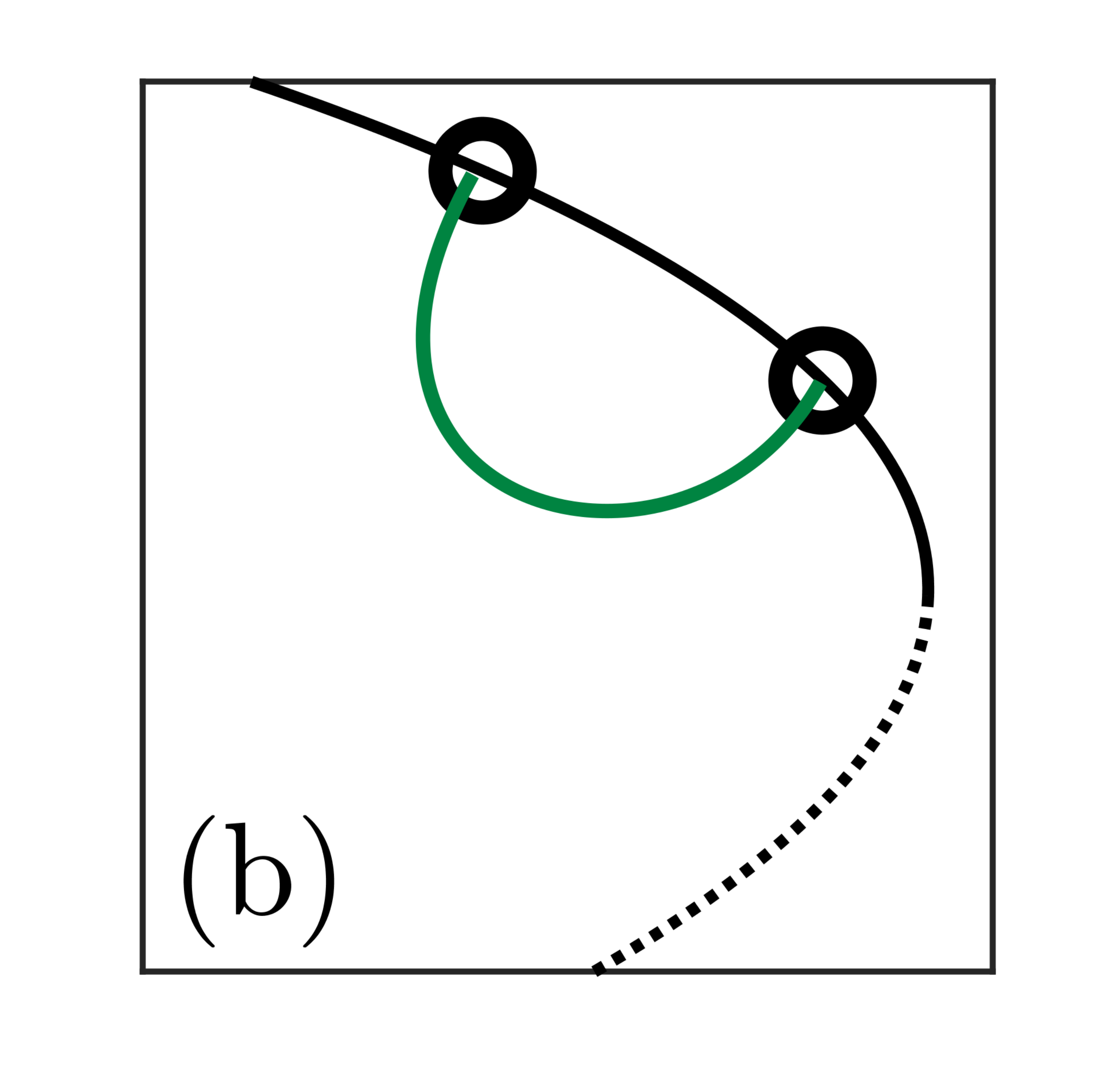}
\vspace{0.1cm}
\includegraphics[width=1.0\textwidth]{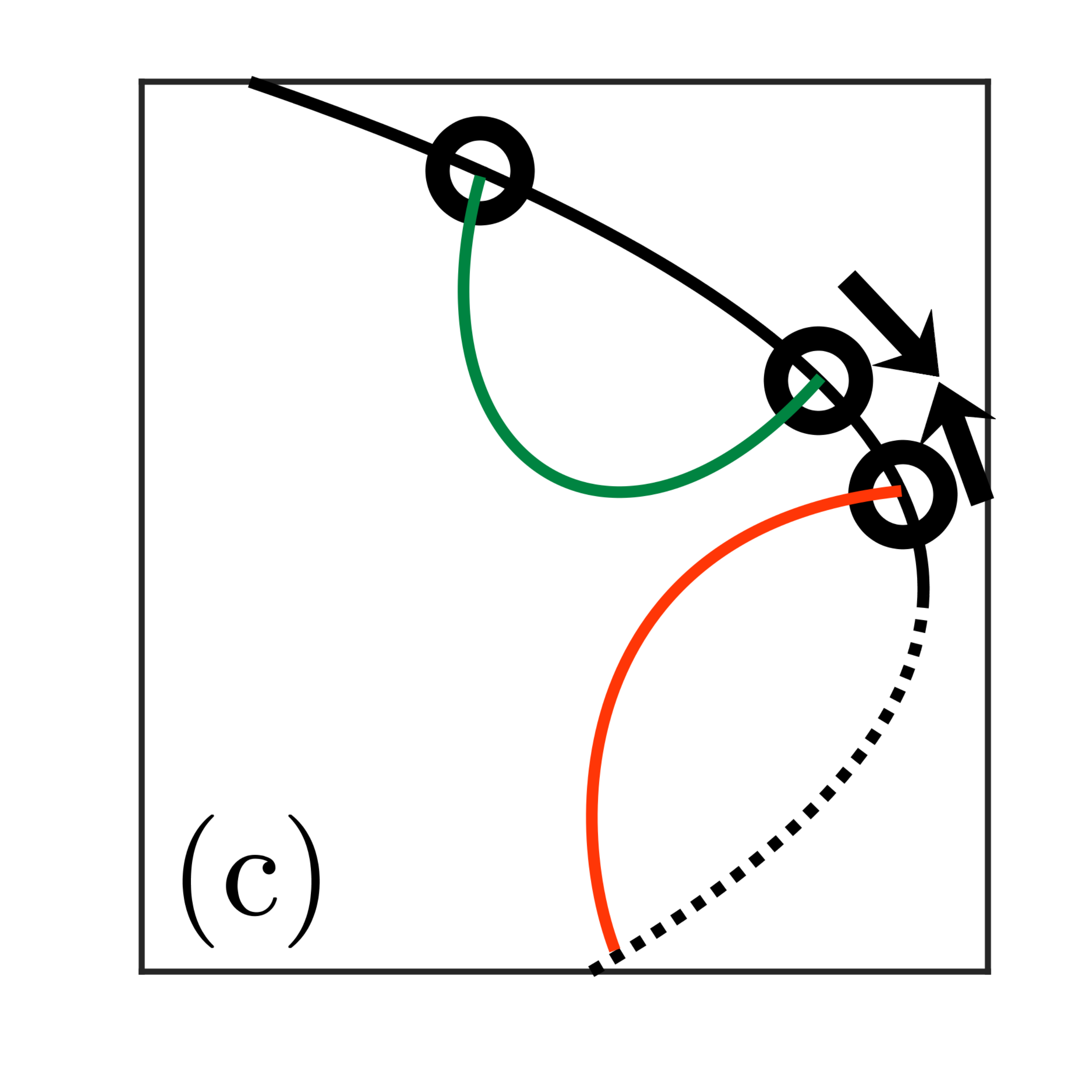}
 \vspace{0.1cm}
\includegraphics[width=1.0\textwidth]{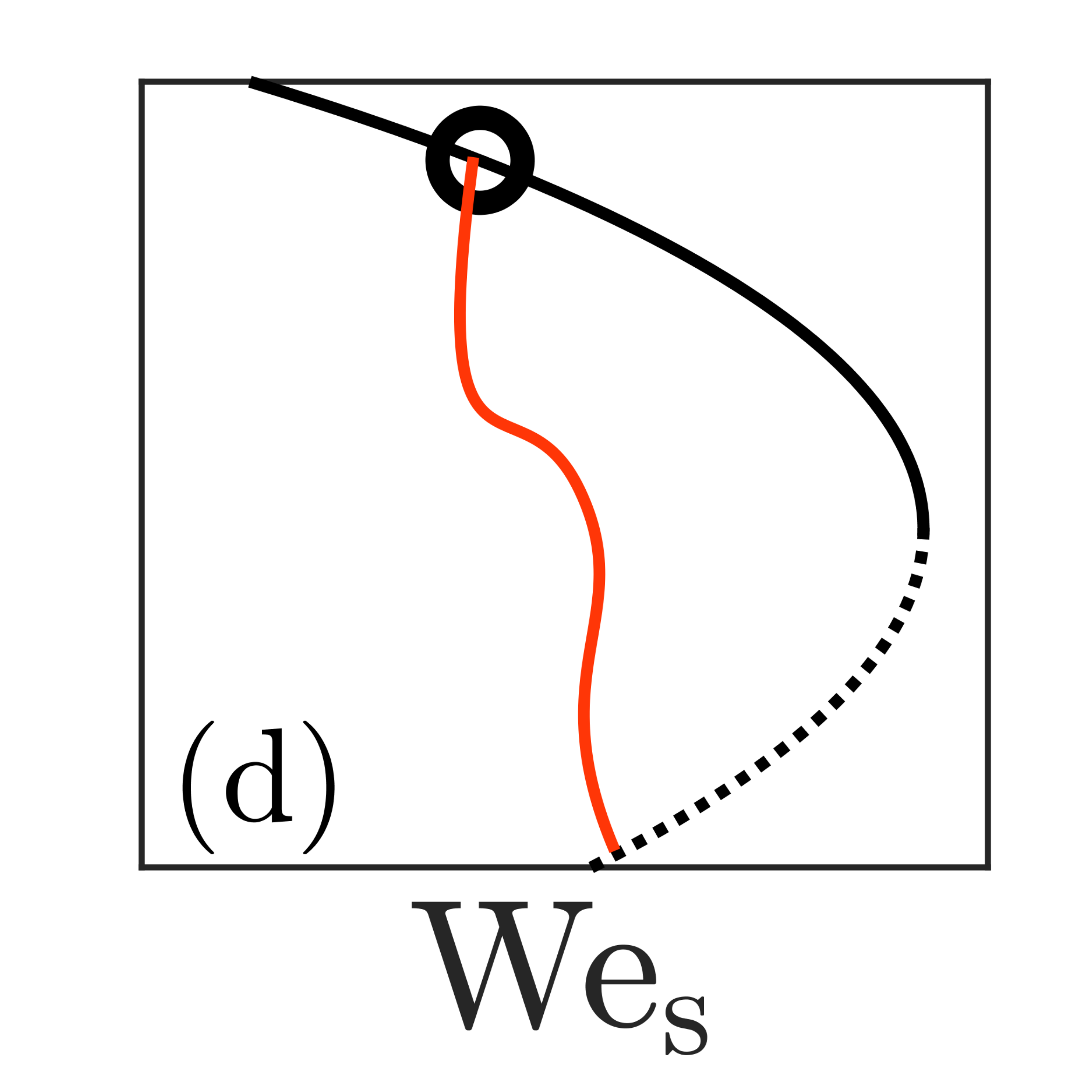}
\end{minipage}
\caption{Panel~(a) provides a zoom of Fig.~\ref{doss:fig:we_ha1_hopfbif_eps_hopfcont} focusing on the loci of a number of Hopf bifurcations, 
while panels (b) to (d) show sketches of the bifurcation diagrams of time-periodic states (TPS) related to different parameter ranges of the red curve in (a). 
Typical parameter values where cases (b) to (d) occur are indicated by vertical dotted lines in (a), marked ``b'', ``c'' and ``d''. 
The occurring changes are described in the main text. In (b) to (d), steady states are given as solid black lines, while green [red] lines indicate branches of TPS that connect two Hopf bifurcations [a Hopf and a homoclinic bifurcation].
}
\label{hp:fig:Bogdanov_Taken_bif_show}
\end{figure}

Before calculating branches of TPS, we begin by discussing the expected behaviour related to the codimension-2 bifurcations where the Hopf bifurcations emerge or vanish. 
Fig.~\ref{hp:fig:Bogdanov_Taken_bif_show}~(a) magnifies the loci of a number of Hopf bifurcations, highlighting in red a line related to particularly intricate behaviour. On the right, panels (b) to (d) sketch different bifurcation diagrams (with $\mathrm{We}_\mathrm{s}$ as control parameter) with decreasing $\epsilon_\mathrm{s}$: {To the right of the saddle-node bifurcation at $\epsilon_\mathrm{s}\approx 1.81$, the branch of steady states is 'bare', i.e.\ bifurcations, which are related to the red branch in (a) do not yet exist. Then, decreasing $\epsilon_\mathrm{s}$, when the saddle-node bifurcation is passed, a branch of TPS is created in a double Hopf bifurcation [Fig.~\ref{hp:fig:Bogdanov_Taken_bif_show}~(b)]. Further decreasing $\epsilon_\mathrm{s}$, at $\epsilon_\mathrm{s}=1.69$ a Bogdanov-Takens bifurcation [red triangle in panel~(a)] is passed. There a branch of TPS appears which connects a Hopf and a homoclinic bifurcation 
[red line in Fig.~\ref{hp:fig:Bogdanov_Taken_bif_show}~(c)]. Continuing to decrease $\epsilon_\mathrm{s}$, two Hopf bifurcations on different branches of TPS approach each other until they finally annihilate at $\epsilon_\mathrm{s}\approx1.66$ in a reverse double 
Hopf bifurcation. Effectively, there, the two branches of TPS are fused into one [red line in Fig.~\ref{hp:fig:Bogdanov_Taken_bif_show}~(d)].} Note that the resulting branch of TPS may still possess a nontrivial structure, e.g., contain saddle-node bifurcations. These are not tracked here. Note that the remaining branch inherits a Hopf bifurcation and a homoclinic bifurcation as its end points.
  
\begin{figure}[hbt]
\begin{minipage}{0.45\textwidth}
\includegraphics[width=1.\textwidth]{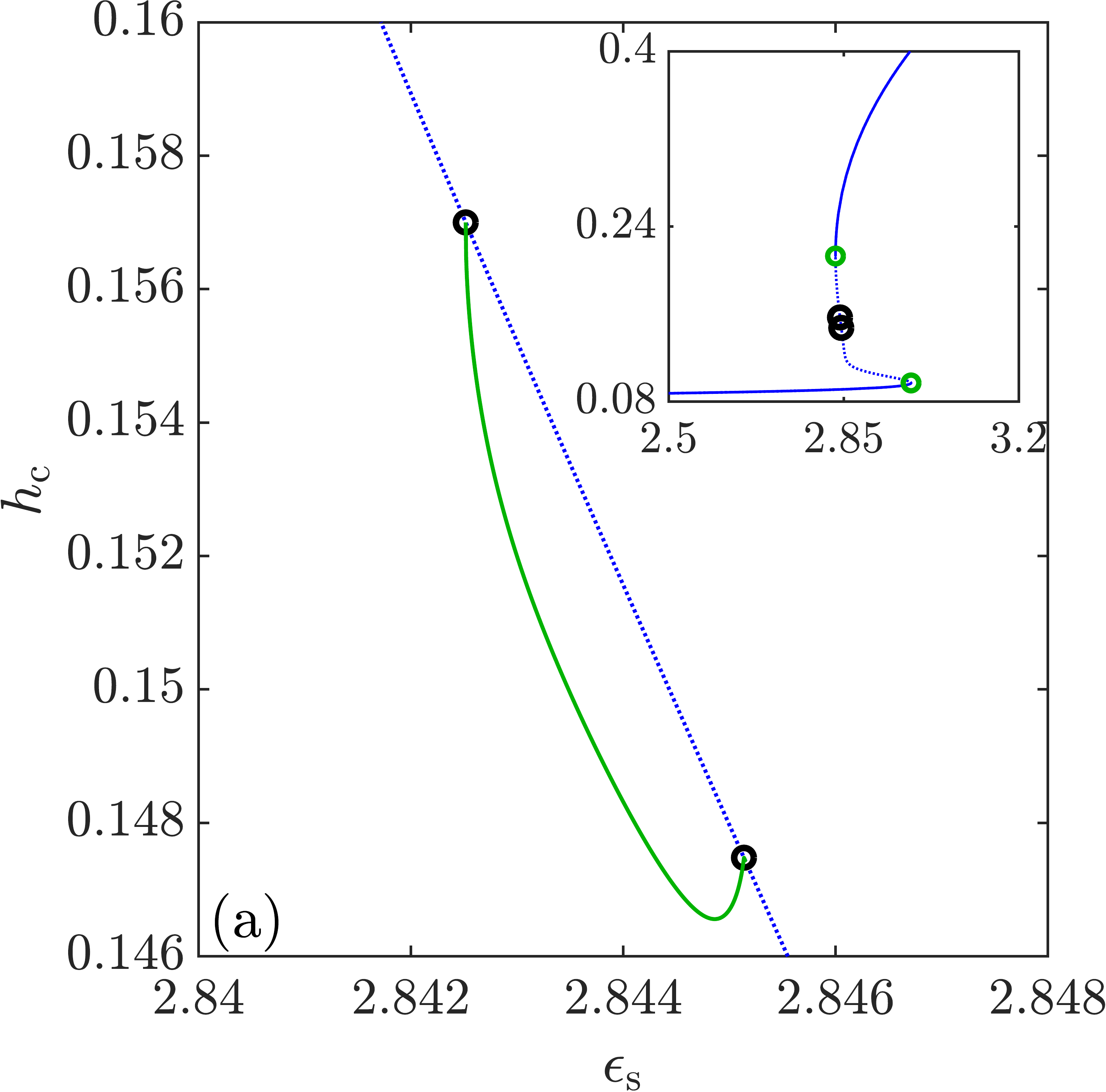}
\end{minipage}
\hspace{0.05\hsize}
\begin{minipage}{0.45\textwidth}
\includegraphics[width=1.\textwidth]{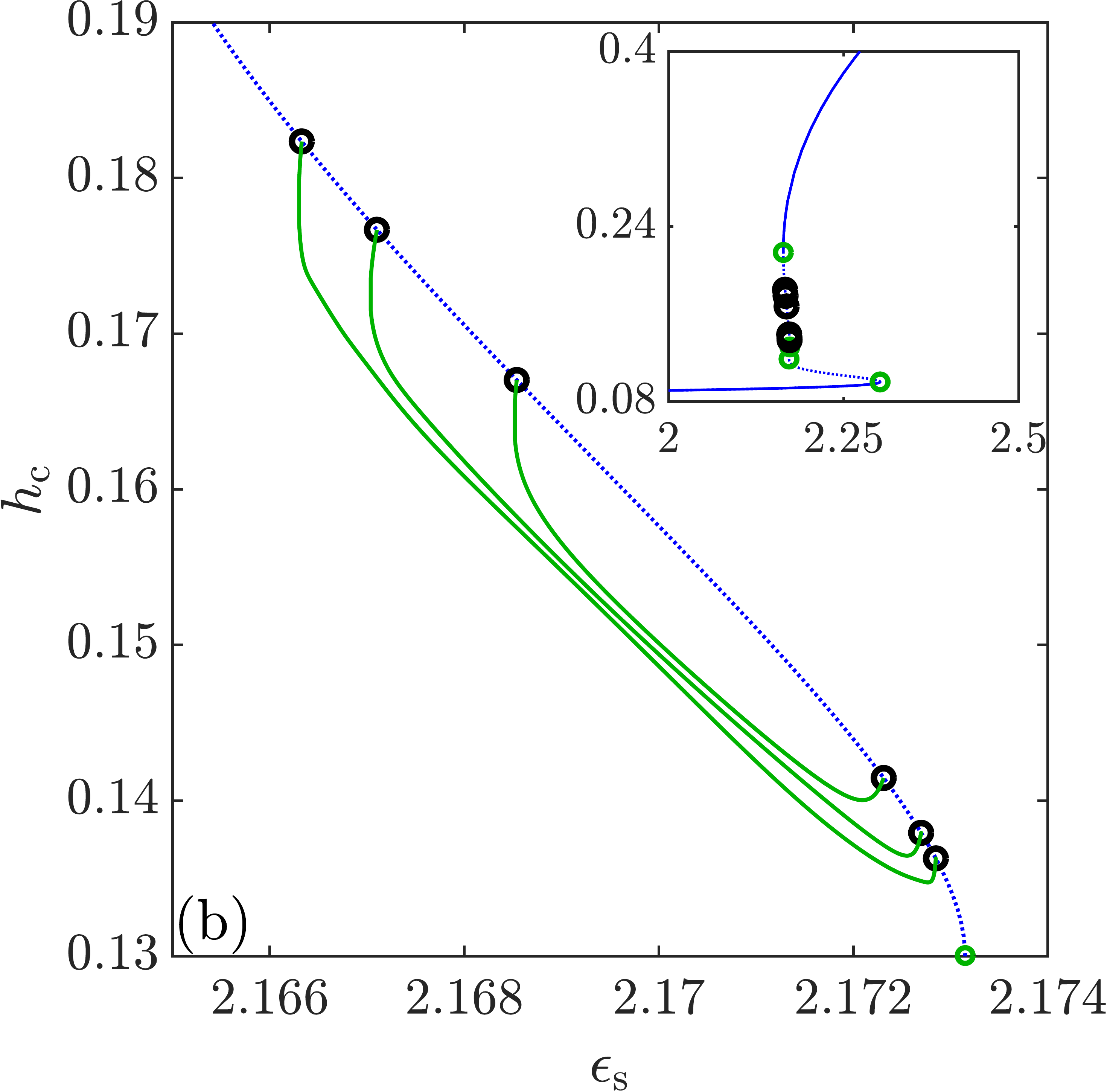}
\end{minipage}
\caption{Panels (a) and (b) give selected parts of the bifurcation diagrams, which show the (time-averaged) coating film thickness $h_c$ as a function of the SAW strength $\epsilon_\mathrm{s}$ at the fixed Weber number values $\mathrm{We}_\mathrm{s}=0.55$ and $\mathrm{We}_\mathrm{s}=0.72$, respectively. Blue and green lines represent steady and time-periodic states, respectively. Hopf bifurcations are marked by black circles. The remaining parameters are as in Fig.~\ref{doss:fig:we_ha1_hopfbif_eps_hopfcont}. The insets give the relevant complete bifurcation curve for the steady states.
}
\label{hp:fig:hopf_branch_ha1_we055_and_we072_hc_eps}
\end{figure}

To study such qualitative changes we numerically determine bifurcation diagrams, which contain steady states and TPS as a function of $\epsilon_\mathrm{s}$ for a sequence of fixed increasing Weber numbers, $\mathrm{We}_\mathrm{s}$. We commence in Fig.~\ref{hp:fig:hopf_branch_ha1_we055_and_we072_hc_eps}~(a) and (b) with two relatively simple diagrams at fixed $\mathrm{We}_\mathrm{s}=0.55$ and $\mathrm{We}_\mathrm{s}=0.72$, respectively. Two and six Hopf bifurcations occur, respectively, and no global bifurcations. 

\begin{figure}[hbt]
\begin{minipage}{0.45\textwidth}
\includegraphics[width=1.\textwidth]{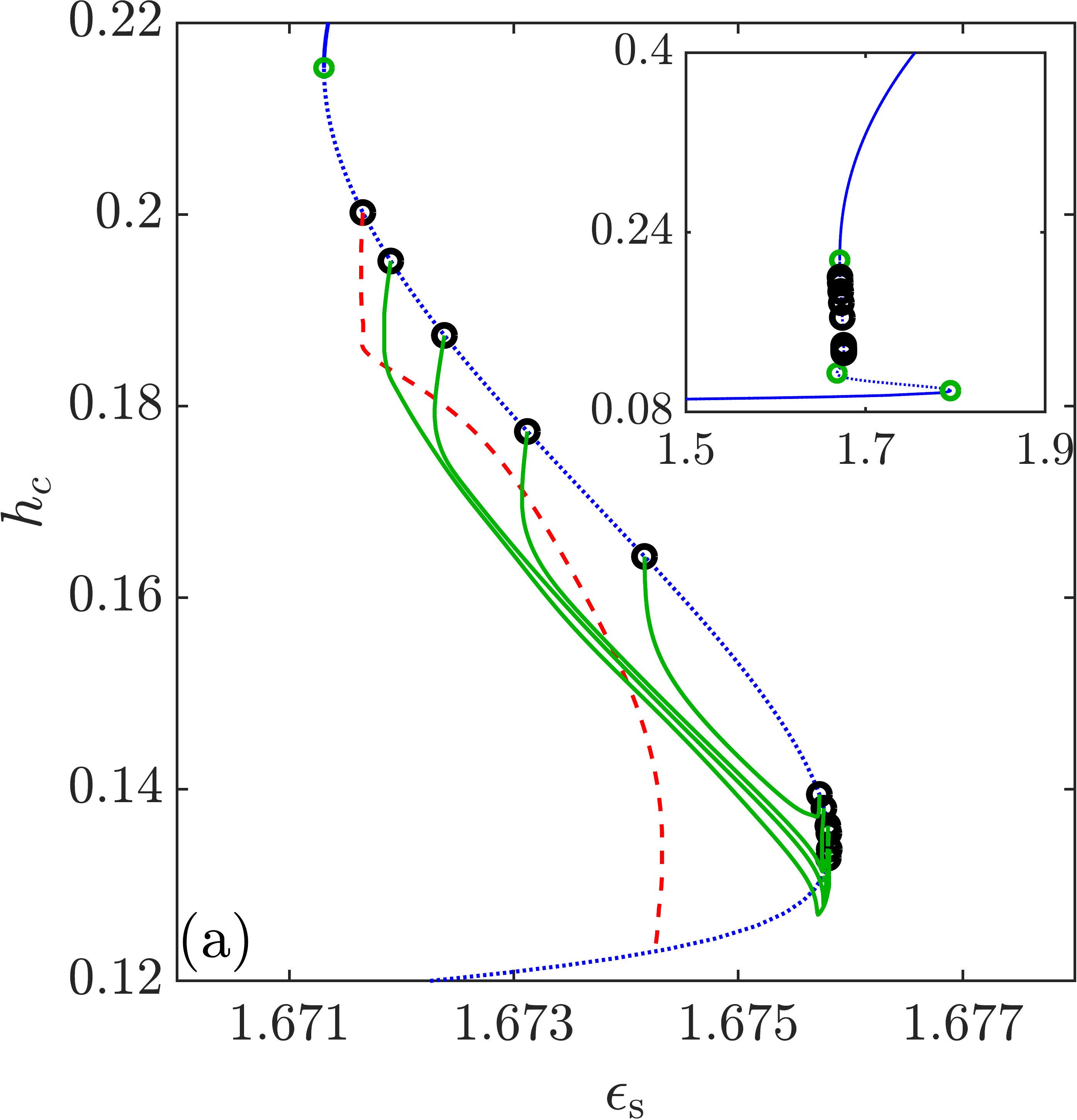}
\end{minipage}
\hspace{0.05\hsize}
\begin{minipage}{0.45\textwidth}
\includegraphics[width=1.\textwidth]{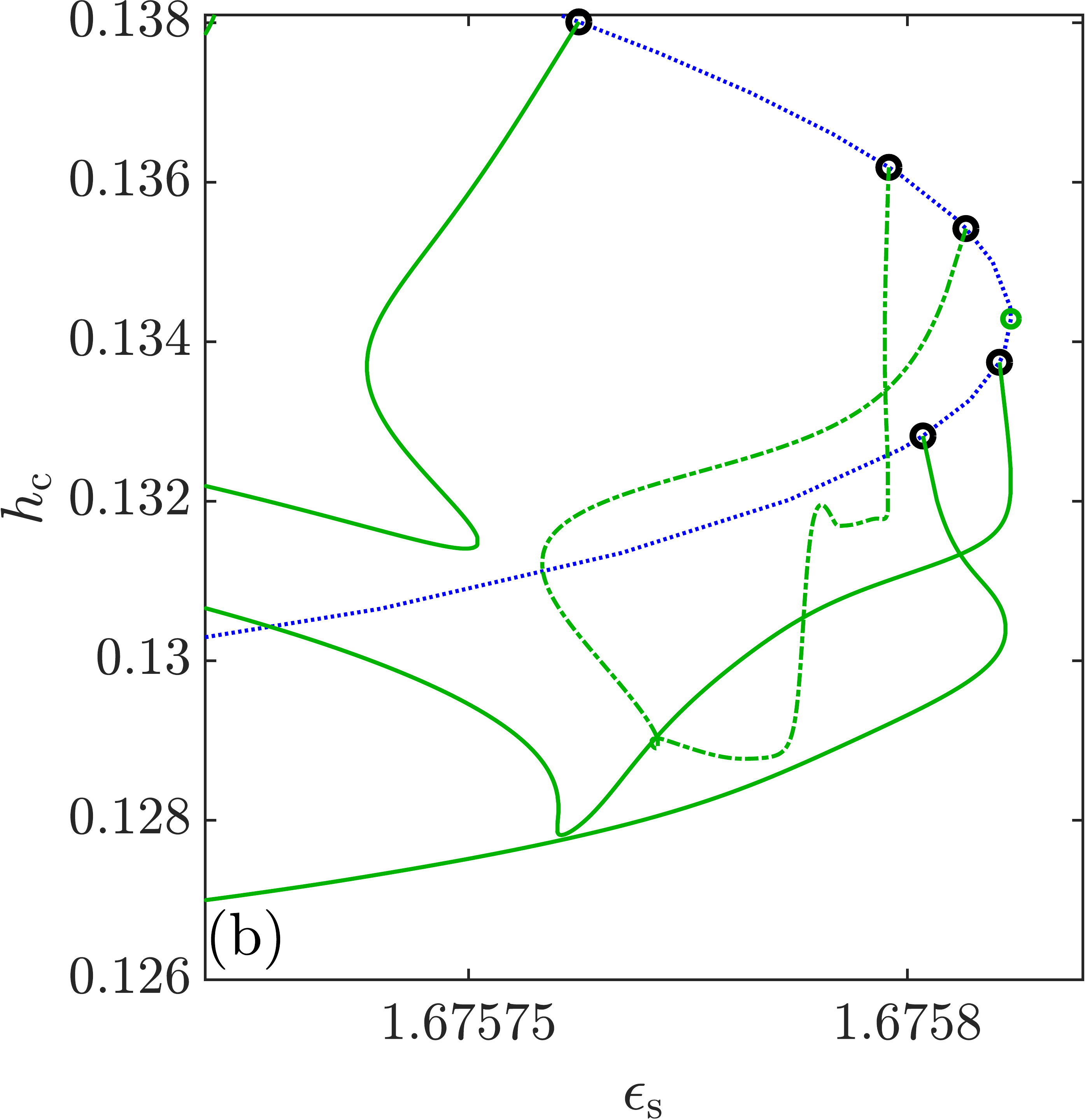}
\end{minipage}
\caption{
Panel~(a) gives a bifurcation diagram showing the (time-averaged) coating thickness $h_c$ in dependence of the SAW strength $\epsilon_\mathrm{s}$ at $\mathrm{We}_\mathrm{s}=0.93$. Remaining parameters, line styles, symbols and inset are as in Fig.~\ref{hp:fig:hopf_branch_ha1_we055_and_we072_hc_eps}.  The branches of TPS, which connect two Hopf bifurcations are shown in green, while the single branch connecting a Hopf and a homoclinic bifurcation is shown as dashed red line. Panel (b) magnifies the region close to the saddle-node bifurcation where panel (a) is very crowded. 
The corresponding periods of the TPS on  $\epsilon_\mathrm{s}$ are presented in Fig.~\ref{hp:fig:hopf_hc_we093_time_periode_zoom}.
}
\label{hp:fig:hopf_hc_we093_bif_inset_zoom}
\end{figure}

\begin{figure}[hbt]
\centering
\includegraphics[width=0.5\textwidth]{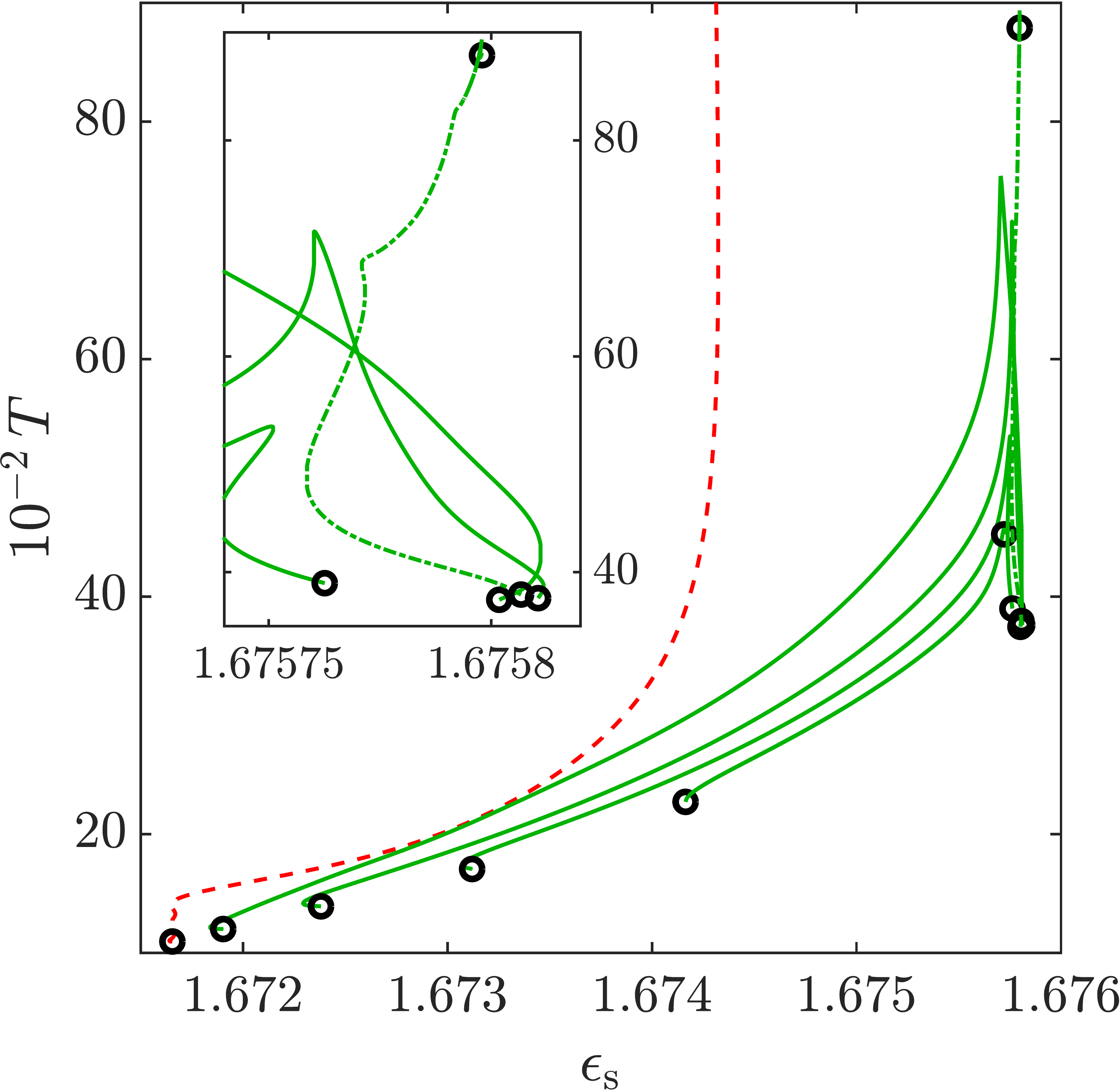}

\caption{
Time period $T$ as a function of $\epsilon_\mathrm{s}$ for all branches of TPS in Fig.~\ref{hp:fig:hopf_hc_we093_bif_inset_zoom}, line styles are identical. The inset provides a magnification.
}
\label{hp:fig:hopf_hc_we093_time_periode_zoom}
\end{figure}

{Increasing $\mathrm{We}_\mathrm{s}$, we observe at $\mathrm{We}_\mathrm{s}=0.93$ an odd number of eleven Hopf bifurcations, see bifurcation diagram in Fig.~\ref{hp:fig:hopf_hc_we093_bif_inset_zoom} and corresponding periods of TPS in Fig.~\ref{hp:fig:hopf_hc_we093_time_periode_zoom}. This implies that a first Bogdanov-Takens bifurcation has occurred and a homoclinic bifurcation is expected to exist. The obtained branches of TPS are consistent with the changes in stability observed when following the branch of steady states with increasing $\epsilon_\mathrm{s}$ from small values: One starts with a stable meniscus state, first passes two destabilizing saddle-node bifurcations [inset of Fig.~\ref{hp:fig:hopf_hc_we093_bif_inset_zoom}~(a)], then a global bifurcation (not affecting stability), two destabilizing Hopf bifurcations and one destabilizing saddle-node bifurcation. After the latter, now with decreasing $\epsilon_\mathrm{s}$ further Hopf bifurcations occur: one destabilizing, one stabilizing, two  destabilizing and five stabilizing.  The resulting steady state has still one unstable eigenvalue that is stabilized at the final saddle-node bifurcation.}

{We emphazise that branches of TPS present a particular numerical challenge when the period becomes large. In Fig.~\ref{hp:fig:hopf_hc_we093_time_periode_zoom} this is the case when approaching the global bifurcation where the period $T$ is expected to diverge.  With the present numerical method we are therefore not able to completely continue all branches of TPS. However, overall, Fig.~\ref{hp:fig:hopf_hc_we093_time_periode_zoom} supports our interpretation. It shows the period $T$ for all TPS branches in Fig.~\ref{hp:fig:hopf_hc_we093_bif_inset_zoom}.  One finds that $T$ for the branch ending at the global bifurcation (red dashed line), indeed seems to diverge. However, also some of the other branches show strongly increasing $T$.}

\begin{figure}[hbt]
  \centering
\begin{minipage}{0.4\textwidth}
\vspace{0.5cm}
\includegraphics[width=.9\textwidth]{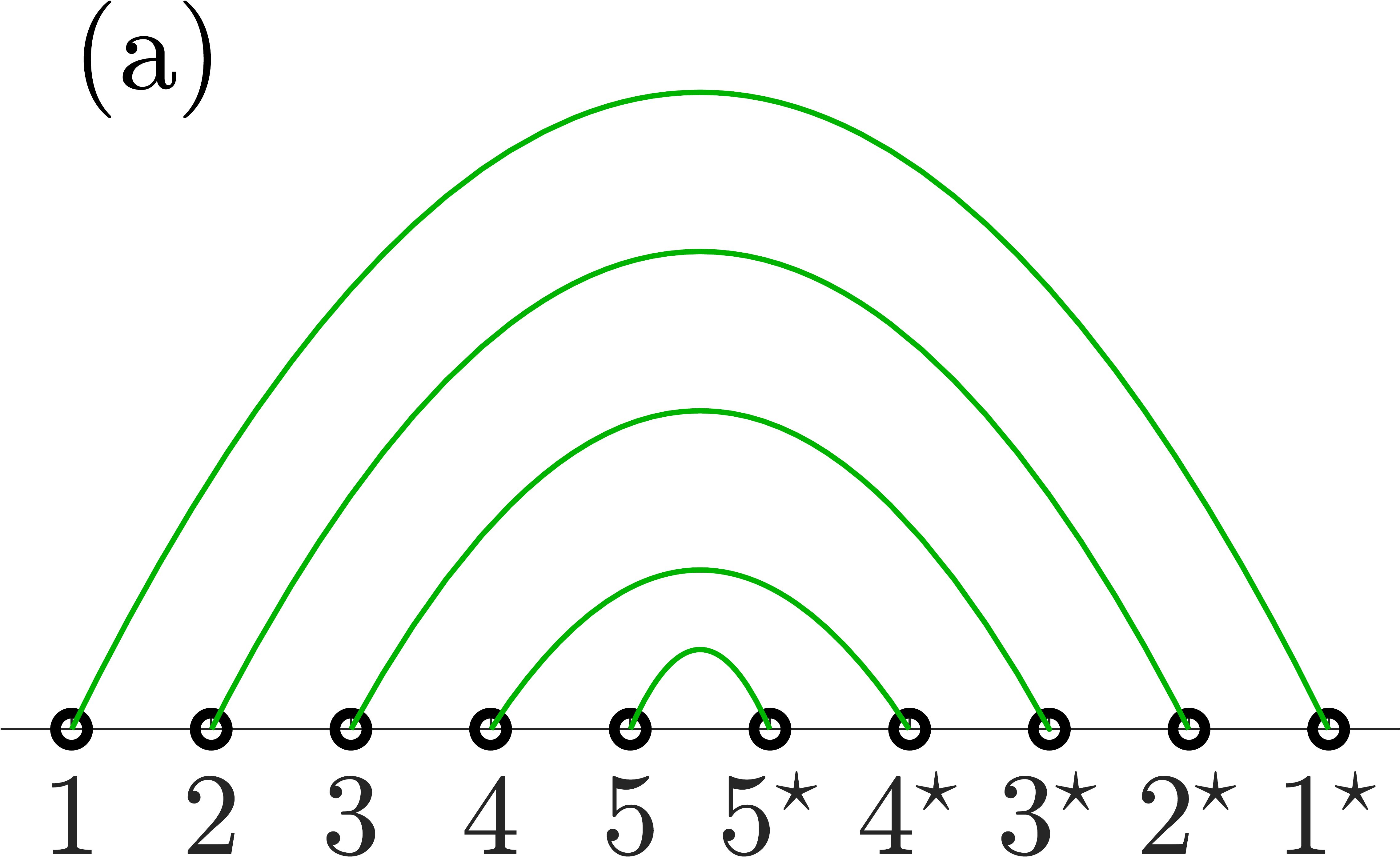}\vspace{0.5cm}
\vspace{0.5cm}
\includegraphics[width=.9\textwidth]{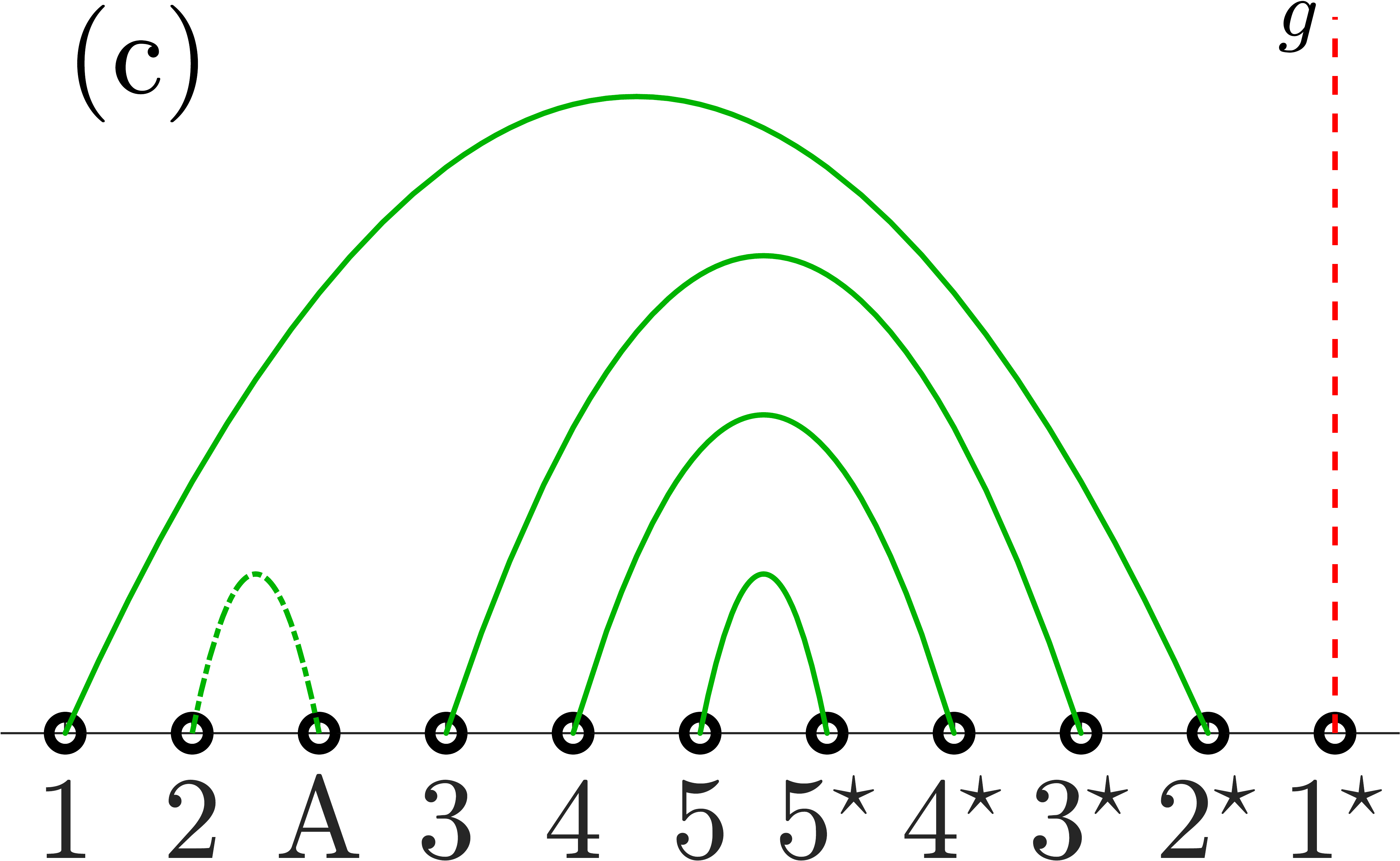}
\vspace{0.5cm}
\includegraphics[width=.9\textwidth]{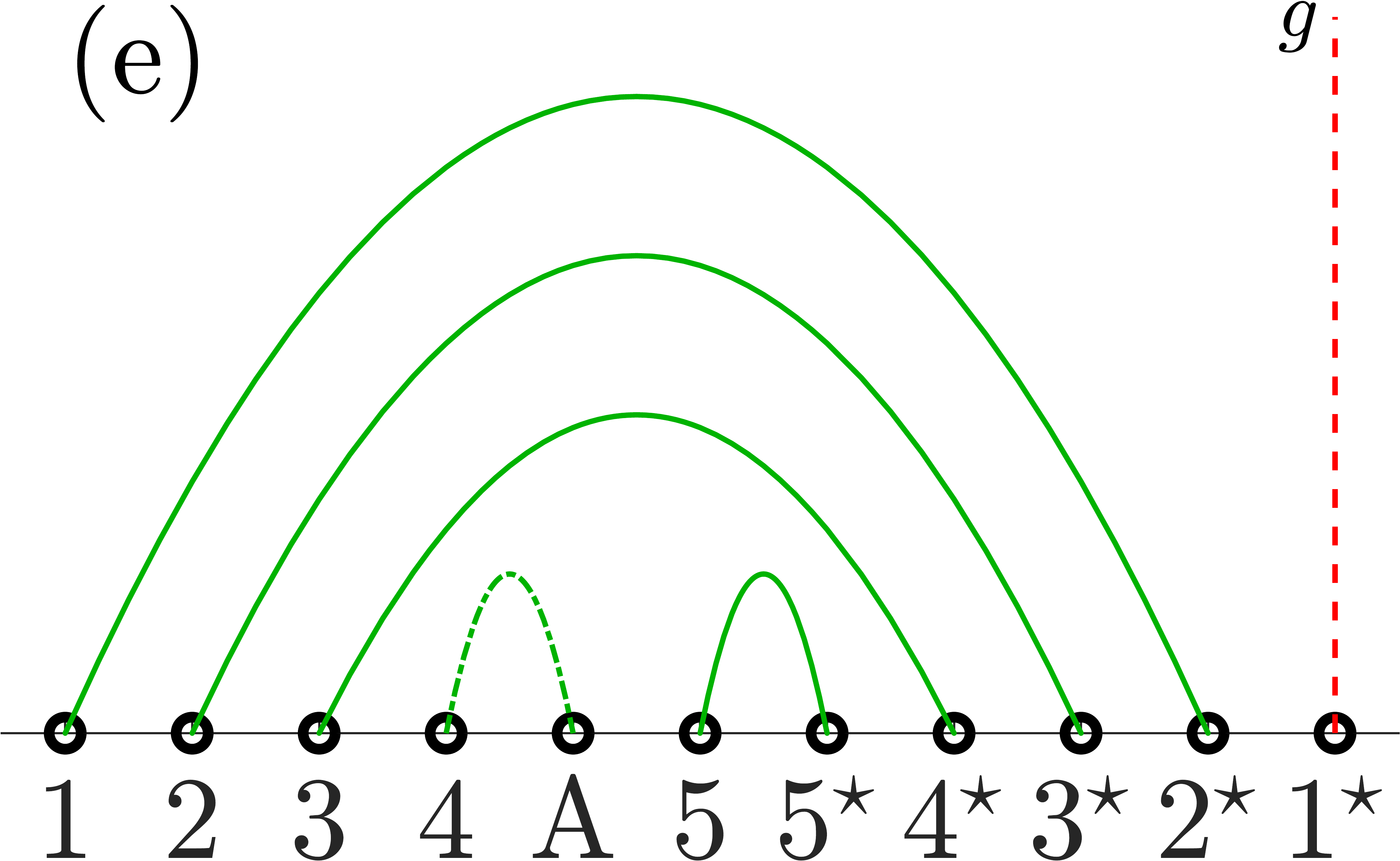}
\end{minipage}
\begin{minipage}{0.4\textwidth}
\vspace{0.5cm}
\includegraphics[width=.9\textwidth]{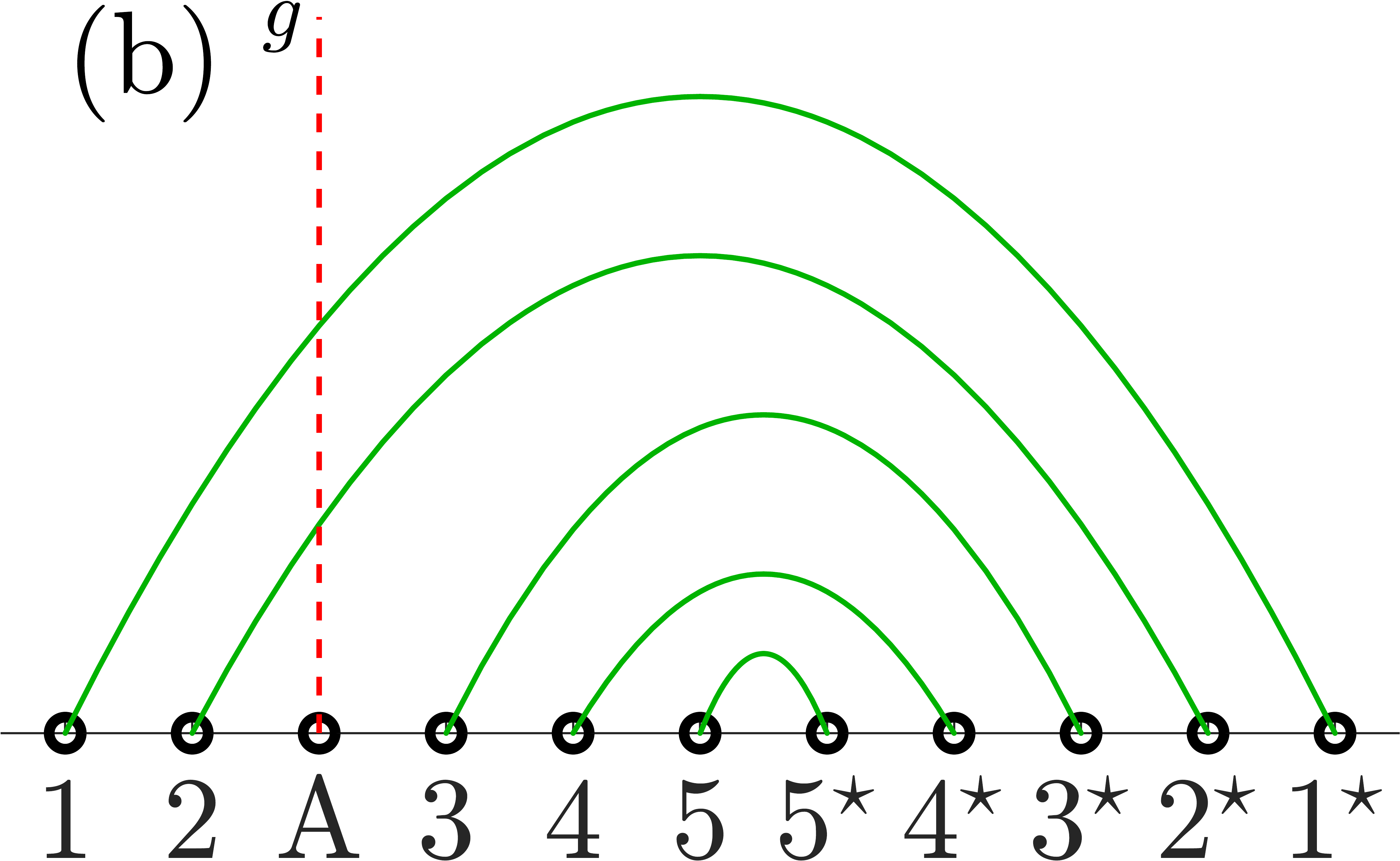}\vspace{0.5cm}
\vspace{0.5cm}
\includegraphics[width=.9\textwidth]{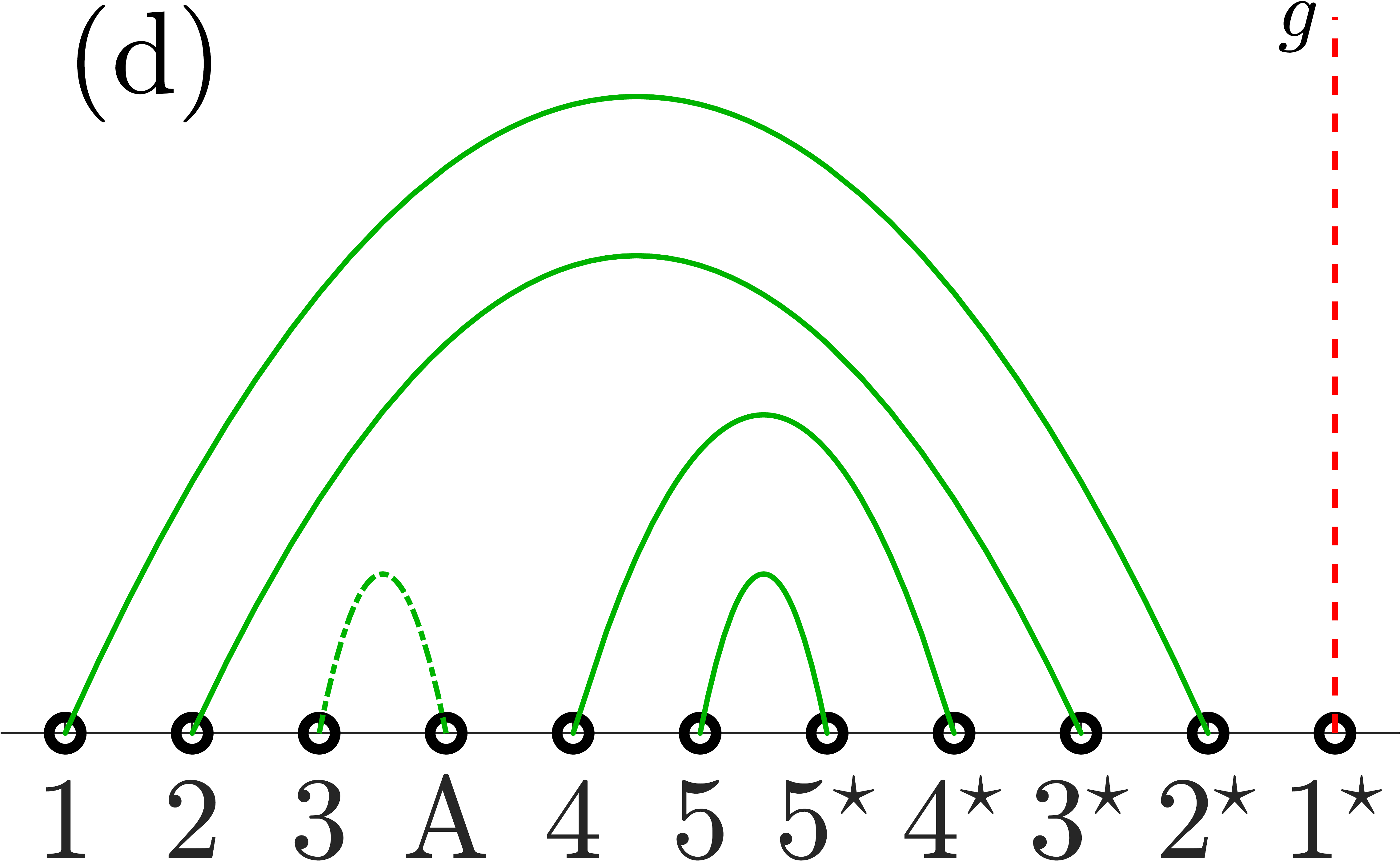}
\vspace{0.5cm}
\includegraphics[width=.9\textwidth]{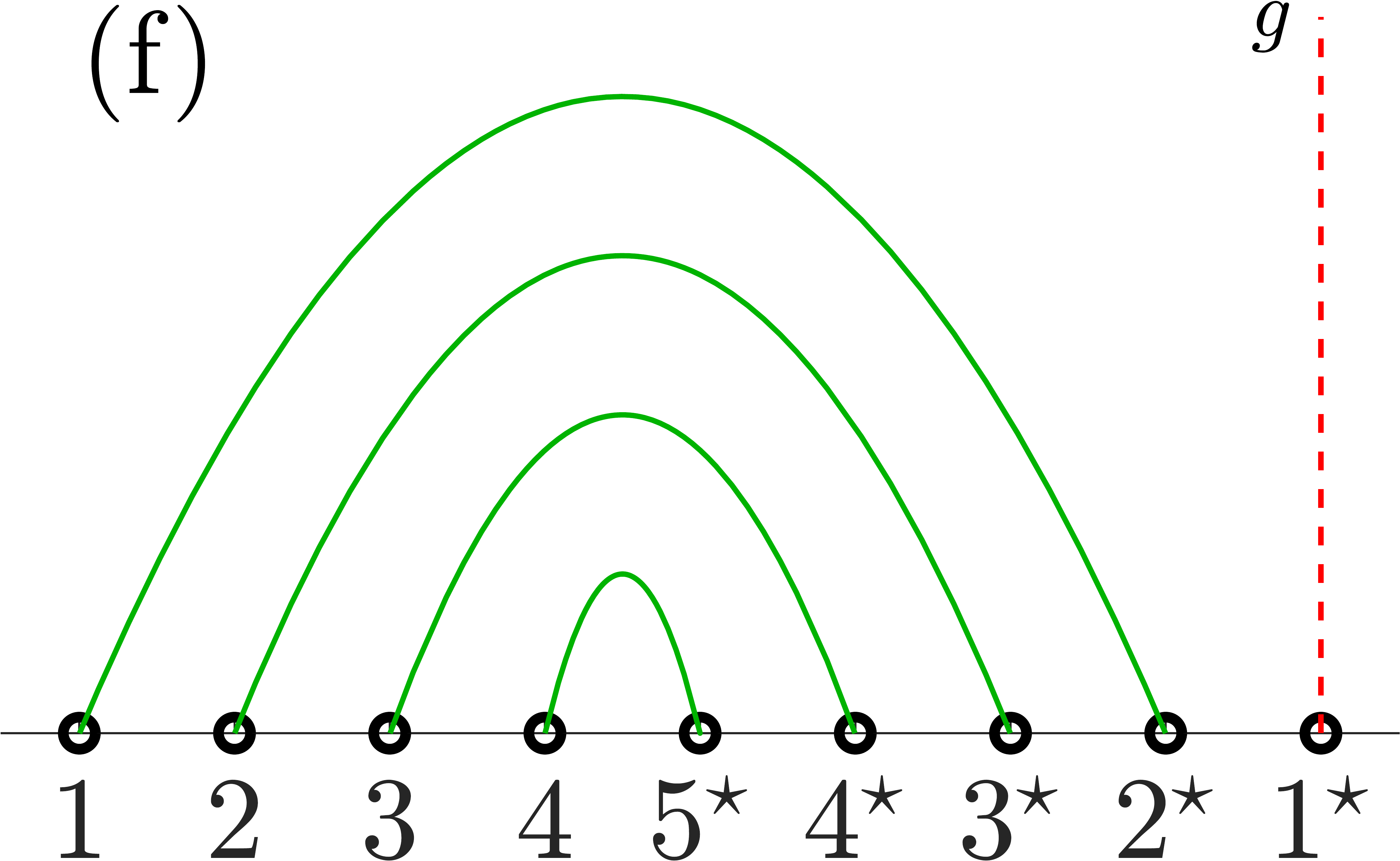}
\end{minipage}
\caption{Illustration of the motion of Hopf bifurcations and the related reconnections of the branches of TPS.  The different steps are schematically represented in parts (a) to (f) with line styles as in Fig.~\ref{hp:fig:hopf_hc_we093_bif_inset_zoom}.  The branch connecting Hopf bifurcation A with another Hopf bifurcation is given as green dot-dashed line. The vertical axes symbolically represents a solution measure, e.g., period $T$. Vertical lines indicate connection to a homoclinic bifurcation. } \label{hp:fig:hopf_illustration_bogdanov} \end{figure}

{For clarity, in Fig.~\ref{hp:fig:hopf_illustration_bogdanov} we schematically illustrate the complicated sequence of branch reconnections that occurs after each Bogdanov-Takens bifurcation.  Fig.~\ref{hp:fig:hopf_illustration_bogdanov}~(a) shows all Hopf bifurcations at $\mathrm{We}_\mathrm{s}=0.9$ (bifurcation diagram not shown) as they occur along the branch of steady states together with their (rather regular) connections. This state defines the numbering of the Hopf bifurcations: there are five pairs ``1-1$^\star$'' to ``5-5$^\star$'' of directly connected pairs of Hopf bifurcations (green lines). Fig.~\ref{hp:fig:hopf_illustration_bogdanov}~(b) gives a case, shortly after a Bogdanov-Takens bifurcation has occurred, adding the TPS branch (red line) connecting the pair ``A-g'', where ``g'' stands for the global bifurcation and ``A'' indicates another Hopf bifurcation. Figs.~\ref{hp:fig:hopf_illustration_bogdanov}~(c) to \ref{hp:fig:hopf_illustration_bogdanov}~(f) illustrate subsequent reconnections that ultimately result in the connection of the outermost Hopf bifurcation 1$^\star$ and the global bifurcation. In Fig.~\ref{hp:fig:hopf_illustration_bogdanov}~(f) all remaining Hopf bifurcations are again well ordered as in Fig.~\ref{hp:fig:hopf_illustration_bogdanov}~(a). Note that the topology of Fig.~\ref{hp:fig:hopf_illustration_bogdanov}~(d) is identical to Fig.~\ref{hp:fig:hopf_hc_we093_bif_inset_zoom}.}

\begin{figure}[hbt]
\begin{minipage}{0.45\textwidth}
\includegraphics[width=1.\textwidth]{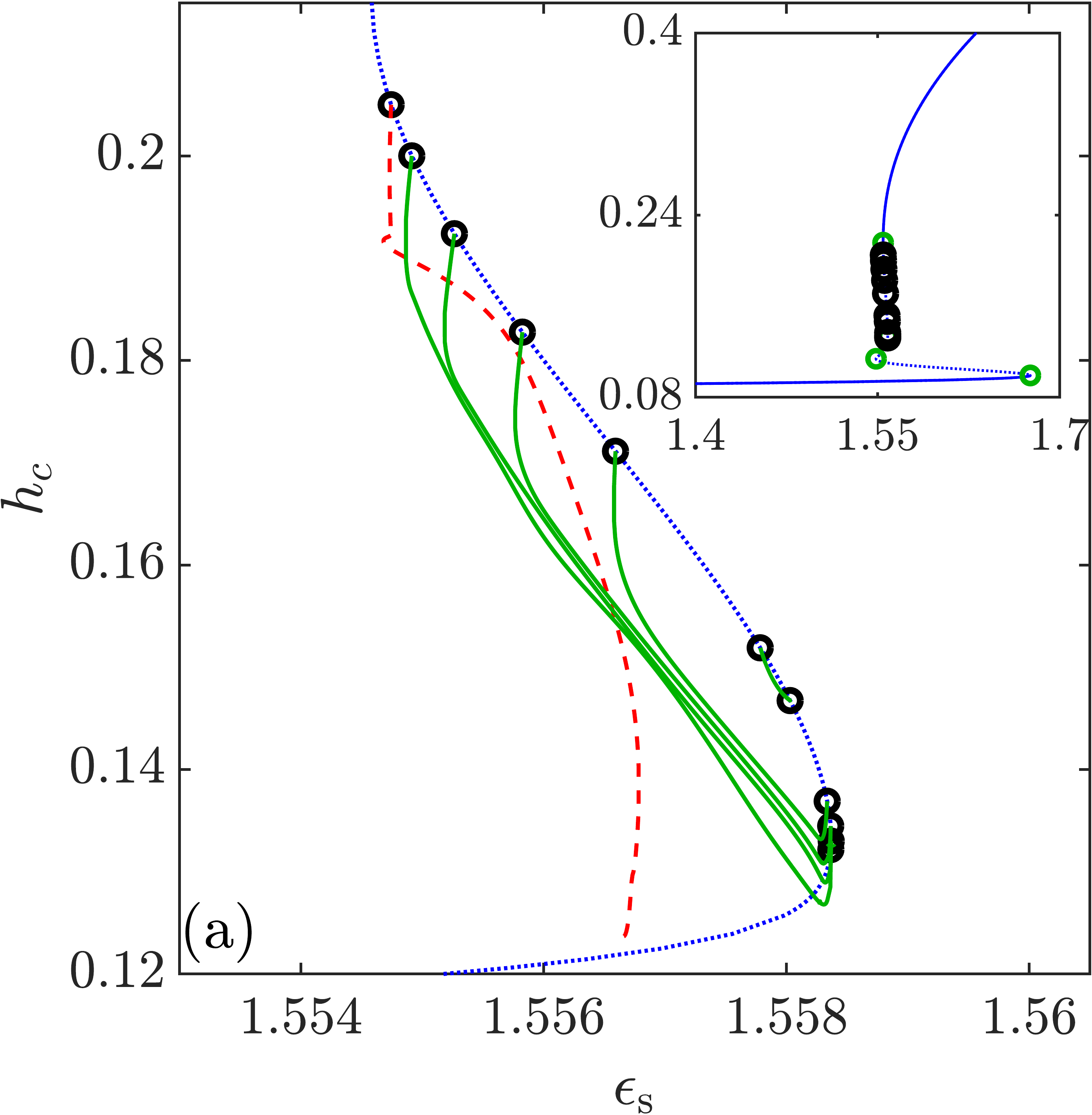}
\end{minipage}
\hspace{0.05\hsize}
\begin{minipage}{0.45\textwidth}
\includegraphics[width=1.\textwidth]{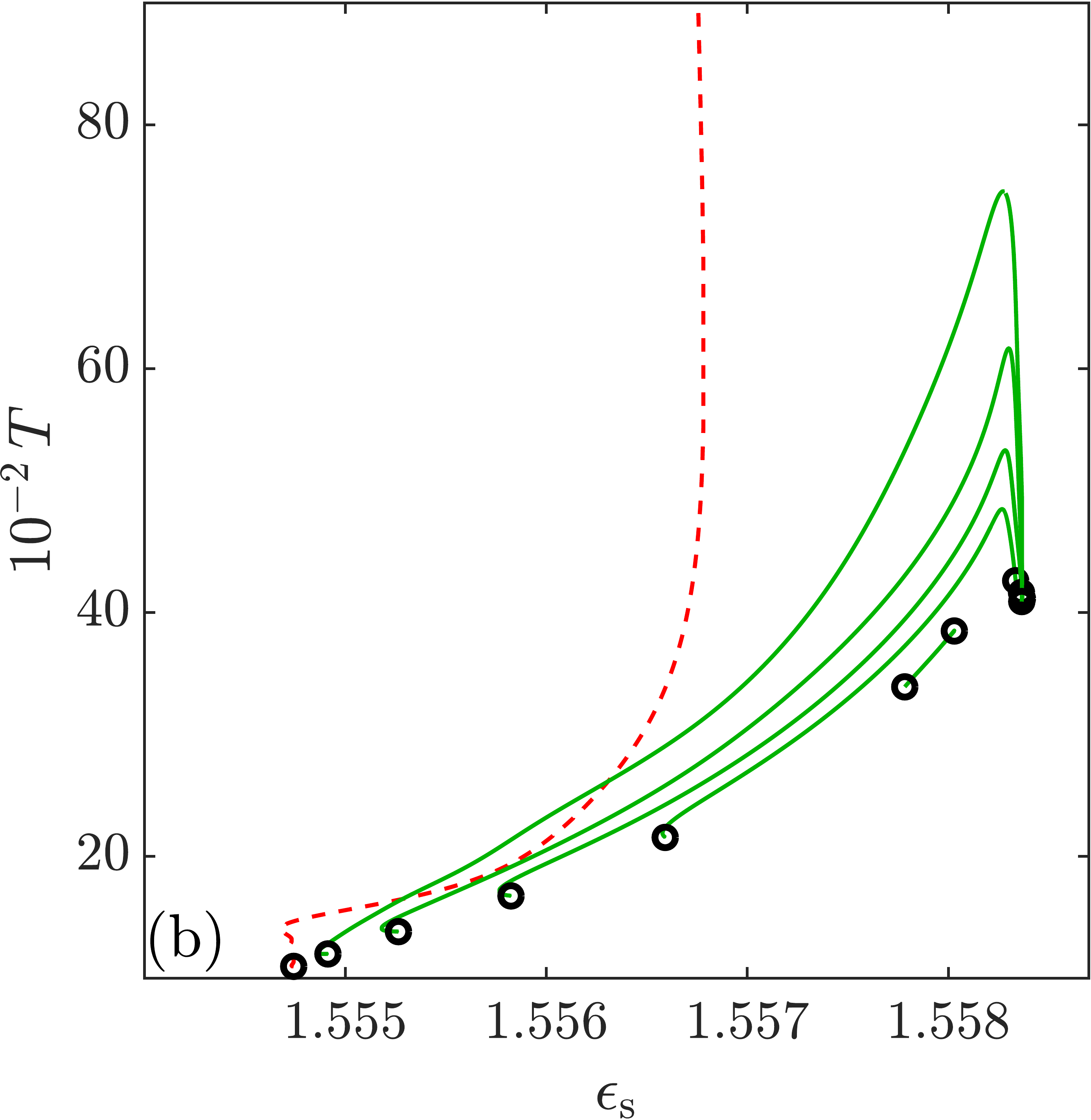}
\end{minipage}
\caption{(a) Bifurcation diagram showing the (time-averaged) coating thickness $h_c$ in dependence of the SAW strength $\epsilon_\mathrm{s}$ at $\mathrm{We}_\mathrm{s}=1.0$. 
Remaining parameters, line styles, symbols and inset as in Fig.~\ref{hp:fig:hopf_hc_we093_bif_inset_zoom}. Panel (b) gives the corresponding periods $T$ of the TPS.
}
\label{hp:fig:hopf_hc_we1_bif_inset_}
\end{figure}

{Overall, the final state in Fig.~\ref{hp:fig:hopf_illustration_bogdanov}~(f) only  differs from the second state in Fig.~\ref{hp:fig:hopf_illustration_bogdanov}~(b) by the position 
of the connection to the global bifurcation. It has moved from the third bifurcation from the left to the rightmost one. A proper calculation of the bifurcation diagram corresponding to Fig.~\ref{hp:fig:hopf_illustration_bogdanov}~(f) is done at $\mathrm{We}_\mathrm{s}=1.0$, see Fig.~\ref{hp:fig:hopf_hc_we1_bif_inset_}~(a). Note that there is a slight mismatch between sketch and bifurcation diagram as the sketch focuses on the reconnections, but omits the emergence in a double Hopf bifurcation on an additional TPS branch visible in Fig.~\ref{hp:fig:hopf_hc_we1_bif_inset_}~(a) at about $\epsilon_\mathrm{s}=1.558$.} Overall, in this cascade of topological changes of the bifurcation diagram, all branches of TPS are reconnected once, so that finally each surviving Hopf bifurcation $n$ (that was initially connected to Hopf bifurcation $(n)^\star$) is connected to bifurcation $(n+1)^\star$. The entire described sequence of events is triggered each time a Bogdanov-Takens bifurcation occurs. Hence, the overall rearrangement of the bifurcation diagram involves a large number of codimension-2 events.

\begin{figure}[hbt]
\begin{minipage}{0.45\textwidth}
\includegraphics[width=1.\textwidth]{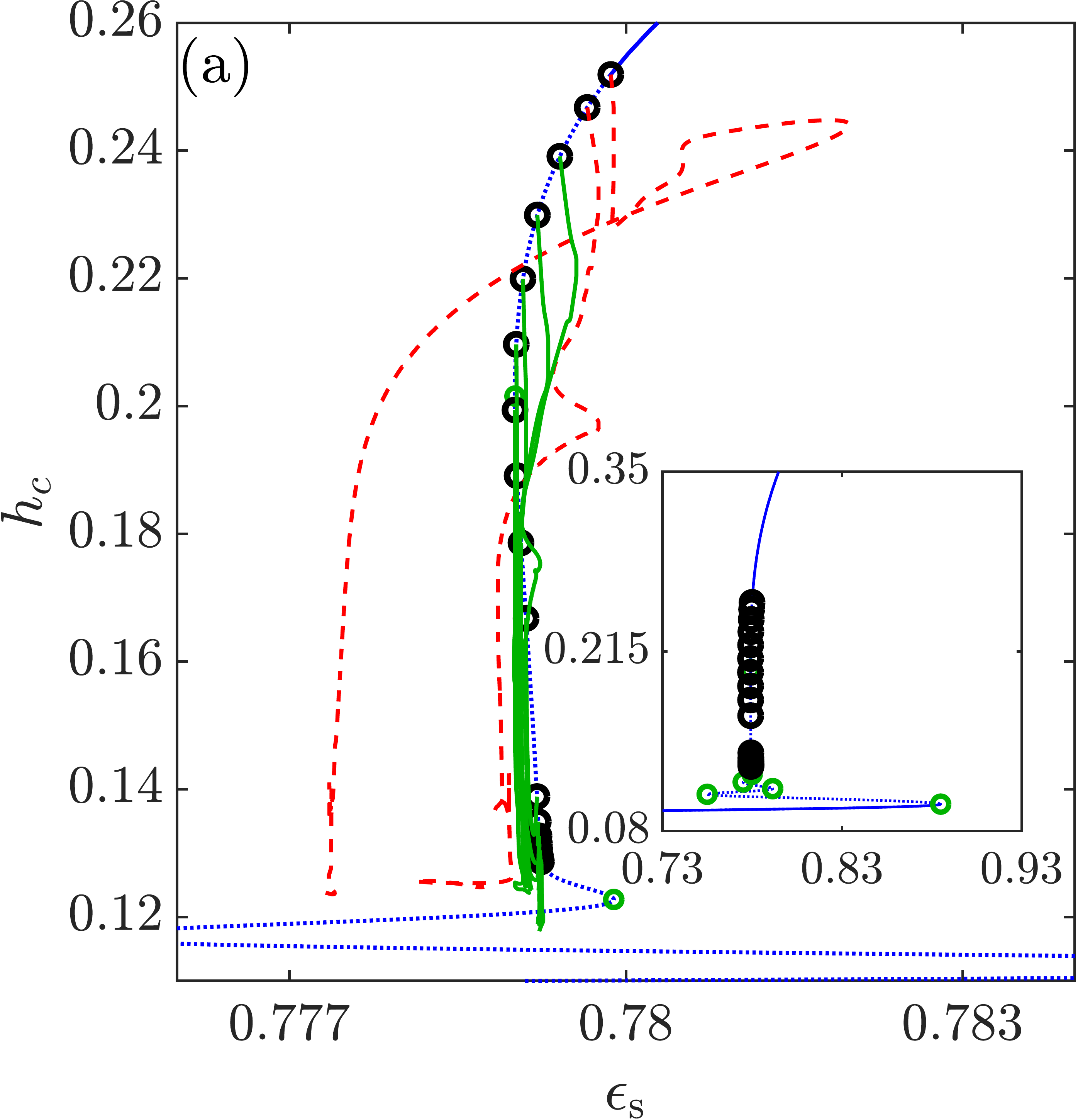}
\end{minipage}
\hspace{0.05\hsize}
\begin{minipage}{0.45\textwidth}
\includegraphics[width=1.\textwidth]{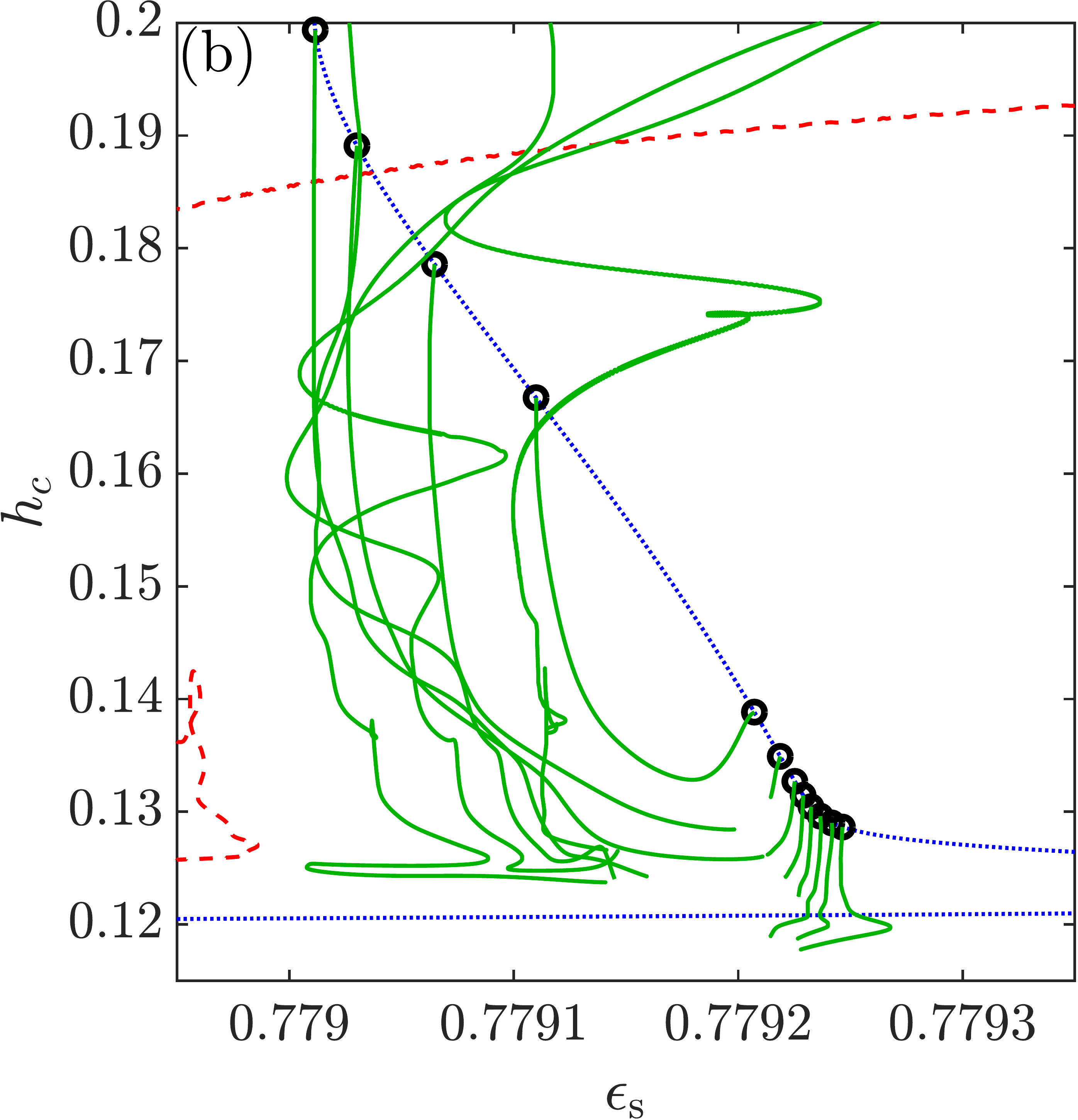}
\end{minipage}
\centering
\caption{Bifurcation diagram, which shows the (time-averaged) coating film thickness $h_c$ as function of the SAW strength $\epsilon_\mathrm{s}$ at $\mathrm{We}_\mathrm{s}=2.0$. 
The full diagram is shown as inset in panel (a) while panels (a) and (b) magnify the most interesting regions. 
  Remaining parameters, line styles, and symbols are as in Fig.~\ref{hp:fig:hopf_hc_we093_bif_inset_zoom}. 
  Fig.~\ref{hp:fig:hopf_hc_we20_time_periode_zoom} shows the corresponding periods of the TPS and selected spacetime plots of TPS.
}
\label{hp:fig:hopf_branch_ha1_we2_hc_eps}
\end{figure}

{We exemplify this, by presenting in Fig.~\ref{hp:fig:hopf_branch_ha1_we2_hc_eps} a bifurcation diagram at larger $\mathrm{We}_\mathrm{s} = 2.0$. There exist 11 branches of TPS. Two connect Hopf and homoclinic bifurcations, the other eight pairs of Hopf bifurcations. The branches involving a global bifurcation have already terminated the sequence of reconnections described above. Inspecting the individual branches we see that they acquired a more complex structure and now feature many more saddle-node bifurcations. This tendency is stronger for the outer branches than for the inner ones. Also the range of periods $T$ of the branches connecting two Hopf bifurcations has increased (not shown). 
This also explains why numerical difficulties stop us from calculating all TPS branches in their full extension [see Fig.~\ref{hp:fig:hopf_branch_ha1_we2_hc_eps}~(b)].}

\begin{figure}[hbt]
\centering
\includegraphics[angle=0,width=0.8\textwidth]{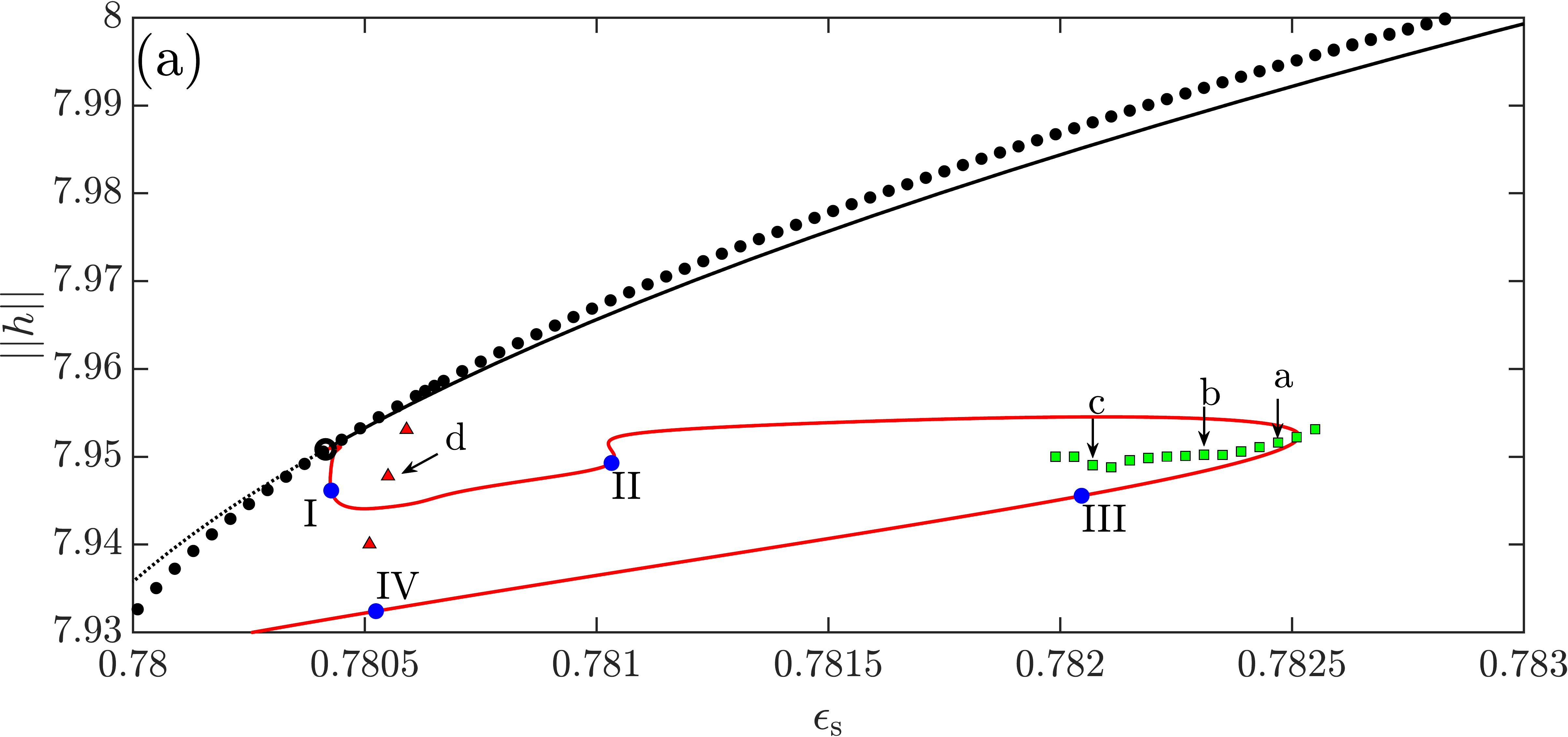}
\hspace{0.05\hsize}
\includegraphics[angle=0,width=.4\textwidth]{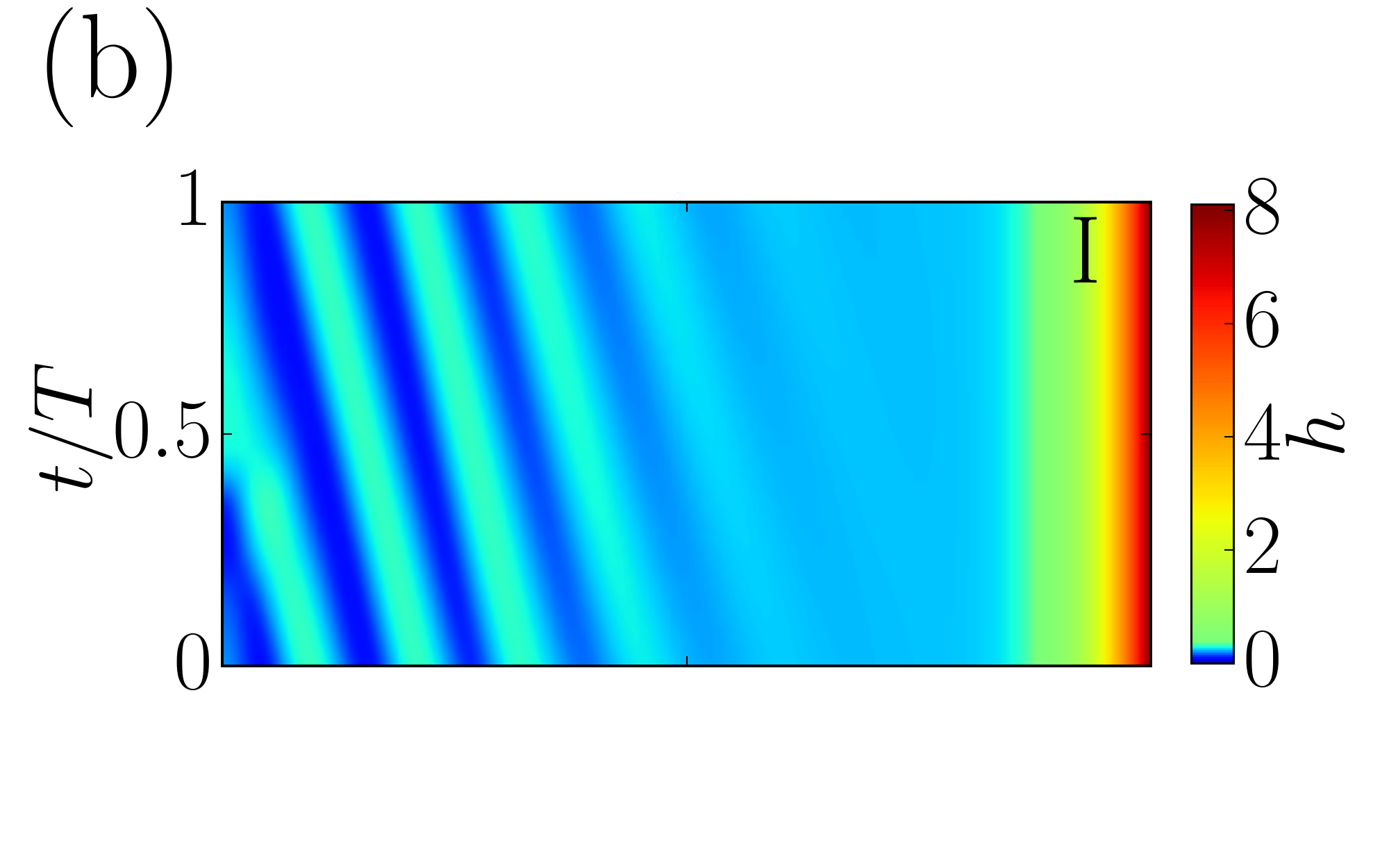}
\hspace{0.05\hsize}
\includegraphics[angle=0,width=.4\textwidth]{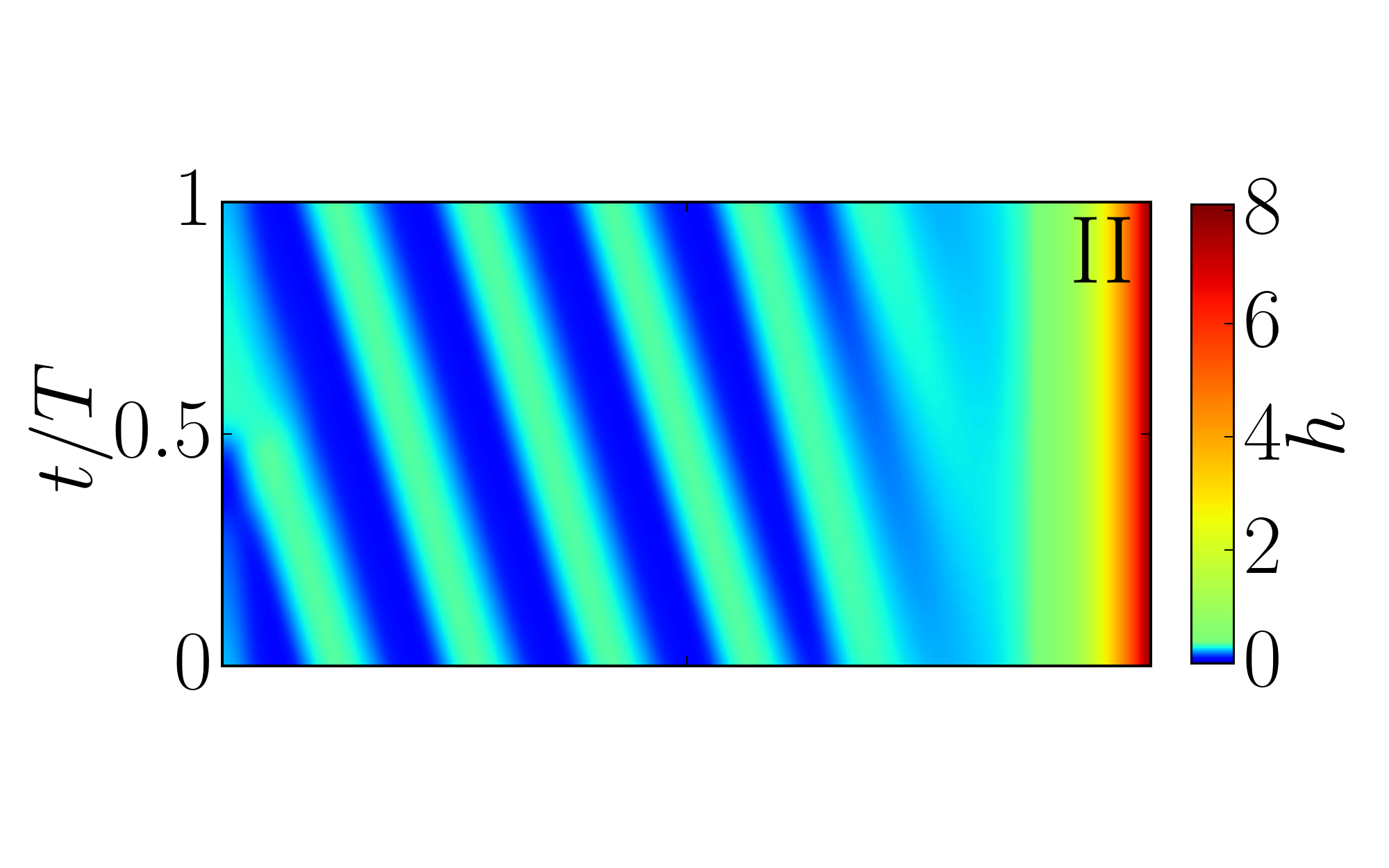}\\[-8ex]
\includegraphics[angle=0,width=.4\textwidth]{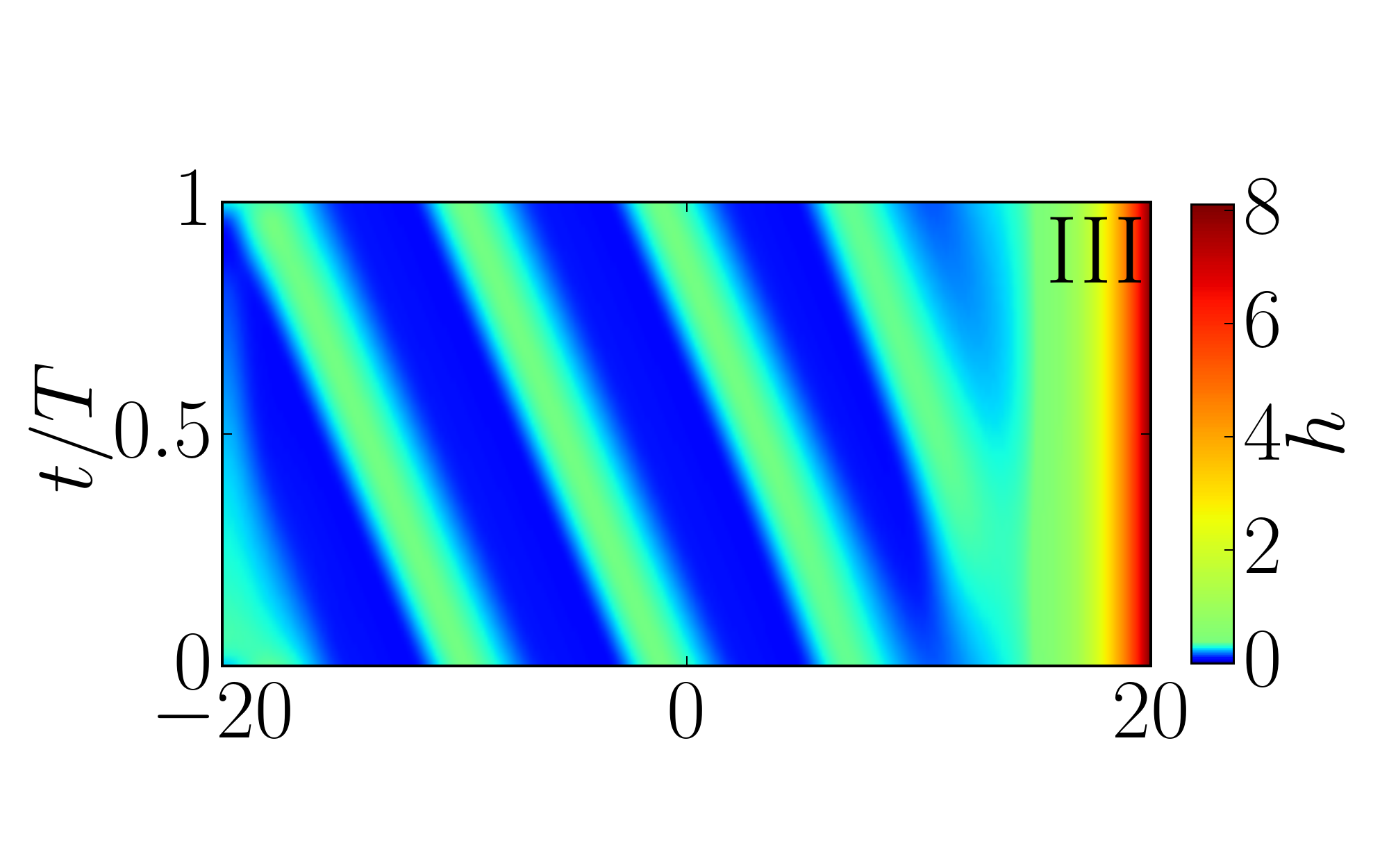}
\hspace{0.05\hsize}
\includegraphics[angle=0,width=.4\textwidth]{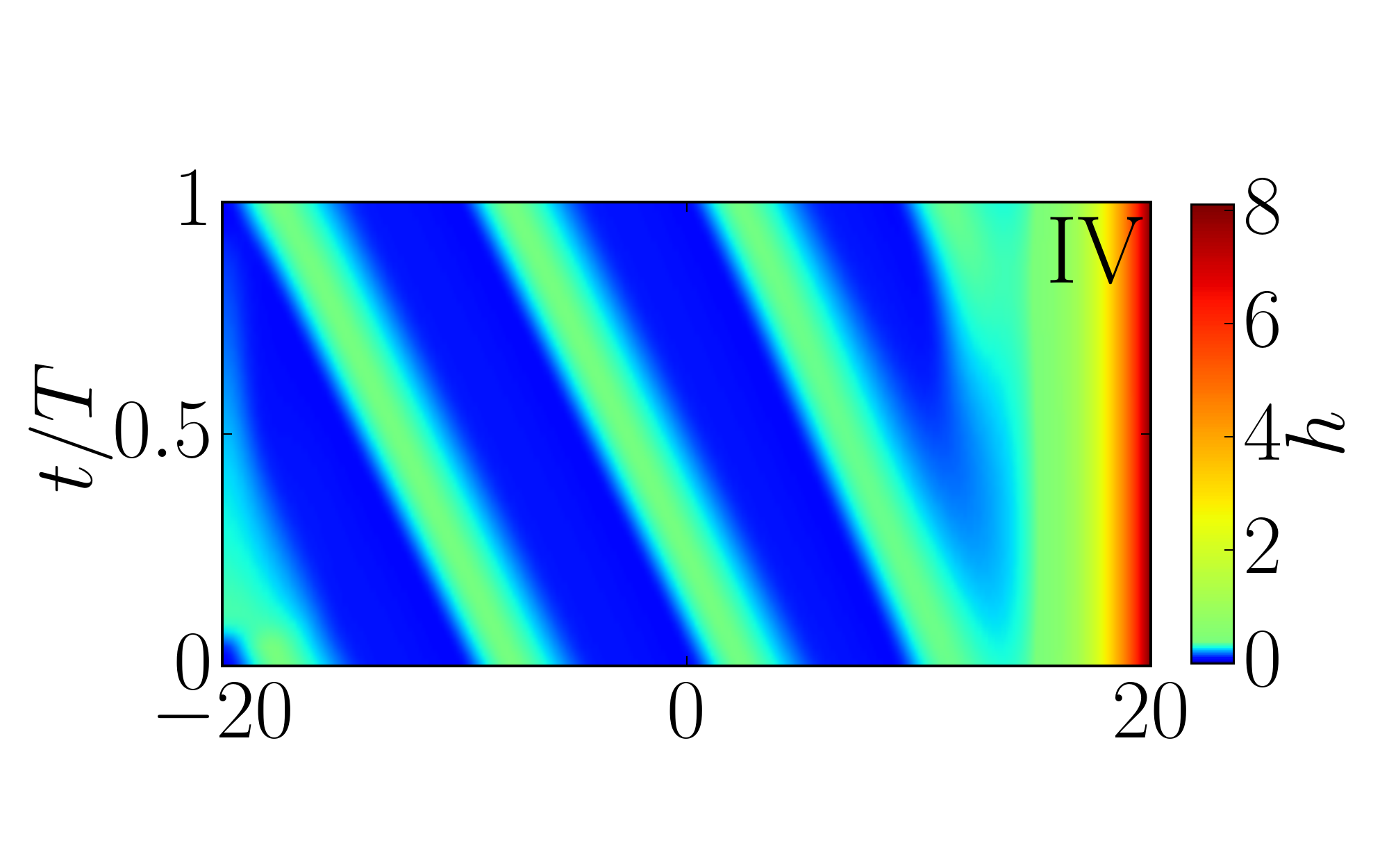}
\caption{{Shown is a magnification of the bifurcation diagram in Fig.~\ref{hp:fig:hopf_branch_ha1_we2_hc_eps} in the region where the final Hopf bifurcation is located using the L$^2$-norm $||h||$ as solution measure.
The black [red] solid lines represent steady [time-periodic] states obtained by continuation. The blue dots marked ``I'' to ``IV'' indicate TPS with periods (I)  $T = 1238$, (II) $T = 1407 $, (III) $T = 1747$, (IV) $T = 2030$ shown as space-time plots in (b).
The remaining symbols represent results of time simulations, in particular, black circles are Landau-Levich film states, red triangles are chaotic states and green squares are time-periodic states. The points marked ``a'' to ``d'' are presented in space-time plots in Fig.~\ref{hp:fig:solutions_direct_numeric}. Remaining parameters are as in Fig.~\ref{hp:fig:hopf_branch_ha1_we2_hc_eps}. }
}
\label{hp:fig:hopf_hc_we20_time_periode_zoom}
\end{figure}

{We now focus on the physically most relevant branch of TPS, namely, the ``outermost'' branch, i.e., the branch that emerges first when decreasing $\epsilon_\mathrm{s}$. The magnification in Fig.~\ref{hp:fig:hopf_hc_we20_time_periode_zoom}~(a) focuses on the region where it (red line) emerges from the branch of steady states (black line). It bifurcates supercritically, folds back shortly thereafter and undergoes a further saddle-node bifurcation at $\epsilon_\mathrm{s}\approx0.7804$. Typical TPS obtained by continuation are presented in Fig.~\ref{hp:fig:hopf_hc_we20_time_periode_zoom}~(b) for loci marked 
by ``I'' to ``IV'' in Fig.~\ref{hp:fig:hopf_hc_we20_time_periode_zoom}~(a). They all show time-periodic states corresponding to ridges that are shed at regular intervals $T$ from the foot structure at the meniscus} and then move at about constant speed along the substrate until they leave the domain on the left hand side. Close to this point they get slightly deformed due to boundary effects that, however, do not influence the main part of their trajectory. Moving along the branch of TPS away from the Hopf bifurcation, where it emerges, the number of drops in the domain becomes smaller as the shedding events that define the period $T$ become less frequent. The foot structure where the ridges emerge does not change much along the branch. This, however, is in marked contrast when inspecting the TPS on other branches (not shown). The difference between the various branches mainly lies in the details of the foot structure like its length and the number of visible undulations.

\begin{figure}[hbt]
\centering
    \includegraphics[width=0.45\textwidth]{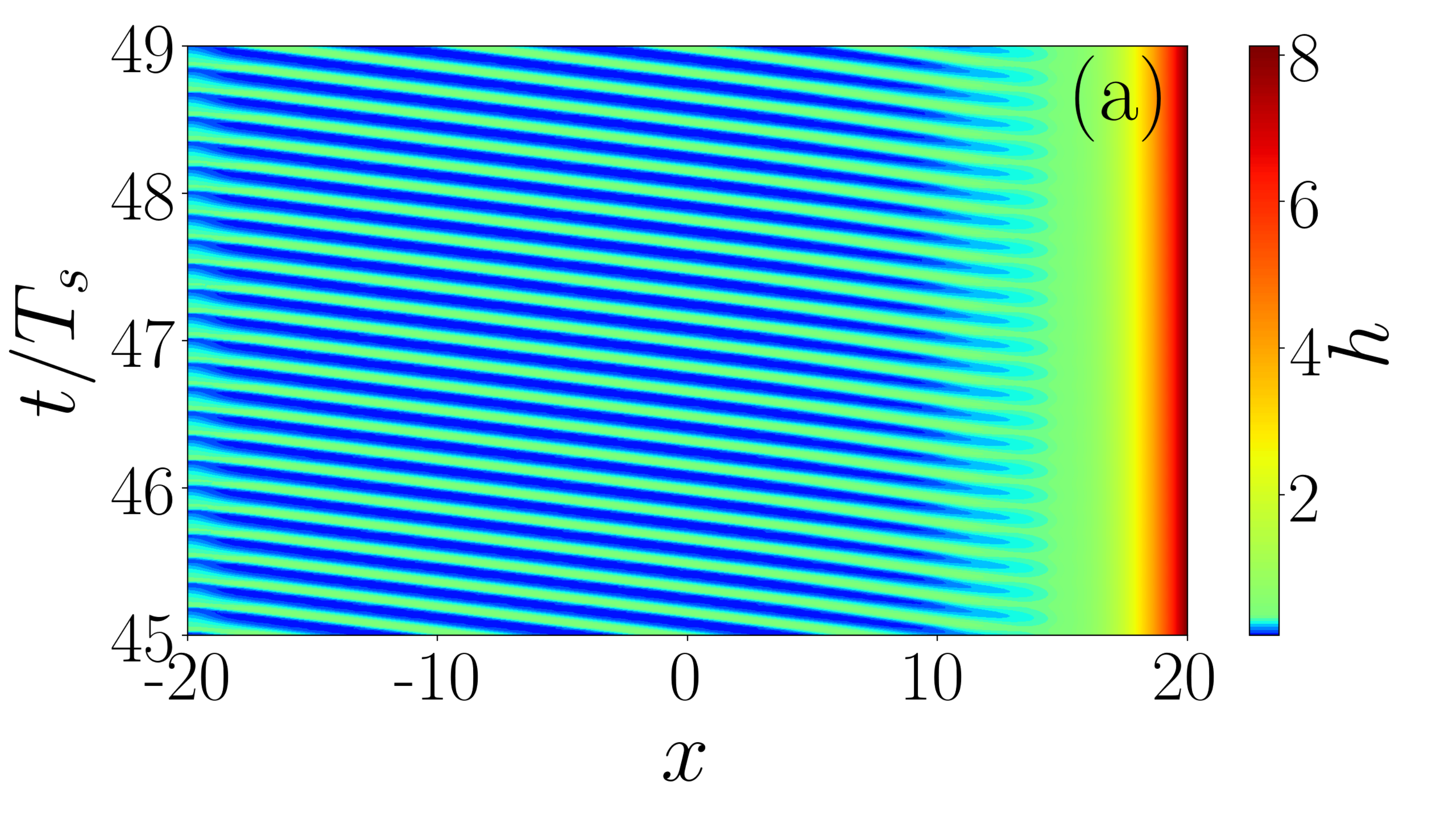}
    \includegraphics[width=0.45\textwidth]{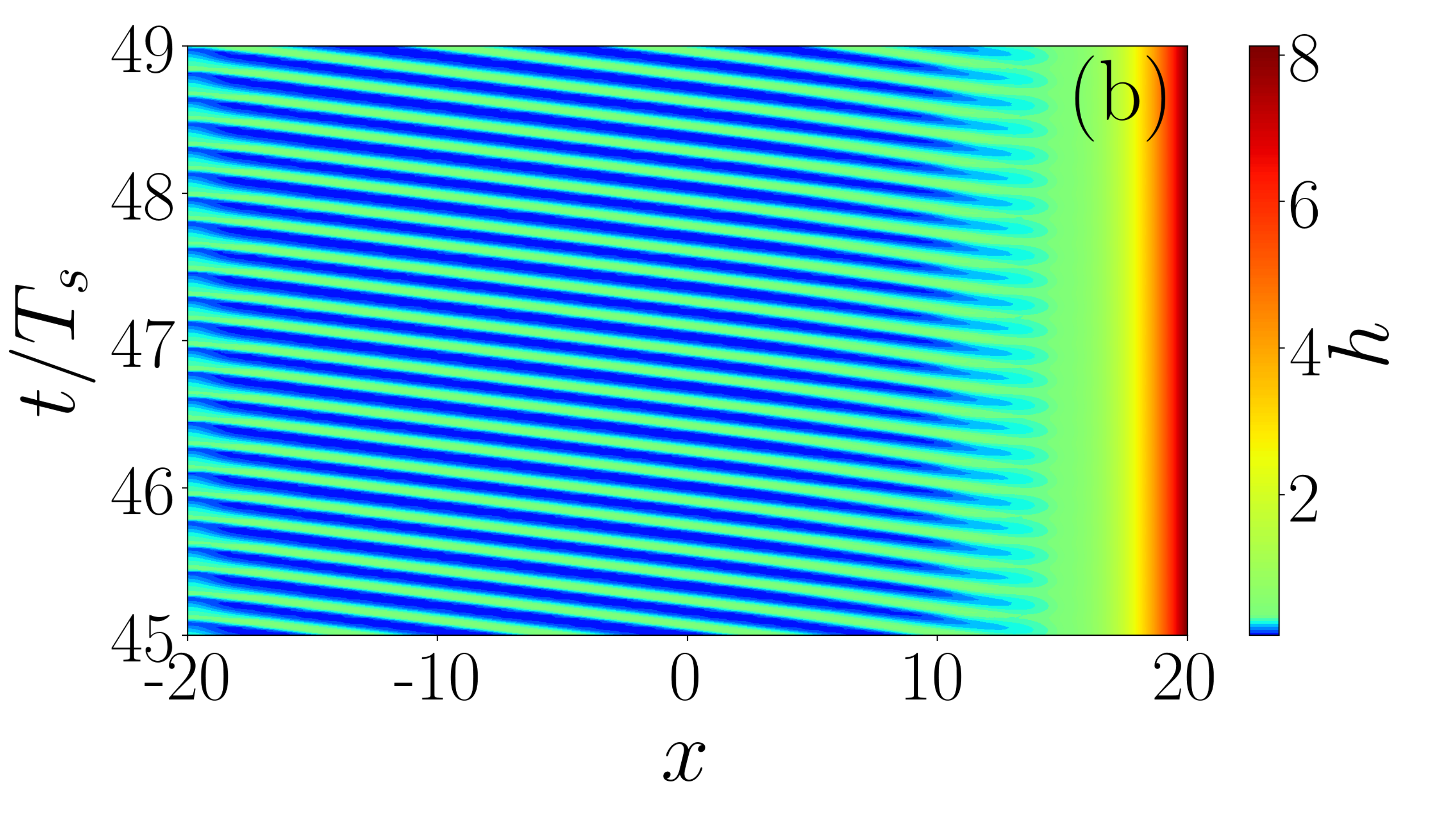}
    \includegraphics[width=0.45\textwidth]{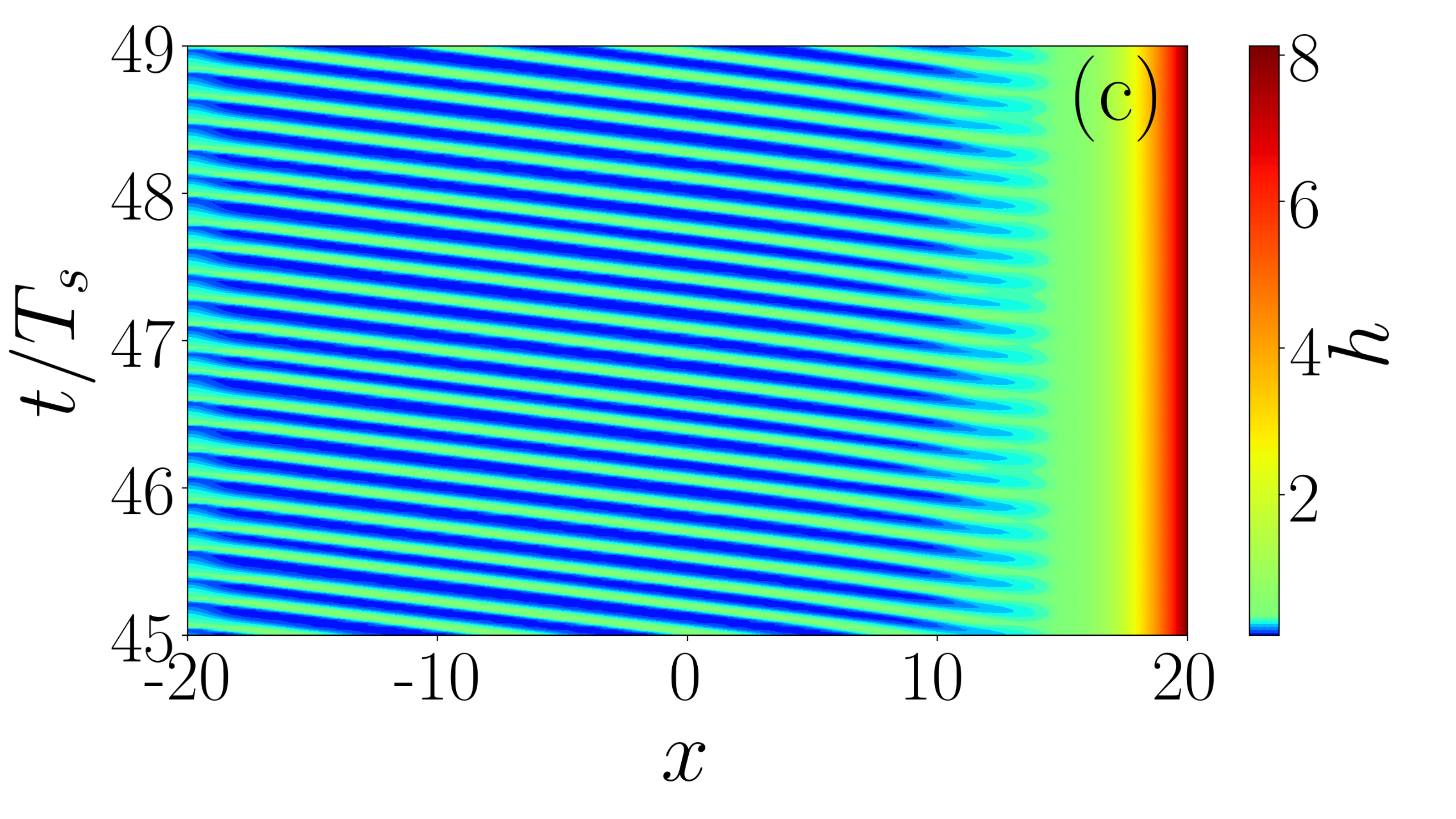}
    \includegraphics[width=0.45\textwidth]{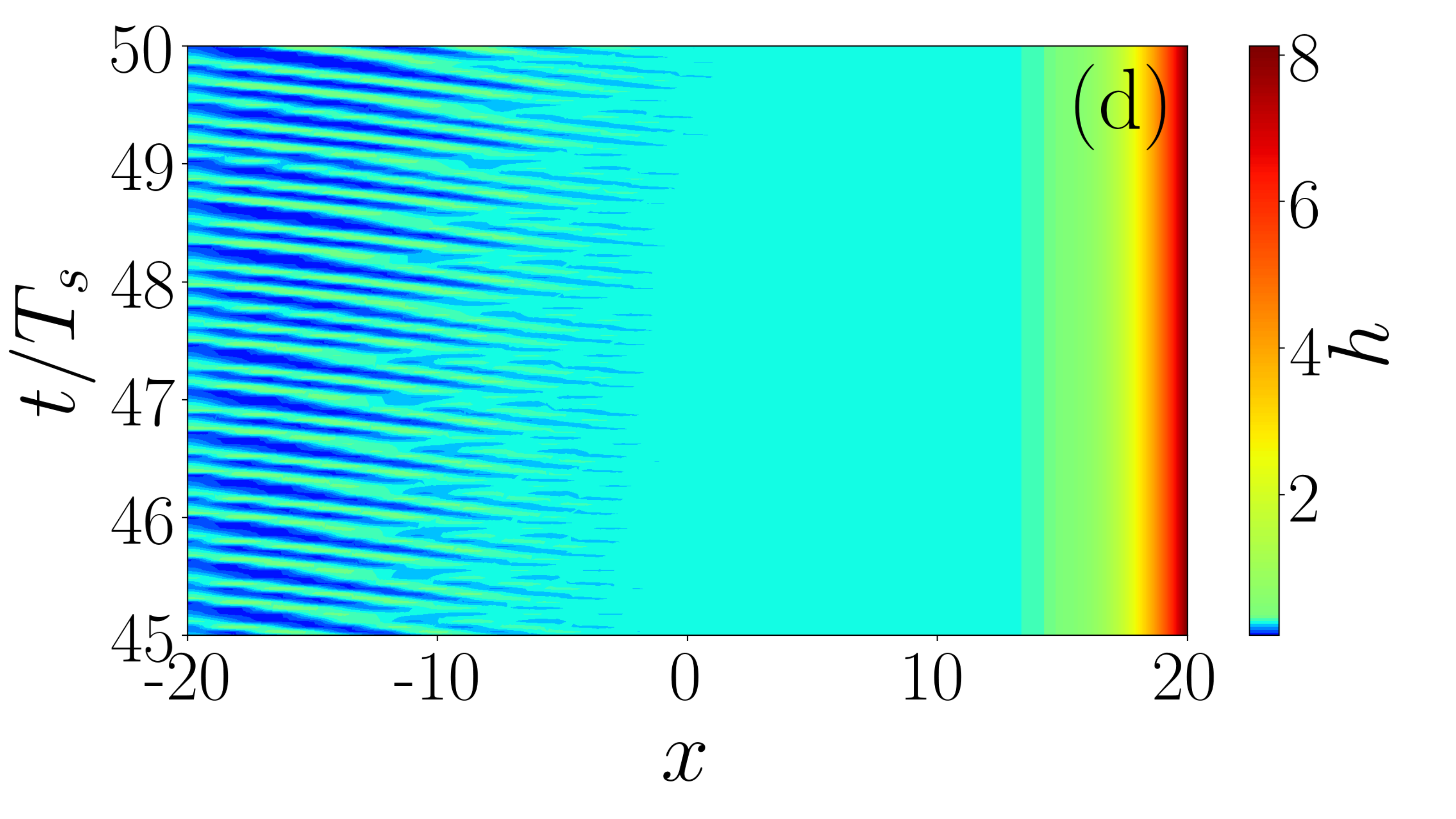}
\caption{
{Panels (a) to (d) present space-time plots of states marked ``a'' to ``d'' in Fig.~\ref{hp:fig:hopf_hc_we20_time_periode_zoom}, respectively. Shown is the behaviour after transients have settled.
They are at (a) $\epsilon_\mathrm{s} = 0.78247$, (b) $\epsilon_\mathrm{s} = 0.78231$, (c) $\epsilon_\mathrm{s} = 0.78207$ and (d) $\epsilon_\mathrm{s} = 0.78055$ and time is scaled by $T_s = 10^4$\,.
Only states in panels (a), (b) and (c) are truly time-periodic states, while (d) shows a chaotic state.}
}
\label{hp:fig:solutions_direct_numeric}
\end{figure}

\begin{figure}[hbt]
  \centering
\includegraphics[width=.485\textwidth]{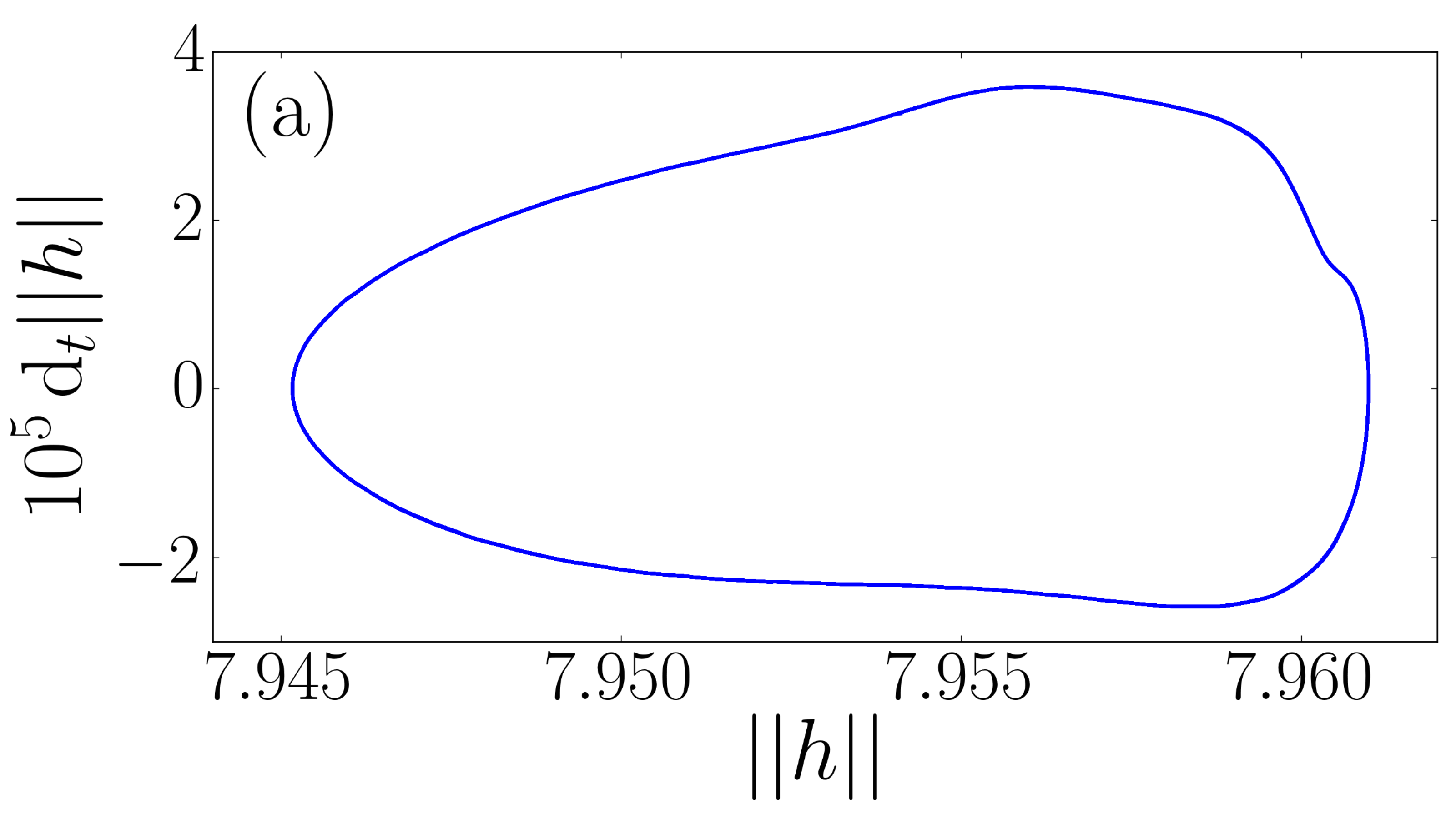}
\includegraphics[width=.485\textwidth]{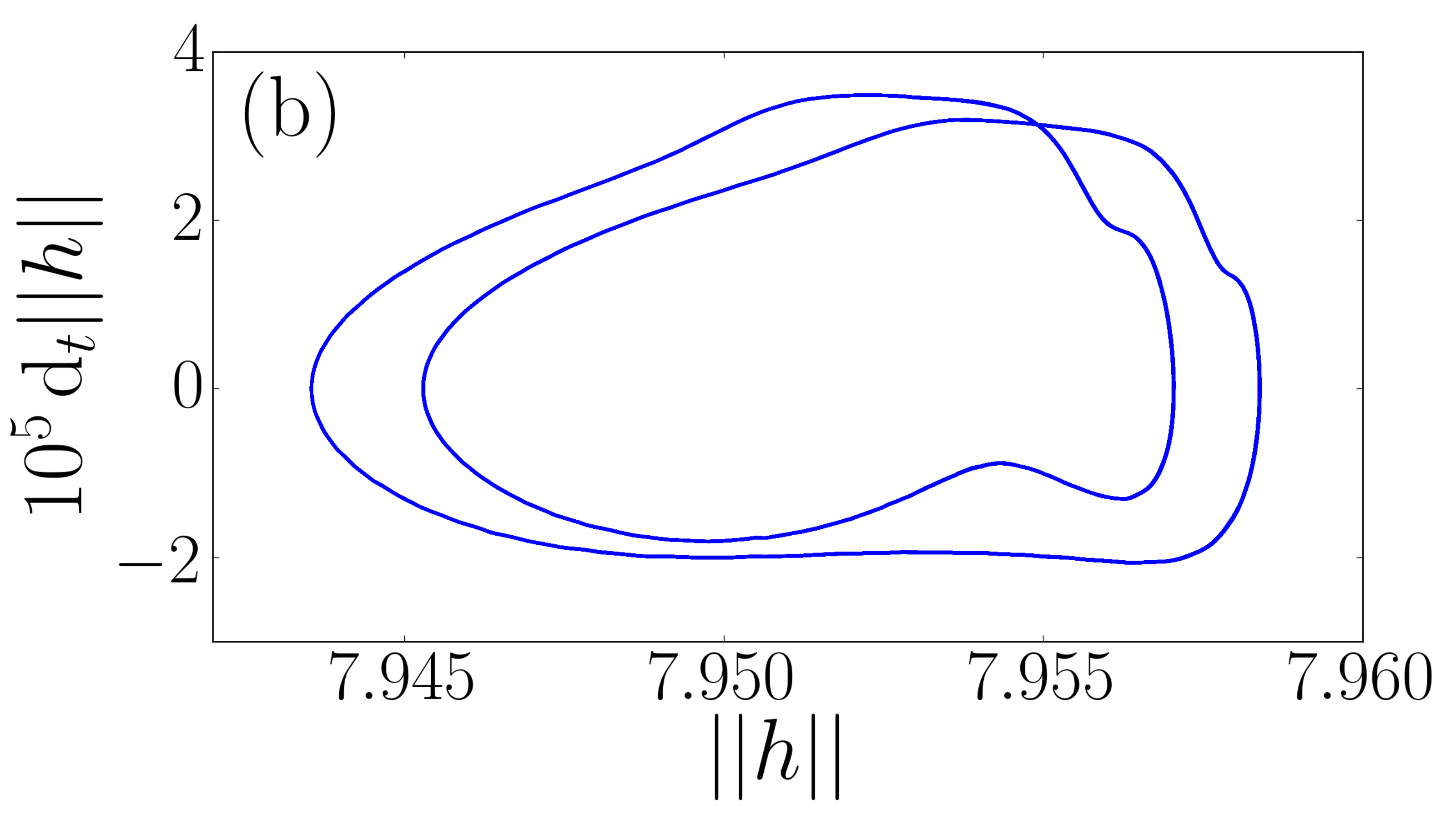}
\includegraphics[width=.485\textwidth]{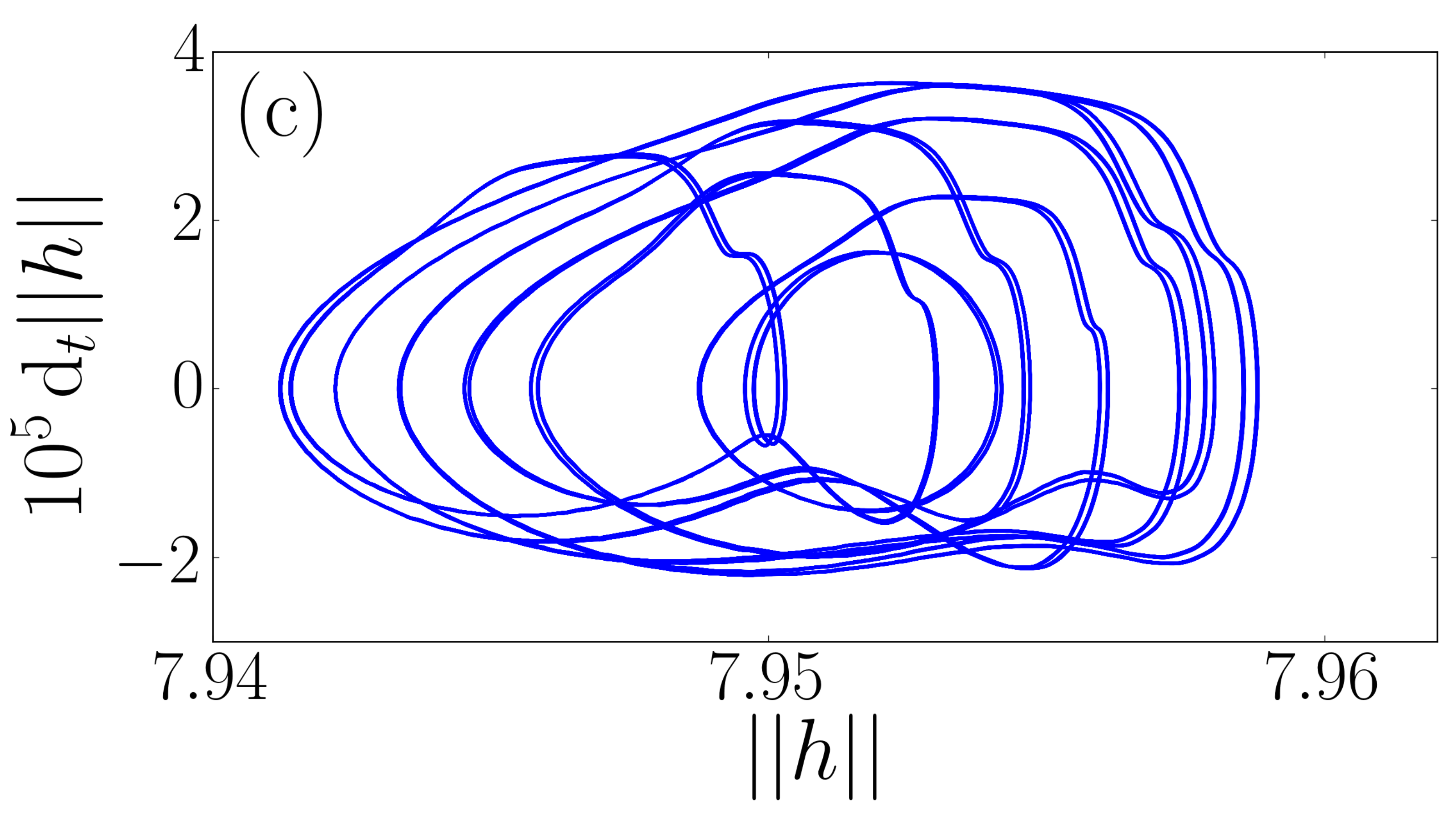}
\includegraphics[width=.485\textwidth]{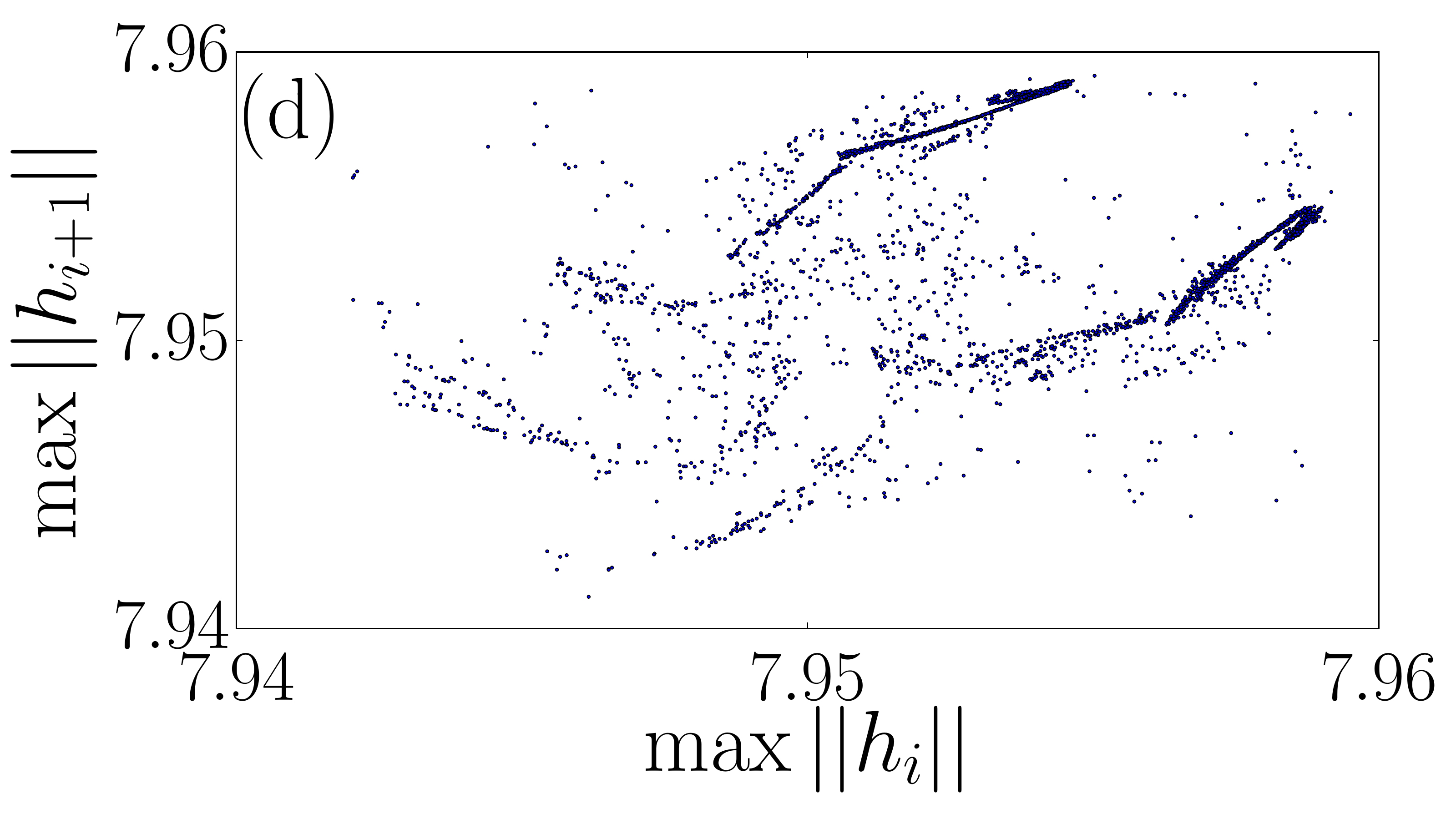}
\caption[Phase-plane plots where showing the time derivative of the L$^2$-norm versus the L$^2$-norm]{
(a) to (d) show phase portraits for the time evolutions in Figs.~\ref{hp:fig:solutions_direct_numeric}~(a) to (d), respectively. In particular, (a) to (c) give the trajectories in a plane spanned by 
  the L$^2$-norm ($||h||$) and its time derivative ($d_t||h||$), while (d) gives a return map where each maximum of the norm $||h_{i+1}||$ is plotted over the previous maximum $||h_{i}||$.}
\label{hp:fig:we_2_phase_plane_l2norm_a_to_d}
\end{figure}

{The TPS~I to~IV shown in Fig.~\ref{hp:fig:hopf_hc_we20_time_periode_zoom}~(b) are obtained by continuation. Using them as inititial conditions in time simulations show that they are actually unstable. 
However, their existence nevertheless indicates the existence of time-periodic transients and may also hint at stable more complicated oscillatory behaviour [cf.~\cite{KoTh2014n}]. 
We investigate this employing a ``primitive continuation scheme'', namely, we perform time simulations starting at various initial states obtained either by continuation or previous time simulations at neighboring parameter values. After transients have settled we obtain the states indicated by symbols in Fig.~\ref{hp:fig:hopf_hc_we20_time_periode_zoom}~(a). Theseare steady Landau-Levich film states (black circles), 
regular TPS (green squares), and irregular TPS that are seemingly chaotic (red triangles). Typical examples marked ``a'' to ``d'' in Fig.~\ref{hp:fig:hopf_hc_we20_time_periode_zoom} are given in the space-time plots in Figs.~\ref{hp:fig:solutions_direct_numeric}~(a) to (d), respectively.}

{Inspecting Fig.~\ref{hp:fig:solutions_direct_numeric} we see that ridges are periodically shed from a foot structure of different length, the latter being most pronounced in Fig.~\ref{hp:fig:solutions_direct_numeric}~(d). The shedding seems regular in Figs.~\ref{hp:fig:solutions_direct_numeric}~(a) to (c), but less so in Fig.~\ref{hp:fig:solutions_direct_numeric}~(d). We also note that most of the corresponding symbols in Fig.~\ref{hp:fig:hopf_hc_we20_time_periode_zoom}~(a) do not coincide with the continuation results.\footnote{This is the case even when taking into account the small systematic mismatch between continuation and time simulation results. It is caused by different underlying discretization schemes and different way of implementing the boundary conditions at the meniscus.}
  The reason becomes clear when analyzing the corresponding phase portraits: Fig.~\ref{hp:fig:we_2_phase_plane_l2norm_a_to_d}~(a) to (c) show trajectories in a plane spanned by  the L$^2$-norm ($||h||$) and its time derivative ($d_t||h||$) that clearly indicate that the visual similarity of Figs.~\ref{hp:fig:solutions_direct_numeric}~(a) to (c) hides the fact that a period doubling bifurcation occurrs when going from 
  Fig.~\ref{hp:fig:solutions_direct_numeric}~(a) [single loop in Fig.~\ref{hp:fig:we_2_phase_plane_l2norm_a_to_d}~(a)] to Fig.~\ref{hp:fig:solutions_direct_numeric}~(b) [double loop in Fig.~\ref{hp:fig:we_2_phase_plane_l2norm_a_to_d}~(b)], while several more occur when going from Fig.~\ref{hp:fig:solutions_direct_numeric}~(b) to Fig.~\ref{hp:fig:solutions_direct_numeric}~(c) [multiple loops in Fig.~\ref{hp:fig:we_2_phase_plane_l2norm_a_to_d}~(c)]. This explains why only the square symbols close to ``a'' in Fig.~\ref{hp:fig:hopf_hc_we20_time_periode_zoom} closely coincide with the continuation results while bend-like deviations occur  when going to ``b'', and again before ``c''. This corresponds to the first steps of a period doubling cascade explaining while most TPS obtained by continuation are unstable.
}
{The irregular time evolution in Fig.~\ref{hp:fig:solutions_direct_numeric}~(d) can not be well represented in the ($||h||$, $d_t||h||$)-plane as it densely fills it. Instead, Fig.~\ref{hp:fig:we_2_phase_plane_l2norm_a_to_d}~(d) gives a return map extracted from the time series of the norms: each maximum $||h_{i+1}||$ in the series is plotted over the previous maximum $||h_{i}||$. The resulting cloud of points indicates that the ridge shedding is chaotic as it resembles a strange attractor. Similar observations are made for line deposition in Langmuir-Blodgett transfer \cite{KGFT2012njp}. This indicates that close to the outermost Hopf bifurcation (where the red triangles, e.g., point ``d'' are located in  Fig.~\ref{hp:fig:hopf_hc_we20_time_periode_zoom}) multistability of chaotic shedding and stable Landau-Levich film deposition exists. This is further supported by the fact that in the simulations resulting in the black dots in Fig.~\ref{hp:fig:hopf_hc_we20_time_periode_zoom} sometimes intermittent bursts of chaotic shedding are observed close to the downstream boundary (not shown).}

%
%%%%%%%%%%%%%%%%%%%%%%%%%%%%%%%%%%%%%%%%%%%%%%%%%%%%%%%%%%%%%%%%%%%%%%%%%%%%%%%%%%%%%%%%%%%%%%%%
\section{Conclusion}
\label{sec:conclusion}
%%%%%%%%%%%%%%%%%%%%%%%%%%%%%%%%%%%%%%%%%%%%%%%%%%%%%%%%%%%%%%%%%%%%%%%%%%%%%%%%%%%%%%%%%%%%%%%%

{We have studied the bifurcation structure corresponding to a meniscus of partially wetting liquid} driven by surface acoustic waves (SAW). {The thin-film model combines elements of two literature models and employs a scaling that allows one to recover their respective scalings as limiting cases.  One is the precursor-based thin-film model for the classical Landau-Levich (or dragged plate) problem for partially wetting liquids that is studied in Refs.~\cite{GTLT2014prl,TsGT2014epje,TWGT2019prf}. The other one is the thin-film model of Ref.~\cite{MoMa2017jfm} for a SAW-driven meniscus without consideration of wettability, i.e., implicitely assuming a wetting liquid. In this way we have been able to} investigate both systems in a common framework, first to reproduce selected results for each of them, and then to look at the behavior of the SAW-driven system in the case of partially wetting liquids in direct comparison to results obtained for the Landau-Levich system. Our investigation has been based on numerical path continuation techniques {that allow one to determine the full bifurcation behaviour of steady and time-periodic states including the unstable states. Additionally, time simulations have been used to illustrate particular behaviour.}

We have focused on one-dimensional substrates, i.e., we have neglected transverse perturbations, and investigated how the various occurring steady and time-periodic states depend on the Weber number and the SAW strength. The full bifurcation structure related to qualitative transitions caused by the SAW has been described with a particular attention on Hopf bifurcations related to the emergence of {transient or sustained time-periodic behaviour. The latter correspond to the regular shedding of liquid ridges from the meniscus.}
The interplay of several of these bifurcations has been investigated by tracking them in selected parameter planes, thereby discussing the codimension-2 bifurcations where they emerge and disappear. In this way, our study has provided detailed information relevant to the entire class of dragged-film and forced-meniscus problems. It well aligns with and extends results on the bifurcation structure obtained for the mentioned Landau-Levich problem for partially wetting liquids \cite{GTLT2014prl,TWGT2019prf} as well as for line deposition in Langmuir-Blodgett transfer \cite{KGFT2012njp,KoTh2014n}. 

Compared to the Landau-Levich problem for partially wetting liquid \cite{GTLT2014prl}, in the SAW-driven case, we have identified the same type of basic states, namely, a meniscus state, where a contact line region separates a meniscus and a precursor film, a foot state, where a foot-like structure protrudes from the meniscus and a Landau-Levich film state where a relatively thick homogeneous film is coating the substrate. Its thickness follows a similar power law  as in the classical Landau-Levich system. 
However, in contrast to the Landau-Levich problem, in the SAW-driven case the foot states show rather strong thickness  modulations along the foot and a more continuous transition towards the precursor layer at the tip of the foot. Note that such modulations can exist in the dragged-plate case, but are much weaker and are centered about a well defined foot thickness  \cite{TsGT2014epje}. {The individual transitions between the three types of steady states occur in a similar manner when either changing the Weber number or the SAW strength.} However, the transitions differ when comparing them to the dragged-plate system. In the present case, the foot states are situated on a {limited snaking part} of the bifurcation curve. Virtually, all of them are unstable, while in the dragged case {exponential snaking continues without limit and the corresponding states consecutively switch between stable and unstable.} The continuous transition between the snaking part of the bifurcation curve and the part on the same curve that corresponds to Landau-Levich films is a feature of the SAW-driven case that has not been observed in the dragged-plate case. There, a snaking curve always diverges at finite driving accompanied by a diverging foot length. This is not the case here.

Overall, the bifurcation structure has turned out to be rather involved, the general tendency being that its complexity increases with increasing Weber number and also with decreasing wettability. More and more saddle-node and Hopf bifurcations appear together with the branches of time-periodic states emerging from the latter. Such branches were recently also observed for the dragged-plate system \cite{TWGT2019prf}. {However, there the emergence of the entire set of branches of such states is not discussed. Our results shown how and where these branches appear. This occurs either through double Hopf bifurcations or Bogdanov-Takens bifurcations} that also create global bifurcations. {We expect these findings to be generic for the class of coating systems where partial wettability plays a role.} On a qualitative level, our results are similar to the ones for Langmuir-Blodgett transfer {involving substrate-mediated phase transition into the liquid-condensed phase of the transfered surfactant \cite{KoTh2014n}. There, the role of wetting energy is taken by the free energy related to the phase transitions between different surfactant phases.} This further strengthens the point made  in \cite{Thie2014acis,WTGK2015mmnp} that the investigation of generic models for such systems as in \cite{GoSc2015arma} is rather important.

{Note that in our study of the SAW-driven meniscus we have restricted our attention to a single representative domain size. All steady states are independent of system size, if an appropriate solution measure is used that does not depend on the size of the meniscus. In contrast, the number and exact location of the Hopf bifurcation points will depend on system size as more spatial modes fit into larger domains.
Taking, for instance, the example of Fig.~\ref{hp:fig:hopf_branch_ha1_we2_hc_eps} at $\mathrm{We}_\mathrm{s}=2$ where 18 Hopf bifurcations occur at the chosen domain size $L=40$. Increasing [decreasing] the size to $L=80$ [$L=20$] 
increases [decreases] the number of Hopf bifurcations to 42 [six]. However, this does not make the results arbitrary as the range where Hopf bifurcations occur is well defined: 
  The location of the rightmost (and arguably the most important) Hopf bifurcation is in such systems well predicted by a marginal stability analysis \cite{KGFT2012njp,TWGT2019prf} independent of domain size.
  Furthermore, also the locus of the very first Hopf bifurcation (encountered when following the branch of steady states starting at small $\epsilon_\mathrm{s}$) is very robust. When going from $L=40$ to $L=20$ or to $L=80$ the variations in these loci are about 0.1\%.}

Our results show that a very finely tuned SAW-driven system could possibly be switched between the deposition of a homogeneous film and the deposition of line or droplet patterns and control by SAW could be used together with other means of control as the usage of prestructured plates \cite{zhu2016branch,Wilczek_2016} or variable plate dragging speed \cite{ly2019effects}. To study the application of SAW control to deposition from solutions or suspensions with volatile solvent \cite{MZAM2016l}, solute dynamics and solvent evaporation have to be incorporated into the model by amending corresponding models reviewed in \cite{CrMa2009rmp,Thie2014acis,Thie2018csa}.  

However, our results also indicate that the parameter region where all these changes occur is very small and might be difficult to access with the present experimental techniques. Therefore, our results should not be taken as predictive in a quantitative way but as a catalogue of qualitative transitions expected in a variety of coating systems.  As discussed in the conclusion of Ref.~\cite{TWGT2019prf}, there exists a number of experimental systems that show related qualitative transitions. These include (i) water drops sliding on an oil film which can be destabilized by an applied voltage and transform into many small oil droplets underneath the water drop \cite{StMu2006prl} (also see \cite{BRML2019sm}); 
(ii) recent experiments on gas bubbles that move through a tube filled with partially wetting liquid where the liquid film between bubble and wall may undergo related instabilities \cite{KSPK2018prf}, and  (iii) the different dynamic regimes of relatively thick viscous liquid films flowing down a cylindrical fiber \cite{JFSZ2019jfm}. For the detailed discussion of these experiments in the context of time-periodic states in the dragged-plate case see the conclusion of Ref.~\cite{TWGT2019prf}. In all these systems, SAW may be employed to stabilize the Landau-Levich(-Bretherton) films.

Finally, we stress again that our investigation has been entirely focused on one-dimensional systems. However, driven contact lines and deposition dynamics often shows transverse instabilities \cite{Thie2014acis}. These have been excluded from our study. A first investigation \cite{LMTG2020pd} has shown, that there exist bifurcations that break the transverse invariance and result in branches of fully two-dimensional states that represent a rich set of structures. However, a systematic study of their properties is a  formidable future endeavour that should also aim at an understanding of time-periodic two-dimensional states occurring in various coating systems.

\section*{Acknowledgement}
\label{sec:acknowledge}

We acknowledge support by the German-Israeli Foundation for Scientific Research and Development (GIF, Grant No. I-1361-401.10/2016), as well as further support by the Deutsche Forschungsgemeinschaft (DFG; Grant No.~TH781/8-1). % Coating
We also thank Sebastian Engelnkemper for the creation of a first tutorial on \textsc{pde2path} usage for dragged-film problems, as well as Tobias Frohoff-H\"ulsmann and Simon Hartmann for support with \textsc{pde2path} and \textsc{oomph-lib} implementations, respectively.

\appendix
\section{Nondimensionalization}
\label{sec:appendix}
Here, we discuss our scaling and relate it to the respective scalings employed in Refs.~\cite{MoMa2017jfm} and \cite{GTLT2014prl}.
Our starting point is the dimensionless thin-film equation given by Eq.~(2.23) of Ref.~\cite{MoMa2017jfm}
\begin{equation}
 \partial_t h = - \partial_x \left[ \frac{h^3}{3\mathrm{We}_\mathrm{s}} \partial_{xxx} h + h v_\mathrm{s}(h) \right].
 \label{appendix:eq:moma_saw_eq_nondim}
\end{equation}
Re-introducing their scales
\begin{equation}
\tau=\frac{L}{\chi U\mathrm{Re}},\qquad \delta,\quad\mathrm{and}\quad L=\frac{\delta}{\epsilon},\qquad
\end{equation}
for time $t$, film thickness $h$, and $x$-coordinate, respectively, and using the dimensionless numbers
\begin{equation}
  \mathrm{We}_\mathrm{s}=\frac{\chi\mathrm{Re}\mathrm{Ca}}{\epsilon^3},\quad
\mathrm{Re}=\frac{\rho \tilde U \delta}{\mu}
  \quad\mathrm{and}\quad \mathrm{Ca}=\frac{\mu \tilde U}{\gamma},
\end{equation}
we obtain the dimensional evolution equation 
\begin{equation}
 \partial_t h = - \partial_x \left[ \frac{h^3}{3\mu} \gamma\partial_{xxx} h + U_w\, h v_\mathrm{s}(h) \right]
 \label{appendix:eq:dim_thin_film_short}
\end{equation}
where now $x$, $t$, and $h$ are dimensional and  $U_w=\rho\chi\delta \tilde U^2/\mu$ is a typical velocity.

The complete dimensional thin-film evolution equation that governs the standard Landau-Levich problem  
for a partially wetting liquid with additional SAW driving and a dragged plate is given by
\begin{align}
 \partial_t h &=- \partial_x \left\{ \frac{  h^3}{3 \mu} \left[ \partial_x  \left(\gamma\partial_{ x  x}h + \kappa\Pi(h) - \rho g h \right)+ \tilde\alpha \rho g \right] + U_w \, h v_{\mathrm{s}} (h) + u h\right\}
 \label{appendix:eq:full_model_dim}
\end{align}
where $\Pi(h)$ is a dimensionless Derjaguin pressure and $\kappa$ the related energy density scale. The parameter $\tilde\alpha$ is the physical inclination angle and $u$ is the velocity of the moving plate. 

Now we introduce the scales $\tau=L/\nu$, $\delta$ and $L=\delta/\epsilon$ for time, film thickness, and $x$-coordinate, respectively. Here $\nu$
is a generic velocity scale that can be specified when comparing to the different scalings used in the literature.  We obtain the nondimensional equation
\begin{align}
 \partial_t h &=- \partial_x \left\{ \frac{  h^3}{3} \partial_x  \left(\frac{1}{\mathrm{We}_\mathrm{s}}\partial_{ x  x}h + \mathrm{Ha}\Pi(h)\right) + G\frac{  h^3}{3}\left(\alpha- \partial_x h \right) +  \epsilon_\mathrm{s} \, h v_{\mathrm{s}} (h) +U_0 h\right\}
 \label{appendix:eq:full_model_nondim}
\end{align}
{where $\alpha=\tilde\alpha/\epsilon$ is the $O(1)$ inclination angle in long-wave scaling and the nondimensional numbers are defined as
\begin{equation}
\frac{1}{\mathrm{We}_\mathrm{s}}=\frac{\epsilon^3\gamma}{\mu\nu},\qquad 
\mathrm{Ha}=\frac{\epsilon\kappa\delta}{\mu\nu},\qquad 
G=\frac{\epsilon\rho g\delta^2}{\mu\nu},\qquad 
\epsilon_\mathrm{s}=\frac{U_w}{\nu},\qquad 
U_0=\frac{u}{\nu}.
\end{equation}
}%
Here, $\mathrm{We}_\mathrm{s}$ is a Weber or an inverse Capillary number, $\mathrm{Ha}$ is a Hamaker number, $G$ is a Gravity or Galileo number, 
$\epsilon_\mathrm{s}$ a velocity ratio corresponding to a SAW strength, and $U_0$ a velocity ratio corresponding to plate dragging. 
Keeping $\nu$ general, i.e., not selecting any leading balance, keeps all five nondimensional numbers in equation~(\ref{appendix:eq:full_model_nondim}) {and corresponds to Eq.~\eqref{eq:LL} of the main text.}

For different specific $\nu$ one recovers various scalings employed in the literature. For instance, using $\nu=\epsilon^3\gamma/\mu$ yields
{\begin{equation}
\mathrm{We}_\mathrm{s}=1,\qquad 
\mathrm{Ha}=\frac{\kappa L}{\epsilon\gamma},\qquad 
G=\frac{\rho g L^2}{\gamma},\qquad 
\epsilon_\mathrm{s}=\frac{U_w\mu}{\epsilon^3\gamma},\qquad 
U_0=\frac{u\mu}{\epsilon^3\gamma}.
\end{equation}
Fixing $\mathrm{Ha}=1$ by choosing $L=\epsilon\gamma/\kappa$ and setting $\epsilon_\mathrm{s}=0$ one recovers Eq.~\eqref{eq:LL} of \cite{GTLT2014prl},
corresponding to the case in the present section~\ref{sec:draggedfilmwithouSAW}.}

{Alternatively, using $\nu=U_w$ gives
\begin{equation}
\frac{1}{\mathrm{We}_\mathrm{s}}=\frac{\epsilon^3\gamma}{\mu U_w},\qquad 
\mathrm{Ha}=\frac{\epsilon\kappa\delta}{\mu U_w},\qquad 
G=\frac{\epsilon\rho g\delta^2}{\mu U_w},\qquad 
\epsilon_\mathrm{s}=1,\qquad 
U_0=\frac{u}{U_w}.
\end{equation}
With $\mathrm{Ha}=G=U_0=0$ one recovers Eq.~(2.23) of \cite{MoMa2017jfm}. Note that there $L$ and $\delta$ are chosen by geometry and SAW-intrinsic length, 
respectively. This corresponds to the case in the present section~\ref{sec:dependenceonthewebernumber}. In the remaining sections we keep the general scaling to facilitate
comparison with the behaviour in the limiting cases of  Refs.~\cite{MoMa2017jfm} and \cite{GTLT2014prl}.}

%%%%%%%%%%%%%%%%
% Bibliography
%%%%%%%%%%%%%%%%
\bibliography{bib_all}

\end{document}